\documentclass{aa}  

\usepackage[hyperindex=true,colorlinks=true,citecolor=blue,linkcolor=blue,breaklinks=true]{hyperref}
\usepackage{amsmath}
\usepackage{graphicx}
\usepackage{txfonts}
\usepackage{natbib,twoopt}
\usepackage{xcolor}
\usepackage{multirow}

\definecolor{red}{rgb}{1.,0.0,0.}

\bibpunct{(}{)}{;}{a}{}{,} 
\newcommandtwoopt{\citeads}[3][][]{\href{http://adsabs.harvard.edu/abs/#3}%
	{\citealp[#1][#2]{#3}}} 
\newcommandtwoopt{\citepads}[3][][]{\href{http://adsabs.harvard.edu/abs/#3}%
	{\citep[#1][#2]{#3}}} 
\newcommandtwoopt{\citetads}[3][][]{\href{http://adsabs.harvard.edu/abs/#3}%
	{\citet[#1][#2]{#3}}}
\newcommandtwoopt{\citeyearads}[3][][]%
{\href{http://adsabs.harvard.edu/abs/#3}{\citeyear[#1][#2]{#3}}}

\usepackage{etoolbox}
\makeatletter
\patchcmd\@combinedblfloats{\box\@outputbox}{\unvbox\@outputbox}{}{%
	\errmessage{\noexpand\@combinedblfloats could not be patched}%
}%
\makeatother

\begin{document}
	
	\title{Multiplicity of Galactic Cepheids from long-baseline interferometry}
	\titlerunning{Multiplicity of Galactic Cepheids from long-baseline interferometry}
	
	\subtitle{IV.~New detected companions from MIRC and PIONIER observations\thanks{Based on observations made with ESO telescopes at Paranal and La Silla observatory under program IDs 091.D-0041, 094.D-0170, 190.D-0237(A) and 091.D-0469(A).}}
	\author{ A.~Gallenne\inst{1,2},
		P.~Kervella\inst{3}, 
		S.~Borgniet\inst{3},
		A.~M\'erand\inst{4},
		G.~Pietrzy\'nski\inst{5},
		W.~Gieren\inst{6,7},
		J.~D.~Monnier\inst{8},
		G.~H.~Schaefer\inst{9},
		N.~R.~Evans\inst{10},
		R.~I.~Anderson\inst{4},
		F.~Baron\inst{8,11},
		R.~M.~Roettenbacher\inst{12,8}
		\and
		P. Karczmarek\inst{13}
	}
	
	\authorrunning{A. Gallenne et al.}
	
	\institute{European Southern Observatory, Alonso de C\'ordova 3107, Casilla 19001, Santiago, Chile
		\and Laboratoire Lagrange, UMR7293, Universit\'e de Nice Sophia-Antipolis, CNRS, Observatoire de la C\^ote dAzur, Nice, France
		\and LESIA, Observatoire de Paris, CNRS UMR 8109, UPMC, Universit\'e Paris Diderot, 5 Place Jules Janssen, F-92195 Meudon, France
		\and European Southern Observatory, Karl-Schwarzschild-Str. 2, 85748 Garching, Germany
		\and Nicolaus Copernicus Astronomical Centre, Polish Academy of Sciences, Bartycka 18, PL-00-716 Warszawa, Poland
		\and Universidad de Concepci\'on, Departamento de Astronom\'ia, Casilla 160-C, Concepci\'on, Chile
		\and Millenium Institute of Astrophysics, Santiago, Chile
		\and Astronomy Department, University of Michigan, Ann Arbor, MI 48109, USA
		\and The CHARA Array of Georgia State University, Mount Wilson CA 91023, USA
		\and Smithsonian Astrophysical Observatory, MS 4, 60 Garden Street, Cambridge, MA 02138, USA
		\and Department of Physics and Astronomy, Georgia State University, Atlanta, GA 30303, USA
		\and Yale Center for Astronomy and Astrophysics, Department of Physics, Yale University, New Haven, CT, 06520 USA
		\and Obserwatorium Astronomiczne, Uniwersytet Warszawski, Al. Ujazdowskie 4, 00-478, Warsaw, Poland
	}

	\offprints{A. Gallenne} \mail{agallenn@eso.org}
	
	
	
	\abstract
	{}
	{We aim at detecting and characterizing the main-sequence companions of a sample of known and suspected Galactic binary Cepheids. The long-term objective is to accurately and independently measure the Cepheid masses and the distances.}
	{We used the multi-telescope interferometric combiners CHARA/MIRC and VLTI/PIONIER to detect and measure the astrometric positions of the high-contrast companions orbiting 16 bright Galactic Cepheids. We made use of the \texttt{CANDID} algorithm to search for the companions and set detection limits from interferometric observations. We also present new high-precision radial velocity measurements which were used to fit radial pulsation and orbital velocities.}
	{We report the detection of the companions orbiting the Cepheids U~Aql, BP~Cir, and S~Mus, possible detections for FF~Aql, Y~Car, BG~Cru, X~Sgr, V350~Sgr, and V636~Sco, while no component is detected around U~Car, YZ~Car, T~Mon, R~Mus, S~Nor, W~Sgr and AH~Vel. For U~Aql and S~Mus, we performed a preliminary orbital fit combining their astrometric measurements with newly obtained high-precision single-line radial velocities, providing the full set of orbital elements and pulsation parameters. Assuming the distance from a period-luminosity (P-L) relation for both Cepheids, we estimated preliminary masses of $M_\mathrm{U\,Aql} = 4.97\pm0.62\,M_\odot$ and $M_\mathrm{S\,Mus} = 4.63\pm0.99\,M_\odot$. For YZ~Car, W~Sgr, V350~Sgr, and V636~Sco, we revised the spectroscopic orbits using new high-precision radial velocities, while we updated the pulsation parameters for BP~Cir, BG~Cru, S~Nor and AH~Vel. Our interferometric observations also provide measurements of the angular diameters, that can be used in a Baade-Wesselink type analysis.}
	{We have now several astrometric detections of  Cepheid companions. When radial velocities of the companions will be available, such systems will provide accurate and independent masses and distances. Orbital parallaxes with an accuracy better than 5\,\% will be particularly useful for a better calibration of the P-L relation. The final Gaia parallaxes will be also particularly helpful for single-line spectroscopic systems, where mass and distance are degenerate. Mass measurements are necessary for a better understanding of the age and evolution of Cepheids.}
	
	
	\maketitle
	
	%
	
	\section{Introduction}
	
	
	Classical Cepheids are mainly known as primary distance indicators in the local Universe thanks to the period-luminosity relation discovered about a century ago \citep{Leavitt_1908__0, Leavitt_1912_03_0}. Moreover, these pulsating intermediate-mass stars provide fundamental constraints for studying pulsation and evolution models \citep[see e.g.][]{Anderson_2016_06_0,Gillet_2014_08_0,Neilson_2014_03_0,Prada-Moroni_2012_04_0,Bono_2006__0}. Now with the actual precise photometric and spectroscopic instruments, finer structures of Cepheids are being revealed \citep[see e.g.][]{Derekas_2017_01_0,Evans_2015_02_0,Anderson_2014_06_0}, providing new information on these standard candles. Yet, despite all instrument improvements, direct measurements of the most fundamental stellar property, the mass, are still missing for most of the Cepheids. It is necessary to obtain an accurate and model-independent estimate of this parameter as the overall lifetime and behaviour of a star is determined primarily by its mass. Measurements of Cepheid masses are even more critical to constrain the long-standing mass discrepancy problem, still not really understood. The most cited scenarios to explain this discrepancy between masses predicted by stellar evolutionary and pulsation models are a mass-loss mechanism during the Cepheid's evolution, convective core overshooting, and rotation during the main-sequence stage \citep{Anderson_2014_04_0,Neilson_2011_05_0,Keller_2008_04_0,Bono_2006__0}. 
	
	Cepheid masses are usually derived through the companion mass which is itself inferred from a mass-temperature relation. The estimated masses are therefore model-dependent. A few dynamical masses of Cepheids were measured in the Large Magellanic Cloud from eclipsing binary systems \citep{Pilecki_2018_07_0,Gieren_2014_05_0,Pilecki_2013_12_0,Pietrzynski_2010_11_0}, and already provide some insight to settle the discrepancy between pulsation and evolution models. In our Galaxy, only the mass of Polaris and V1334~Cyg has been measured through the combination of astrometric and spectroscopic measurements of its close companion \citep{Gallenne_2018_11_0,Evans_2018_08_1,Evans_2008_09_0}. Otherwise, all companions were detected from the orbital effects on the radial velocities (RVs) and the variability of the systemic velocity, which provided the first information about the orbits. However, only massive and close components can be detected this way \citep[see e.g.][]{Moore_1929_02_0,Abt_1959_11_0,Szabados_1989_01_0,Szabados_1991_01_0}. Spectra from the International Ultraviolet Explorer (IUE) also provided useful information such as the spectral type of some companions \citep[see e.g][]{Bohm-Vitense_1985_09_0,Evans_1992_01_0}, but this also suffers from observational bias as only UV bright companions can be detected. However, UV wavelengths seem to be the best option to detect lines from the orbiting companions as most of them are main-sequence B stars, although the task is complicated as they are often fast rotators with very broad lines. At longer wavelengths, the Cepheid brightness outshines the companion, making challenging any detection with the current available instruments. 
	
	Recently, \citet{Kervella_2018__0} searched for close-in companions from the signature of their orbital motion on the proper motion vector, i.e. by comparing the Hipparcos and Gaia proper motions (PMs) to the mean PM. This work revealed a significant number of new candidate companions, and indicated a high binarity fraction (close to 100\,\%). For some of them with a known spectroscopic orbit, the combination of the PM and the distance (assumed from a P-L relation) allowed the determination of the full set of orbital elements. With an assumed mass for the Cepheids, it also provides approximate masses of their companions.
	
	Another issue in detecting Cepheid companions is the angular separation ($\lesssim 50$\,mas) which makes difficult the detection from high-contrast imaging instruments. Adaptive-optics (AO) works in the infrared, and the flux ratio, $f$, at these wavelengths between the companion and the Cepheid is only a few percent ($f < 1$\,\% in most cases), making impossible to directly detect faint components located within 100\,milliarcseconds (mas). A few wider companions, however, can be spatially resolved with or without AO, as demonstrated by \citet{Gallenne_2014_07_0} and \citet{Evans_2013_10_0}. A possible solution to detect such faint components in the separation range $50 < r < 100$\,mas would be to use aperture masking, which can yield contrast $\Delta K \sim 6$\,mag at $\lambda /D$ \citep{Kraus_2012_01_0}. The best technique for $r \lesssim 50$\,mas is long-baseline interferometry (LBI), which can reach contrast $\Delta H \sim 6$\,mag. \citet{Gallenne_2015_07_0,Gallenne_2014_01_0,Gallenne_2013_04_0} proved the efficiency of LBI by detecting close faint companions of Cepheids, down to $\Delta H = 5.3$\,mag \citep[$f = 0.8$\,\%,][]{Gallenne_2014_01_0}. \citet{Gallenne_2015_07_0} also detected a companion as faint as $f = 0.22$\,\%, which is the faintest companion detected so far by interferometry, but still needs to be confirmed.
	
	The advantage of spatially resolving binary Cepheids lies in the combination of the astrometric and spectroscopic measurements which provides an independent and reliable way to determine stellar masses and distances. But the mass and the distance are degenerate parameters if the system is not a double-line spectroscopic binary. This is the case for most of the binary Cepheids as usually RV measurements are made in the $V$ band, and only provide RVs of the Cepheid (the primary). As mentioned previously, RV measurements in the UV are more favourable to detect lines of the companions, and combining such observations with astrometry will provide both the Cepheid mass and distance. However, broad features (many are fast rotators) and blended lines in the spectra of the companions complicate the analysis, and prevent the determination of accurate RVs. In a recent work, we succeeded in determining RVs of the companion orbiting the Cepheid V1334~Cyg using the Space Telescope Imaging Spectrograph (STIS) on board the Hubble Space Telescope (HST). This work provides the most accurate distance for a Cepheid at a 1\,\% accuracy level, and masses with 3\,\% accuracy \citep{Gallenne_2018_11_0}. For systems where RVs measurements of the companions are still not possible, the future Gaia parallaxes will allow us to break the degeneracy between mass and distance, and we will then be able to combine interferometry with single-line velocities to estimate dynamical masses of many Cepheids.
	
	We are engaged in a long-term interferometric observing program to detect and follow-up close faint companions orbiting the brightest Galactic Cepheids. This program already provided new information for four Cepheids and their respective companions \citep{Gallenne_2015_07_0,Gallenne_2014_01_0,Gallenne_2013_04_0}. We are also engaged in an ultraviolet and visible spectroscopic observing campaign which aims at detecting the companion lines and measured contemporaneous high-precision RVs.
	
	In this fourth paper, we report new interferometric observations of 16 Galactic Cepheids. We used the multi-telescope combiners CHARA/MIRC and VLTI/PIONIER which offer the best $(u, v)$ coverage and sensitivity to detect faint companions in both the northern and southern hemisphere. The observations and data reduction are described in Sect.~\ref{section__observation_and_data_reduction}. In Sect.~\ref{section__companion_search_and_sensitivity_limit}, we used the \texttt{CANDID}\footnote{Available at \url{https://github.com/amerand/CANDID}} tool \citep{Gallenne_2015_07_0} to search for companions using all available observables and then derive the detection sensitivity. We discuss our results for each individual binary Cepheid in Sect.~\ref{section__discussion}, in which we also analyse new high-precision RV measurements. We conclude in Sect.~\ref{section__conclusion}.

	\section{Observation and data reduction}
	\label{section__observation_and_data_reduction}
	
	To spatially resolve faint companions, we need a high-precision multi-telescope recombiner. High-precision because the variations in the signal caused by a faint component have a small amplitude, while using several telescopes gives much more simultaneous measurements to properly cover the $(u,v)$ plane, and improve the observing efficiency considerably. For this purpose we used the CHARA/MIRC and VLTI/PIONIER combiners, allowing us to observe Cepheids in both hemispheres.
	
	\subsection{Northern interferometric observations}
	
	\begin{table*}[]
		\centering
		\caption{Journal of the observations.}
		\begin{tabular}{ccccccc} 
			\hline
			\hline
			Star	&	UT 						 &  Instrument &		Configuration	& $N_\mathrm{sp}$	& $N_{V^2}$,	$N_{CP}$	 & Calibrators	\\
			\hline
			\object{U~Aql}	&	2012~Jul.~27	& 	MIRC	&	S1-S2-E2-W1-W2		&	8 &	1088, 662		&	\object{HD185124}, \object{HD198001} 		\\
			&	2013~Sep.~14	& 	MIRC	&	S1-S2-E2-W1-W2		&	8 &	560, 511		&	\object{HD184985}, \object{HD196870} 		\\
			&	2016~Jul.~20	 & 	MIRC	&	S1-S2-E2-W1-W2		&	8 &	1337, 1594		&	\object{HD177067}, \object{HD178218}, \object{HD188844} 	 		\\
			\object{FF~Aql}	&	2012~Jul.~26	&	MIRC &	S1-S2-E1-E2-W1-W2	& 8&	1496,	2000		&	\object{HD166230}, \object{HD182807} 		\\ 
			\object{U~Car}	&	2016~Mar.~07	&	PIONIER &	A0-G1-J2-J3		& 6	&	1018, 483		&	\object{HD96068}, \object{HD98897}, \object{HD100078}	 		\\
			&	2017~Mar.~21	&	PIONIER &	A0-G1-J2-K0		& 6	&	1950, 1561		&	\object{HD98692}, \object{HD98732}, \object{HD99048}	 		\\			
			&	2017~Jun.~27	&	PIONIER &	A2-D0-J3-K0		& 6	&	583, 475		&	\object{HD90074}, \object{HD90980}, \object{HD98897}	 		\\				
			\object{Y~Car}    	&	2016~Mar.~06	& 	PIONIER	& 	A0-G1-J2-J3	&	6	&	573, 362	&	\object{HD93307}, \object{HD92156}, \object{HD89839} 		\\
			\object{YZ~Car}    	&	2016~Mar.~05	& 	PIONIER	& 	A0-G1-J2-J3	&	6	&	898, 458	&	\object{HD90246}, \object{HD89517}, \object{HD85253} 		\\
			&	2016~Mar.~06	& 	PIONIER	& 	A0-G1-J2-J3	&	6	&	817, 524	&	\object{HD97744}, \object{HD94256} 		\\
			\object{BP~Cir}	&	2015~Feb.~17	&	PIONIER &	A1-G1-J3-K0		& 6	&	1620, 1080		&	\object{HD130551}, \object{HD132209}, \object{HD121901} 	 \\
			\object{BG~Cru}	&	2016~Mar.~04	&	PIONIER &	A0-G1-J2-J3			& 6	&	807, 556		&	\object{HD110532}, \object{HD110924}, \object{HD115669} 	\\			
			\object{T~Mon}	&	2016~Dec.~27	&	PIONIER &	A0-G1-J2-J3		& 6	&	323, 240		&	\object{HD43299}, \object{HD45317}	 		\\
			&	2017~Jan.~06		&	PIONIER &	A0-G1-J2-K0 	& 6	&	360, 240		&		 		\\
			\object{R~Mus}	&	2016~Mar.~07	&	PIONIER &	A0-G1-J2-J3		& 6	&	1060, 705		&	\object{HD109761}, \object{HD101805}, \object{HD125136} 	 		\\
			\object{S~Mus}		&	2015~Feb.~16	&	PIONIER &	 A1-G1-J3-K0		&	6	 &	1980,	1320 &	\object{HD107013}, \object{HD109761}, \object{HD102969}  		\\
			&	2016~Mar.~05	& 	PIONIER	& 	A0-G1-J2-J3		&	6	&	1046, 689	&	\object{HD112124}, \object{HD105939}, \object{HD107720} 		\\
			&	2017~Mar.~04	& 	PIONIER	 & 	A0-G1-J2-J3		&	6	&	1440, 960	&	 	\object{HD102534}, HD105939, HD107720		\\
			\object{S~Nor}	&	2016~Mar.~05	&	PIONIER &	A0-G1-J2-J3		& 6	&	1251, 833		&	\object{HD147075}, \object{HD148679}, \object{HD151005} 	 		\\
			&	2017~May~27	&	PIONIER &	B2-D0-J3-K0		& 6	&	872, 584		&	\object{HD146247}, \object{HD145361}, \object{HD148679} 	 		\\
			&	2017~Jun.~22	&	PIONIER &	A0-G1-J2-J3		& 6	&	1912, 1302		&	\object{HD147422}, \object{HD144230}, \object{HD145883} 	 		\\
			\object{W~Sgr}    	&	2017~Jun.~23	& 	PIONIER	& 	A0-G1-J2-J3	&	6	&	1094, 1595	&	\object{HD162415}, \object{HD163652}, \object{HD169236} 		\\				
			\object{X~Sgr}    	&	2017~May~27	& 	PIONIER	& 	B2-D0-J3-K0	&	6	&	540, 360	&	\object{HD157919}, \object{HD156992}, \object{HD166295} 		\\
			\object{V350~Sgr} &	2013~Jul.~10	&	PIONIER &	A1-G1-J3-K0		& 3	&	630, 392		&		\object{HD174774}, \object{HD171960}	\\			
			&	2013~Jul.~14	&	PIONIER &	D0-G1-H0-I1		& 3	&	821, 546		&	 		\\	
			&	2017~Jun.~22	&	PIONIER &	A0-G1-J2-J3		& 6	&	409, 201		&	\object{HD174423}	 		\\	
			\object{V636~Sco}	&	2013~Jul.~10	&	PIONIER &	 A1-G1-J3-K0			&	3	 &	322,	311		&	\object{HD154486}, \object{HD159941} 		\\
			&	2015~Feb.~16	& 	PIONIER	& 	A1-G1-J3-K0		&	6	&	936, 237	&	\object{HD151337}, \object{HD159217}, \object{HD160113} 		\\
			&	2016~Mar.~06	& 	PIONIER	& 	A0-G1-J2-J3		&	6	&	1183, 307	&	\object{HD149835}, \object{HD152272}, \object{HD155019} 		\\
			&	2016~Mar.~07	& 	PIONIER	& 	A0-G1-J2-J3		&	6	&	672, 312	&	 		\\
			&	2017~May~27	& 	PIONIER	& 	B2-D0-J3-K0	&	6	&	377, 582	&	\object{HD154250}, \object{HD159285}, \object{HD155019} 		\\
			\object{AH~Vel}	&	2016~Dec.~28	& 	PIONIER	&	A0-G1-J2-J3	&	6 &	191, 195		&	\object{HD66080}, \object{HD70195} , \object{HD73075}		\\
			&	2017~Jan.~01	& 	PIONIER	&	A0-G1-J2-J3	&  6 &	1044, 696		&		\\
			&	2017~Jan.~27	& 	PIONIER	&	A0-G1-J2-J3	&	6 &	828, 552	&	\\
			\hline
		\end{tabular}
		\tablefoot{$N_{V^2}$ and $N_{CP}$: total number of squared visibilities and closure phases. $N_\mathrm{sp}$: number of spectral channels. Adopted calibrator diameters are listed in Table~\ref{table_calibrators}.
		}
		\label{table__journal}
	\end{table*}
	
	The observations were performed with the Michigan InfraRed Combiner (MIRC) installed at the CHARA array \citep{ten-Brummelaar_2005_07_0}, located on Mount Wilson, California. The CHARA array consists of six 1\,m aperture telescopes with an Y-shaped configuration (two telescopes on each branch), oriented to the east (E1, E2), west (W1,W2) and south (S1, S2), providing a good coverage of the $(u, v)$ plane. The baselines range from 34\,m to 331\,m, providing a high angular resolution down to $\sim 0.5$\,mas in $H$. The MIRC instrument \citep{Monnier_2004_10_0,Monnier_2010_07_0} is an image-plane combiner which enables us to combine the light coming from all six telescopes in $K$ or $H$. MIRC also offers three spectral resolutions ($R = 42, 150$ and 400), which provide 15 visibility and 20 closure phase measurements across a range of spectral channels.
	
	We observed the Cepheids FF~Aql (HD~176155, $P_\mathrm{puls} = 4.47$\,d) and U~Aql (HD~183344, $P_\mathrm{puls} = 7.02$\,d) with five and six telescopes. We used the $H$-band filter with the lowest spectral resolution, where the light is split into eight spectral channels. Table~\ref{table__journal} lists the journal of our observations. We followed a standard observing procedure, i.e. we monitored the interferometric transfer function by observing a calibrator before and/or after our Cepheids. The calibrators, listed in Table~\ref{table__journal}, were selected using the \textit{SearchCal}\footnote{Available at \url{http://www.jmmc.fr/searchcal}.} software \citep{Bonneau_2006_09_0,Bonneau_2011_11_0} provided by the Jean-Marie Mariotti Center (JMMC).
	
	We reduced the data using the standard MIRC pipeline \citep{Monnier_2007_07_0}, which consists of computing the squared visibilities and triple products for each baseline and spectral channel, and to correct for photon and readout noises. Squared visibilities are estimated using Fourier transforms, while the triple products are evaluated from the amplitudes and phases between three baselines forming a closed triangle.
	
	\subsection{Southern interferometric observations}
	
	We used the Very Large Telescope Interferometer \citep[VLTI ;][]{Haguenauer_2010_07_0} with the four-telescope combiner PIONIER \citep[Precision Integrated Optics Near-infrared Imaging ExpeRiment,][]{Le-Bouquin_2011_11_0} to measure squared visibilities and closure phases of the southern binary systems. PIONIER combines the light coming from four telescopes in the $H$ band, either in a broad band mode or with a low spectral resolution, where the light is dispersed across six spectral channels (three before December 2014). The recombination provides simultaneously six visibilities and four closure phase signals per spectral channel.
	
	Our observations were carried out from 2013 to 2017 using the 1.8\,m Auxiliary Telescopes with the largest available configurations, providing six projected baselines ranging from 40 to 140\,m. PIONIER was setup in \emph{GRISM} mode, i.e. the fringes are dispersed into six spectral channels (three before December 2014). As for MIRC, we monitored the interferometric transfer function with the standard procedure which consists of interleaving the science target by reference stars. The calibrators were also selected using the \textit{SearchCal} software, and are listed in Table~\ref{table__journal}, together with the journal of the observations.
	
	The data have been reduced with the \textit{pndrs} package described in \citet{Le-Bouquin_2011_11_0}. The main procedure is to compute squared visibilities and triple products for each baseline and spectral channel, and to correct for photon and readout noises.
	
	\section{Companion search and sensitivity limit}
	\label{section__companion_search_and_sensitivity_limit}
	
	We used the \texttt{CANDID} code developed by \citet{Gallenne_2015_07_0}, which is a set of \texttt{Python} tools allowing us to search systematically for companions and estimate the detection limit using all interferometric observables. Briefly, the first main function of \texttt{CANDID} performs a 2D grid of fit using a least-squares algorithm. At each starting position, the companion position, its flux ratio and the angular diameters (if components are spatially resolved) are fitted. \texttt{CANDID} also includes a tool to estimate the detection level of the companion in number of sigmas (assuming the error bars in the data are uncorrelated). Uncertainties on the fitted parameters are estimated using a bootstrapping function. From the distribution, we took the median value and the maximum value between the 16th and 84th percentiles as uncertainty for the flux ratio and the angular diameter. For the fitted astrometric position, the error ellipse is derived from the bootstrap sample (using a principal components analysis). The second main function incorporates a robust method to set a $3\sigma$ detection limit on the flux ratio for undetected components, which is based on an analytical injection of a fake companion at each point in the grid. We refer the reader to \citet{Gallenne_2015_07_0} for more details about \texttt{CANDID}.
	
	Because of spectral smearing across one spectral channel, we searched for companions with a maximum distance to the main star of 50\,mas. The spectral smearing field of view (FoV) is defined by $\lambda^2/(B\,\Delta \lambda)$, where $\lambda$ is the wavelength of the observations, $\Delta \lambda$ the width of the spectral channels, and $B$ the interferometric baseline. For our observation, we have limited our search within 50\,mas, although the \texttt{CANDID} algorithm includes an analytical model to avoid significant smearing.
	
	In the following, we search for companions using either all observables (i.e. the squared visibilities $V^2$, the bispectrum amplitudes $B_\mathrm{amp}$, and the closure phases $CP$) or only the closure phases. As explained by \citet{Gallenne_2015_07_0}, the $CP$ is more sensitive to faint off-axis companions (although depending on its location and the $(u,v)$ coverage) and is also less affected by instrumental and atmospheric perturbations than the other observables. Fitting all of the observables can improve the detection level because we add more information, but it can also affect the results, depending on the magnitude of the biases altering the $V^2$ data. The detection of a companion is claimed if the significance level is $> 3\sigma$ and consistent between observables.
	
	Although \texttt{CANDID} has implemented two methods to derive the sensitivity limits, we only listed the ones given by the injection method, which has been proven to be more robust for biased data \citep[see][]{Gallenne_2015_07_0}.
	
	In this section we present the results of the companion search for each individual stars. A detailed discussion is presented in Sect.~\ref{section__discussion} for detected and non-detected components.

	\begin{sidewaystable*}[]
	\centering
	\caption{Final best-fit parameters.}
	\begin{tabular}{ccccccccccccc} 
		\hline
		\hline
		\#  &  Star		&	MJD	& $\phi$		&	$\theta_\mathrm{UD}$ 	&	$f$	&	$\Delta \alpha$	&	$\Delta \delta$ & $\sigma_\mathrm{a}$ & $\sigma_\mathrm{b}$ & $\sigma_\mathrm{PA}$ & \multicolumn{2}{c}{$n\sigma$} \\
		&			&	(day)		&				&	(mas)	&	(\%)	&	(mas)	&	(mas)	& (mas) & (mas) & ($^\circ$) &		All	&	$CP$		 \\
		\hline
		1	&U~Aql					&	56135.316	& 0.11	&	$0.706\pm0.042$					&	$0.74\pm0.15$	&		$-4.855$	&	$-0.874$ & 0.111 & 0.068 & 99.7	&  2.8  &   3.1  \\ 
		2	&							&	56549.230	& 0.04	&	$0.792\pm0.027$					&	$0.44\pm0.09$		&		3.798		&	$-8.948$ & 0.075 & 0.056 & $-52.7$ &  5.8  &  6.1	\\
		3	&							&	57589.297	& 0.11	&	$0.786\pm0.042$					&	$0.73\pm0.32$		&		$-2.709$		&	6.564 & 0.161 & 0.048 & 68.2 &	2.9   &   2.2  \\
		4	&FF~Aql					&	56134.252	& 0.11	&	$0.834\pm0.014$					&	$0.56\pm0.29$	&	1.167	&	8.787 & 0.100 & 0.090 & 40.8	&  1.9  &  1.2	\\ 
		5	& U~Car					&	57455.099 &		0.85	&	$0.909\pm0.011$				&	--									&		--		&	-- & -- & -- &	--	&  1.3							&   2.8  \\ 
		6	&								&	57834.138 &	 0.61 &	$0.948\pm0.024$				&	--									&		--		&		-- & -- & -- & --		&  1.2 				&   1.1  \\ 		
		7	&								&	57901.017 &	0.35 &	$0.825\pm0.037$					&	--									&		--		&			--  & -- & -- & --		&  1.1 				&   2.2  \\ 	
		8	& Y~Car					&	57454.082		& 0.42	&	$0.19\pm0.10$				&  $0.94\pm0.10$	&		2.440	&  0.392 & 0.323 & 0.070  & 68.2	&  6.3		&   2.4  \\ 
		9	& YZ~Car					&	57453.111	&	0.13	&	$0.367\pm0.036$				&  --	&		--	&  -- & -- & -- & --	&  2.7							&   1.3  \\ 
		10	& 								&	57454.157		& 0.19	&	$0.372\pm0.033$				&  --	&		--	&  --  & -- & -- & --	&  2.1							&   2.9  \\ 
		9+10	& 							&	--			& &	$0.370$				&  --	&		--	&  --	&  --			& -- & -- & -- &   2.7  \\ 
		11	&BP~Cir					&	57071.327	& 0.59	&	$0.254\pm0.125$				&	$3.24\pm0.21$			&		11.097	&	34.525 & 0.058 & 0.053 & $-16.5$ &  $>50$  &  $>50$  \\
		12	&BG~Cru				&	57452.399		& 0.11 &	$0.694\pm0.028$					&	$0.53\pm0.12$	&	1.981		&	6.352 & 0.238 & 0.097 & $-66.8$	&  2.3 &   4.5  \\ 
		13	&T~Mon					&	57750.185	&	0.79	&	$0.904\pm0.016$				&	--							&		--		&		--	& -- & -- & --	&  2.9 							&   2.8  \\ 
		14	&							&	57760.259	& 0.16 &	$0.703\pm0.093$				&	--									&		--		&		--	& -- & -- & --	&  2.3 			&   1.3  \\ 		
		13+14	&						&		--	& & $0.803$					&	--									&		--		&									--		&  -- 	& -- & -- & --	&   2.0  \\ 
		15	&R~Mus					&	57455.169	& 0.31	&	$0.457\pm0.023$				&	--									&		--		&		--	& -- & -- & --	&  0.9 			&   1.6  \\ 
		16	& S~Mus	   				&		57070.277	& 0.17	&	$0.560\pm0.034$	&	$0.86\pm0.03$		& 	0.648	&	2.297 & 0.117 & 0.026  & 99.6	&  21.2				&   $> 50$  \\ 
		17	&								&	57453.206 &	0.81 &	$0.599\pm0.027$		& $1.09\pm0.09$	     &   3.037	& $-0.804$ & 0.108 & 0.060 &	68.5	&  25	&   22  \\ 		
		18	&								&	57817.174 &	0.49 &	$0.677\pm0.033$			&  $0.80\pm0.11$	 &	$-0.496$	&	$-2.394$ & 0.218 & 0.133 & 55.2	& 5.7	&   7.4 \\ 
		19	&S~Nor					&	57453.297		& 0.35	&	$0.639\pm0.020$				&	--									&		--		&		-- & -- & -- &	--	&  1.8 							&   2.2  \\ 
		20	&								&	57901.172		& 0.26	&	$0.661\pm0.039$				&	--							&		--		&		-- & -- &-- &	--	&  1.7 							&   1.5  \\ 		
		21	&								&		57927.113 & 0.92	&	$0.511\pm0.040$					&	--					&		--		&	--	& -- & -- & --	&  2.3 							&   1.3  \\ 
		20+21	&								&			&	& $0.586$						&	--									&		--		&			--		&  -- 	& -- & -- & -- &   1.3  \\ 
		22	& W~Sgr					&	57928.170		& 0.23	&	$1.079\pm0.023$				&	--						&		--		&		--  & -- & -- & --		&  2.7							&   1.7  \\ 
		23	& X~Sgr					&	57901.305		& 0.96	&	$1.250\pm0.026$				&  $0.43\pm0.10$	&		$-11.801$	&  $-7.704$ & 0.628 & 0.122 & 58.3	&  2.4		&   5.8  \\ 
		24	& V350~Sgr				&	56484.247	& 0.90	&	$0.452\pm0.039$				&	$0.62\pm0.12$		& 	0.680 &	2.949 & 0.470 & 0.126  & 59.6	&  4.1	&   1.5  \\ 
		25	&								&	56488.126 &	0.65 &	$0.515\pm0.044$					& $0.50\pm0.09$	&	0.563		& 2.845 & 0.454 & 0.181  & $-50.8$	&  5.4 		&   4.5  \\ 		
		24+25	&						&		--			&	&	$0.480$								&  $0.55\pm0.11$	 &	0.862	&	2.483 & 0.184 & 0.115  & $-25.1$	&  --							&   4.1 \\ 
		26	&								&	57927.193 &	0.86 &	$0.428\pm0.022$							&	--			&		--				&		-- & -- & -- & --		&  1.5 							&   1.5  \\ 	
		27	& V636~Sco				&	56484.157	& 0.65	&	$0.533\pm0.012$				&	--						& 	-- 	& -- & -- & --	&	--							&  1.4							&   2.3  \\ 
		28	&								&	57070.380 &	0.90 &	$0.530\pm0.012$					& --	&--			&--		& -- & -- & --&  1.3							&   2.5  \\ 		
		29	&								&	57454.321 &	0.38 &	$0.583\pm0.012$						&  --	 &	--	&	-- & -- & -- & --	& 1.3							&   2.9 \\ 
		30	&								&	57455.328	& 0.53 &	$0.566\pm0.016$						&  --	 &	--	&	-- & -- & -- & -- & 2.2							&   1.2 \\ 
		29+30	&						&	--					& &	$0.575$								&  --			 &	--	&	--	& --		& -- & -- &--					&   1.7 \\ 
		31	&								&   57901.416	& 0.16 &	$0.53\pm0.10$		&	$1.04\pm0.58$	&		$-6.594$	&  4.591 & 0.488 & 0.039 &	38.4	&  3.8 				&   3.6  \\ 	
		32	&AH~Vel					&	57751.103	& 0.21	&	$0.343\pm0.051$				&	--									&		--		&			--	&-- & --&--	&  1.0 			&   2.3  \\ 
		33	&								&	57755.265	& 0.20	&	$0.456\pm0.038$			&	--						&		--		&		-- & -- & -- &		&  0.8 		&   1.2  \\ 			
		34	&							&		57781.183 &	0.33 &	$0.408\pm0.049$					&	--				&		--		&		--	& -- & -- &	&  1.4 		&   1.0  \\ 					
		32+33+34	&					&						--	& &		$0.402$				&	--									&		--		&		--		&  -- 	& -- & -- & -- &   1.4  \\ 		
		\hline
	\end{tabular}
	\tablefoot{MJD: modified Julian date. $\phi$: pulsation phase, calculated with the ephemeris listed in Table~\ref{table_orbit_uaql} to \ref{table_orbit_v636sco} for U~Aql, FF~Aql, YZ~Car, BP~Cir, BG~Cru, S~Mus, S~Nor, W~Sgr, V350~Sgr, V636~Sco and AH~Vel, and from \citet{Samus_2017_01_0} for U~Car, Y~Car, T~Mon, R~Mus and X~Sgr. $\theta_\mathrm{UD}$: uniform disk angular diameter. $f$, $\Delta \alpha$, $\Delta \delta$: $H$-band flux ratio and relative astrometric position of the companion. $\sigma_\mathrm{a}, \sigma_\mathrm{b}, \sigma_\mathrm{PA}$: uncertainty of the astrometric position expressed by the error ellipse with major and minor axes and the position angle measured from north through east. $n\sigma$: detection level in number of sigma when fitting all observables or only the $CPs$. The listed astrometric positions correspond to the fit with the highest detection level.
	}
	\label{table__fitted_parameters}
\end{sidewaystable*}

\begin{table*}[!h]
	\centering
	\caption{$3\sigma$ average contrast limits in the $H$ band using all observables or only the closure phases.}
	\begin{tabular}{c|c|cc|cc|cc|c} 
		\hline
		\hline
		&											&	\multicolumn{4}{c|}{$\Delta H$ (mag)}							&		& \\
		&											&	\multicolumn{2}{c|}{All}	&	\multicolumn{2}{c|}{$CP$} 									&	\multicolumn{2}{c|}{Sp. Type}	& \multirow{2}{*}{Instrument}\\
		&											&	$r < 25$\,mas      & $r <50$\,mas 		&	$r < 25$\,mas      & $r < 50$\,mas 	&	upper limit		&	\\
		\hline
		\multirow{3}{*}{U~Aql} & 2012-07-27 &	4.6					&	4.3			& 4.7	&	4.4	 &	B6V	& B3V	&	\multirow{3}{*}{MIRC}	\\
		& 2013-09-14 &	5.8					&	5.5					& 5.5	&	5.5	& B9V	& B9V	&		\\
		& 2016-07-20	&	5.5					&	5.4				 	& 5.6			&	5.2 	& B9V	& B9V	&	 \\
		\hline
		FF~Aql &	2012-07-26	&	5.3  & 4.8	& 5.7  & 5,3 	&	A3V	& A0V&	MIRC\\
		\hline
		\multirow{3}{*}{U~Car } & 2016-03-07&	5.7					&	5.6			& 6.0	&	6.0	 &	B2V	&B2V	&		\multirow{3}{*}{PIONIER}	\\
		&2017-03-21 &	4.9					&	4.6							& 5.5	&	5.2	& B1V	& B1V	&		\\
		&2017-06-27	&	5.1					&	4.8				 	& 5.8			&	5.7 	&	B2V	& B2V	&	 \\
		\hline
		Y~Car	&2016-03-06		&	5.1				&	5.0				& 4.0	&	4.0 	 	&	A0V	&A0V&	PIONIER\\
		\hline
		\multirow{3}{*}{YZ~Car}	&2016-03-05		&	5.0				&	5.3				& 4.6	&	4.4 	 	&	B2V	&B2V&	\multirow{3}{*}{PIONIER} \\
		&2016-03-06		&	5.2				&	5.4				& 4.2	&	4.4	 	&	B2V	&B3V&	\\
		& combined		&	--				&	--				& 4.8	&	4.7  	&	B2V	&B2V& \\	
		\hline
		BP~Cir &   2015-02-17		& 5.0					&	4.7				 & 4.9				&	4.6 	&	A6V & A5V		& PIONIER\\
		\hline
		BG~Cru & 2016-03-04 &	5.1					&	4.9				 & 5.3				&	5.3 	&	A2V & A2V		&	PIONIER \\
		\hline
		T~Mon & combined			&		--					&	--				 & 4.9					&	4.7 		&		B1V &B1V		&PIONIER\\
		\hline
		R~Mus &2016-03-07			&	4.8					&	4.4				 & 5.1				&	5.0 		&		B8V & B8V		&	PIONIER \\
		\hline
		\multirow{3}{*}{S~Mus} & 2015-02-16		&	4.6				&	4.2				& 6.0	&	5.8	 	&	B9V	& B9V&	\multirow{3}{*}{PIONIER} \\
		&2016 Mar. 05		&	5.2				&	5.2				& 5.1	&	5.0	 	&	B6V	&B6V&	\\
		&2017 Mar. 04		&	5.0				&	4.8				& 5.0	&	4.6	 	&	B6V	&B3V& \\
		\hline
		\multirow{2}{*}{S~Nor} &2016-03-05			&	5.1					&	4.8				 & 5.3			&	5.1		&	B6V & B5V		&	\multirow{2}{*}{PIONIER} \\
		
		&2017 combined	&	--					&	--				 & 5.6					&	5.3 	&	B7V & B5V		&	 \\
		\hline
		W~Sgr	&2017-06-23	&	4.8				&	4.5				& 5.2	&	5.1	 	&	B8V	&B8V&	PIONIER\\
		\hline
		X~Sgr	&2017-05-27		&	4.7				&	4.4				& 5.4	&	5.3 	 	&	B8V	&B8V&PIONIER\\
		\hline
		\multirow{2}{*}{V350~Sgr} & 2013 combined	&	--					&	--			& 5.3	&	5.0 	 &	B9V	& B8V	&		\multirow{2}{*}{PIONIER}	\\
		&2017-06-22 &	4.8					&	4.9							& 4.3	&	4.8	& B8V	& B8V	&	\\
		\hline
		\multirow{4}{*}{V636~Sco} &2013-07-10				&	5.1					&	4.9			& 5.0	&	4.6	 &	B8V	&B8V	&		\multirow{4}{*}{PIONIER}	\\
		&2015-02-02 &	5.3					&	5.1							& 5.6	&	5.6 	& B9V	& B9V	&	\\
		&2016 combined	&	--					&	--				 	& 5.8			&	5.8	&	B9V & B9V	&	 \\
		&2017-05-27	&	5.3					&	5.2				 	& 5.3			&	5.2  &	B8V & B8V	& \\
		\hline
		AH~Vel & combined	&	\multicolumn{2}{c|}{--}	& 5.5	&	5.3 &   A1V& A0V &	PIONIER\\
		\hline
	\end{tabular}
	\label{table__limits}
	\end{table*}

	\subsection{U~Aql}
	
	This 7.02\,d period Cepheid has been studied for decades using visible spectroscopy \citep[see e.g.][]{Sanford_1930__0, Szabados_1989_01_0, Wilson_1989_04_0, Bersier_2002_06_0}. It is a well known spectroscopic binary with an orbital period of 1856.4\,d, first discovered by \citet{Slovak_1979_12_0}, and then confirmed with the determination of the orbital elements by \citet{Welch_1987_07_0}. Observations using the IUE also detected the presence of a hot companion \citep{Bohm-Vitense_1985_09_0, Evans_1992_04_0}, from which a spectral type of B9.8V has been estimated. This would correspond to a flux ratio $\sim 0.45$\,\%\footnote{Converted using the grid of \citet{Pecaut_2013_09_0} (see also \citealt{Pecaut_2012_02_0} and \url{http://www.pas.rochester.edu/~emamajek/EEM_dwarf_UBVIJHK_colors_Teff.txt}) and the average Cepheid magnitudes \citep{Storm_2011_10_0,van-Leeuwen_2007_08_0}.} (mean value as it also depends on the pulsation phase of the Cepheid). According to the orbital elements and the estimate of the orbital inclination by \citet[][$i = 74^\circ$, using the mass function and the assumed masses $7\,M_\odot$ for U~Aql and $2\,M_\odot$ for its companion]{Welch_1987_07_0}, the projected semi-major axis is expected to be $\sim 9$\,mas. This value is consistent with the $9.5$\,mas estimate of \citet{Kervella_2018__0} using the values of the spectroscopic orbit and the proper motion vectors (and assuming a mass of $5.2\,M_\odot$ for the Cepheid). The available baselines at CHARA allow us to spatially resolve such separation.
	
	We obtained three measurement epochs spread over four years, and a companion is detected at each epoch. For the first observations, the seeing conditions were poor at $\sim 1.3$\arcsec. We noticed that including $B_\mathrm{amp}$ leads to the detection of a companion which is too bright ($f \sim 2.5$\,\%), and is not consistent with the detection using $CP+V^2$ or only $CP$, which give the exact same position and flux ratio. We therefore decided to not use this observable which is more affected by the atmospheric conditions.
	The detections using the $CPs$ only and $CP+V^2$ give similar levels, $3.1\sigma$ and $2.8\sigma$ respectively. For the second epoch, the atmospheric conditions were better ($\sim 0.7\arcsec$), and a companion is detected at more than $5\sigma$ in all observables. For the last epoch, the detection level ranges between 1.5 and $2.9\sigma$. Although the significance level is low, the detection is well localized and the derived flux ratio seems consistent with our previous observations. The values are listed in Table~\ref{table__fitted_parameters}. Although additional observations are necessary to confirm, a tentative orbital fit is discussed in Sect.~\ref{section__discussion}.
	
	As previously explained, we estimate the $3\sigma$ sensitivity limits with \texttt{CANDID} (i.e. by injecting fake companions with various flux ratio at all azimuth), to detect a possible third component by analytically removing the detected companion for all epochs \citep[see][]{Gallenne_2015_07_0}. In Table~\ref{table__limits}, we calculated conservative values corresponding to the mean plus the standard deviation for the given radius ranges $r < 25$\,mas and $r < 50$\,mas. We found no companion with a flux ratio higher than 0.6\,\% within 50\,mas.

	We also measured the angular diameter for a uniform disk model, its value is reported in Table~\ref{table__fitted_parameters}. We estimated the uncertainty by using the conservative formalism of \citet{Boffin_2014_04_0}:
		\begin{equation}
		\sigma^2_{\theta_\mathrm{UD}} = N_\mathrm{sp}\sigma_\mathrm{stat}^2 + \delta \lambda^2\theta_\mathrm{UD}^2
		\end{equation}
		where $N_\mathrm{sp}$ is the number of spectral channels. The first term takes into account that the spectral channels are almost perfectly correlated, and $\delta \lambda$ = 0.0025 or 0.0035, coming from the fact that the absolute wavelength calibration is precise at a 0.25\,\% level for MIRC \citep{Monnier_2012_12_0} and 0.35\,\% for PIONIER \citep{Gallenne_2018_08_0,Kervella_2017_01_1}. $\sigma_\mathrm{stat}$ is the statistical error from the bootstrapping technique (bootstrap on the modified Julian date of the calibrated data, with replacement) using 1000 bootstrap samples. We then took from the distributions the median and the maximum value between the 16\,\% and 84\,\% percentiles as the uncertainty (although the distributions were roughly symmetrical). We used this formalism for all Cepheids in this paper. Note that the precision of the angular diameter measurement depends on how far down the visibility curve is sampled, so smaller stars will have lower precision.
	

	\subsection{FF~Aql}

	\begin{figure*}[!t]
		\centering
		\resizebox{\hsize}{!}{\includegraphics[width = \linewidth]{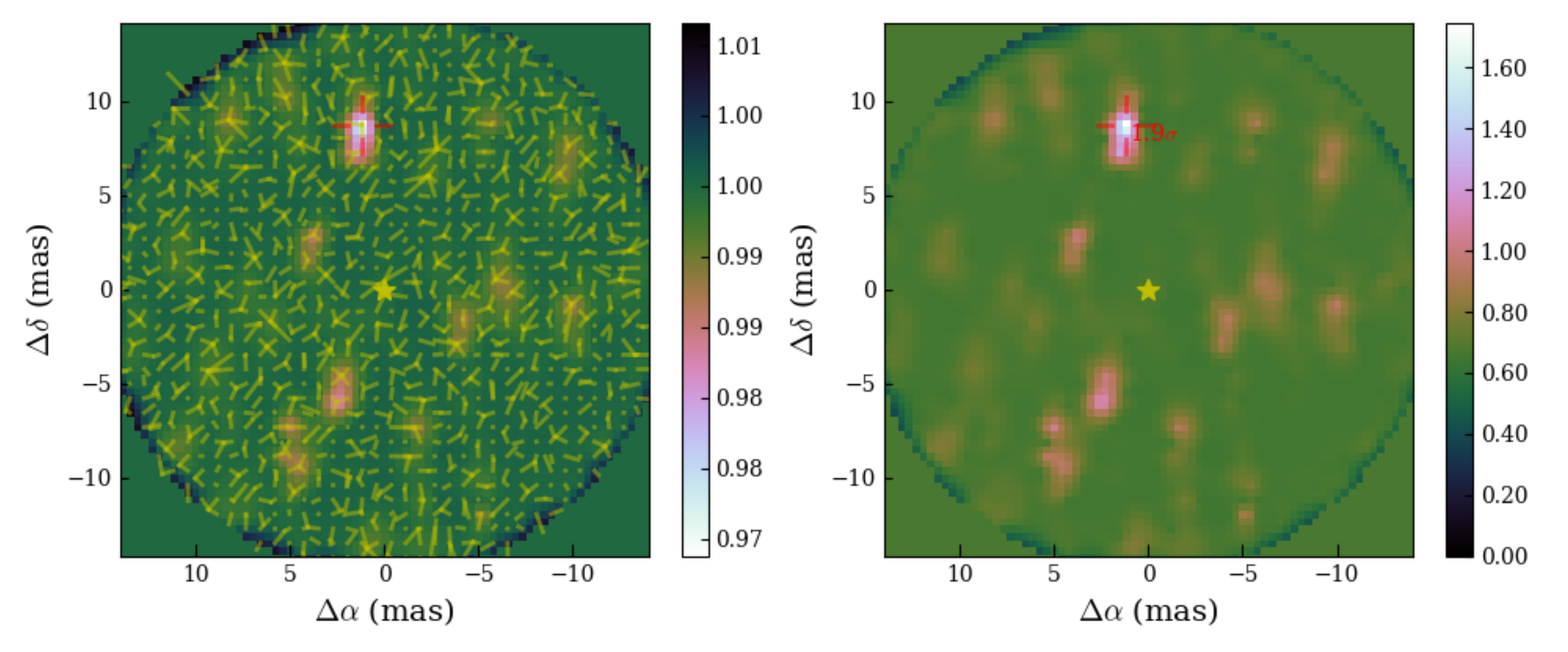}}
		\caption{$\chi_r^2$ map of the local minima (left) and detection level map (right) of FF~Aql using all observables. The yellow lines represent the convergence from the starting points to the final fitted position \citep[for more details see][]{Gallenne_2015_07_0}. The maps were re-interpolated in a regular grid for clarity. The yellow star denotes the Cepheid.}
		\label{image__chi2_ffaql}
	\end{figure*}
	\begin{figure}[!h]
		\centering
		\resizebox{\hsize}{!}{\includegraphics{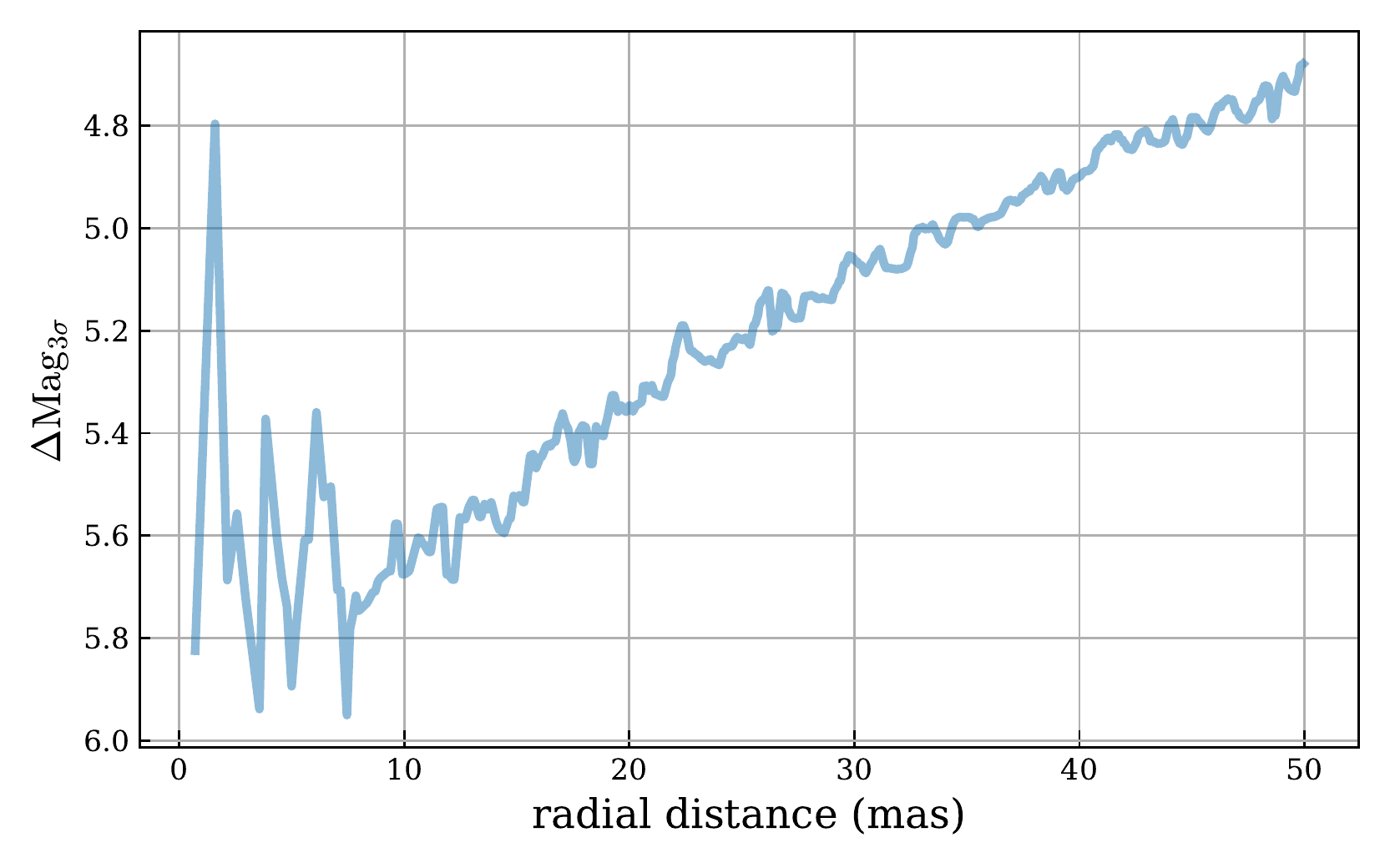}}
		\caption{$3\sigma$ contrast limit for  FF~Aql.}
		\label{image__limit_ffaql}
	\end{figure}
	
	This Cepheid is possibly a member of a quadruple system, with a spectroscopic and two wide companions. While the spectroscopic component is well known \citep[see e.g.][]{Abt_1959_11_0,Szabados_1977_01_0,Evans_1990_05_0,Benedict_2007_04_0}, the existence of the wide companions possibly located at $\sim 0.2$\arcsec\ and $\sim 6.5$\arcsec\ is more uncertain \citep[for a more detailed discussion of these components see][]{Gallenne_2014_07_0}. From our data, these possible wide companions will not be detected, first because of the spectral smearing FoV, and because of the Gaussian single-mode fiber transmission profile. The spectroscopic component was first discovered by \citet{Abt_1959_11_0}, who derived an orbital period of 1435\,days. Additional RVs later enabled us to slightly revise the orbit \citep{Evans_1990_05_0,Gorynya_1998_11_0}.
	
	Our observations are more suitable to detect this spectroscopic companion, which has a semi-major axis $a = 12.8$\,mas. This has been derived by \citet{Benedict_2007_04_0} from binary astrometric perturbations in the Fine Guidance Sensor (FGS) data on board HST. \citet{Evans_1990_05_0} estimated the companion spectral type to be between A9V and F3V, which give an approximate expected flux ratio in the $H$ band in the range 0.2-0.4\,\%. Fig.~\ref{image__chi2_ffaql} shows the 2D detection maps given by \texttt{CANDID}, and we see that our MIRC observations did not show any signature of this companion with a detection level $> 3\sigma$. We can see a marginal detection at a $1.9\sigma$ level using all observables, while at $1.2\sigma$ with the closure phases only. However, although it is not statistically significant, the detection is well localized and the measured flux ratio $f \sim 0.6$\,\% seems fairly consistent with what expected. Additional data are necessary to confirm this detection though. This position of the companion is reported in Table~\ref{table__fitted_parameters}, together with the detection levels.
	
	We also derived the $3\sigma$ sensitivity limits as previously explained, they are listed in Table~\ref{table__limits}. In Fig.~\ref{image__limit_ffaql} we show this contrast upper limit at $3\sigma$ using all observables. A maximum contrast of 1:190 is reached within 25\,mas. A more detailed discussion is presented in Sect.~\ref{section__discussion}.

	\subsection{U~Car}

	This long-period Cepheid (38.8\,d) did not benefit from extensive spectroscopic studies in the past, which is why its binarity was revealed only by \citet{Bersier_2002_06_0}. However, he only detected a 10\,km~s$^{-1}$ offset compared to the previous observations of  \citet{Coulson_1985__0}, and there is no orbital solution yet. This companion was also not detected by \citet{Evans_1992_01_0} from IUE observations, but she derived a spectral type limit for the companion: it has to be later than A1V. This would correspond to a very high contrast of $\Delta H \gtrsim 8.7$\,mag. Note also that \citet{Kervella_2018__0} did not detect a significant anomaly in the PM vectors, suggesting a very long period or a very low mass companion.

	We performed three observations with PIONIER spread over approximately one year (see Table~\ref{table__journal}). Due to the very high contrast of the companion, we do not detect it, our highest detection level being $2.8\sigma$ (see Table~\ref{table__fitted_parameters}). According to our contrast limit listed in Table~\ref{table__limits}, we were only sensitive to companions with a contrast lower than $\sim 6$\,mag with the best data set.

	In Table~\ref{table__fitted_parameters} we also report the angular diameters for a uniform-disk model for each epoch of observations.

	\subsection{Y~Car}

	This short-period double-mode Cepheid is known to be a member of a binary system from the Fourier analysis of the RVs performed by \citet{Stobie_1979_12_0}, who could only estimate a period ranging from 400-600\,d because of an incomplete orbital coverage. The full spectroscopic elements were later determined by \citet{Balona_1983_06_0}, and measured an orbital period of $993 \pm 11$\,d. This was then refined by \citet{Petterson_2004_05_0} to $993 \pm 2$\,d with new RV measurements. Its spectral type was identified to be B9V by \citet{Evans_1992_02_0} using the IUE satellite. From HST/GHRS measurements, \citet{Bohm-Vitense_1997_09_0} measured its orbital velocity amplitude to estimate its mass to be $3.8 \pm 1.2\,M_\odot$. \citet{Kervella_2018__0} derived a similar mass from the analysis of the PM vectors. However, \citet{Evans_2005_08_0} later obtained HST/STIS spectra of this hot companion and found a large variation in the RV in a short time scale, being $7\,\mathrm{km~s^{-1}}$ change in 4\,days. They interpreted this as the companion being itself a short-period binary system, the brighter component being the B9V star.
	
	Such companion would correspond to $\Delta H \sim 4.9$\,mag, which is detectable with the current interferometric recombiner. We observed the Y~Car system with VLTI/PIONIER on March 2016 (see Table~\ref{table__journal}). We have a likely detection at $6\sigma$ using all observables, while it is not detected at more than $2.4\sigma$ with the CPs only (see Table~\ref{table__fitted_parameters}). Our estimated contrast of $\sim 5$\,mag is in agreement with what expected, however, additional observations will be necessary to firmly conclude about this detection.
	
	We analytically removed this possible companion to estimate the detection limit of a third component, they are listed in Table~\ref{table__limits}. We can exclude any brighter component with $\Delta H \lesssim 5.1$\,mag.

	\begin{figure*}[]
	\centering
	\resizebox{\hsize}{!}{\includegraphics[width = \linewidth]{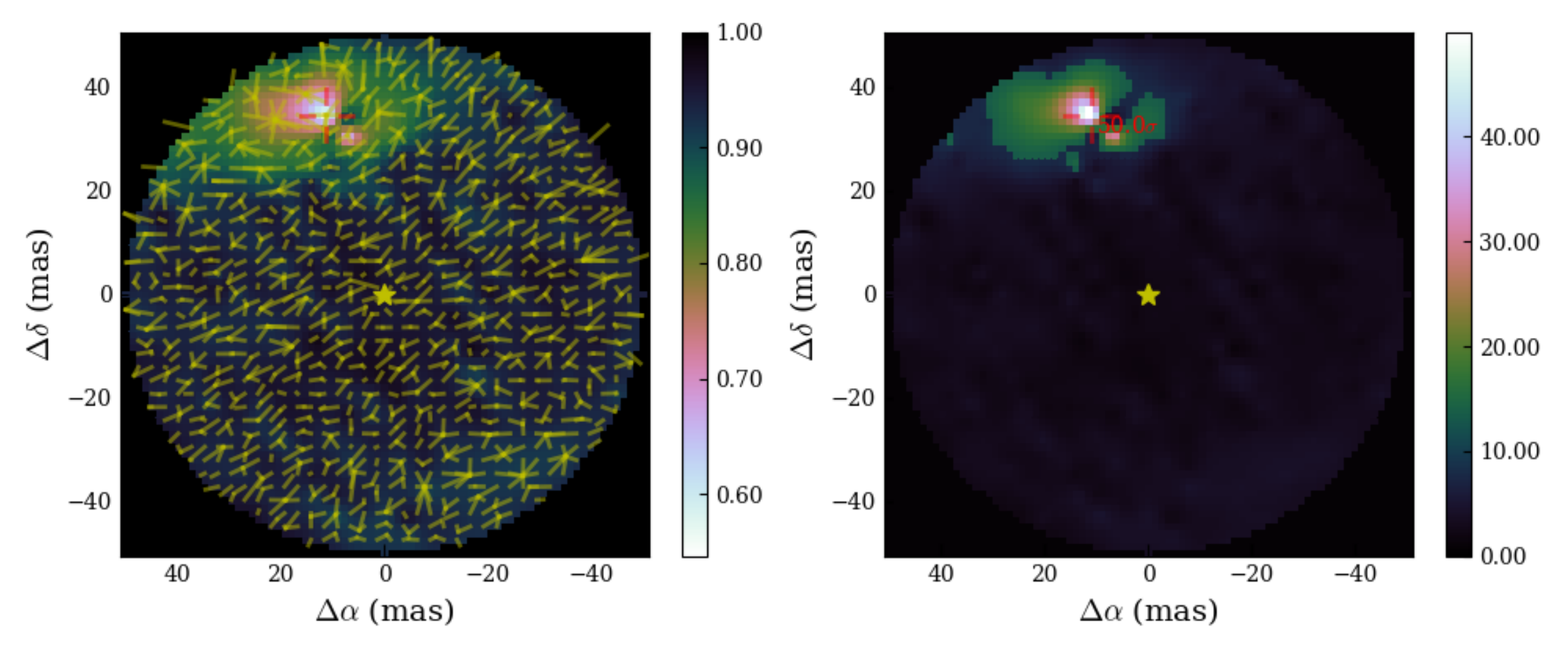}}
	\caption{Same as Fig.~\ref{image__chi2_ffaql} but for BP~Cir, using all observables.}
	\label{image__chi2_bpcir}
	\end{figure*}

	\subsection{YZ~Car}
	
	From RV measurements, \citet{Coulson_1983_12_0} discovered that this long-period Cepheid (18.2\,d) belongs to a binary system. He derived the preliminary spectroscopic orbital parameters, giving an orbital period of 850\,d, and suggested a spectral type between B5V and A5V for this companion. \citet{Evans_1993_09_0} later tightened the range of spectral type to B8V-A0V from ultraviolet spectra of the companion. Additional RVs were obtained by \citet{Bersier_2002_06_0}, but he did not revise the orbital elements. A new orbit was fitted by \citet{Petterson_2004_05_0} combining the RVs of \citet{Coulson_1983_12_0} with new measurements, and updated the orbital period to 657\,d. However, this is not consistent with Coulson's value, neither with the last determination by \citet{Anderson_2016_10_0} who derived $P_\mathrm{orb} = 830$\,d, in better agreement with \citet{Coulson_1983_12_0}. We discuss the possible origin of the disagreement in Sect.~\ref{section__discussion}. Analysis of the PM vectors \citep{Kervella_2018__0} provides an approximate mass of $1.9\pm0.3\,M_\odot$, which is consistent with the expected range of spectral types.
	
	This companion would correspond to a flux ratio in the range 0.1-0.3\,\%, which is challenging to detect in interferometry. The PIONIER instrument was used on March 2016 and we observed YZ~Car two consecutive nights (see Table~\ref{table__journal}). Unfortunately, the companion is not detected, even combining the two datasets. According to our sensitivity limits (Table~\ref{table__limits}), we were only able to detect companions with flux ratio $> 0.75$\,\% in the best case.

	\subsection{BP~Cir}

	BP~Cir is a first overtone Cepheid with a pulsation period of 2.40\,days. Its companion was first suggested by \citet{Balona_1981_12_0} because the Cepheid appeared bluer in his surface brightness-color relation, although he did not detect orbital motion in their RV measurements. Recent and more accurate velocity measurements by \citet{Petterson_2004_05_0} showed more scatter than expected in the $\gamma$-velocity, but it was not sufficient to obtain reliable orbital solutions, suggesting a long orbital period. Such long orbit is consistent with the non-detection in the PM vectors of \citet{Kervella_2018__0}.

	The blue companion was detected from IUE spectra \citep{Arellano-Ferro_1985_10_0,Evans_1994_11_0}, and a B6V star best matches the observed spectra. Using the distance $d = 850$\,pc from the $K$-band P-L relation for first overtone pulsators \citep[][non-canonical model]{Bono_2002_07_0}, we would expect a $H$-band flux ratio around 3.2\,\%.

	Our PIONIER observations clearly reveal this companion with a detection level higher than $50\sigma$ in all observables. The detection maps are presented in Fig.~\ref{image__chi2_bpcir}, and clearly show the secondary at $\rho \sim 36.29$\,mas with a flux ratio $f \sim 3.2$. Our fitted values are reported in Table~\ref{table__fitted_parameters}.

	The companion was then analytically removed to derive the detection limit for a possible third component. We can exclude any companion with a flux ratio higher than 1.3\,\% within 50\,mas (see Table~\ref{table__limits}).

	\subsection{BG~Cru}
	
	BG~Cru is a suspected spectroscopic binary with an orbital period between 4000 and 6650 days \citep{Szabados_1989_01_0}. Orbital elements are still unknown due to low-precision RVs and a lack of long term measurements. \citet{Evans_1992_01_0} did not detect this companion either from IUE spectra, and she set an upper limit on its spectral type to be later than A1V. From PM vector anomaly, \citet{Kervella_2018__0} also flagged this Cepheid as possible binary.
	
	We observed this target with VLTI/PIONIER on 2016 March 4th with the largest available quadruplet. A candidate companion is detected with a moderate detection level. With all observables, we have a detection at $2.3\sigma$, while we have $4.5\sigma$ with only the closure phases. Furthermore, the projected separation and flux ratio are consistent with each other, which add more confidence about this detection. The values are listed in Table~\ref{table__fitted_parameters}. This possible detection needs to be confirmed with additional interferometric observations.
	
	The average $3\sigma$ sensitivity limits are reported in Table~\ref{table__limits} (after removing this possible detection). We reached a maximum contrast of 1:130, i.e. a dynamic range of 5.3\,mag in the $H$ band.

	\subsection{T~Mon}

	This long-period Cepheid (27.0\,d) has an orbiting hot component discovered by \citet{Mariska_1980_12_0} from IUE observations, from which they inferred the spectral type to be A0V. The binarity was then confirmed from long-term variations in the RVs \citep{Coulson_1983_06_0,Gieren_1989_06_0}, but due to the very long orbital period and low-precision RVs, different orbital parameters are found in the literature. \citet{Gieren_1989_06_0} estimated a period of 175\,yr, which is within the range given later by \citet{Evans_1999_10_0}, while the more recent estimate seems to be around $\sim 89$\,yr \citep{Groenewegen_2008_09_0}. The spectral type of the companion was slightly revised by \citet{Evans_1994_06_0} to be B9.8V. The contrast would correspond to $\Delta H \sim 8.1$\,mag. It is worth mentioning that this companion is also suspected to be itself a binary \citep{Evans_1999_10_0}. The small PM anomaly detected by \citet{Kervella_2018__0} confirms the long orbital period.

	The PIONIER observations of T~Mon are listed in Table~\ref{table__journal}. We have two observations separated by one month from each other. Neither individual epoch provides significant detection larger than $2.9\sigma$ (see Table~\ref{table__fitted_parameters}). Combining them decrease the detection level to $2\sigma$ (see Table~\ref{table__limits}). According to our detection limits, our dataset is sensitive to a contrast $< 4.9$\,mag, and therefore explains that we do not detect this companion.
	
	\subsection{R~Mus}
	
	Some evidence that this Cepheid is a spectroscopic binary was given by \citet{Lloyd-Evans_1982_06_0}. This was also the conclusion of \citet{Szabados_1989_01_0} from variations in the systemic velocity. Since then, long term monitoring of the RV of R~Mus is still missing, which results in a still unknown orbital period. \citet{Evans_1992_01_0} did not detect this companion from IUE spectra, and set an upper limit to A0V for its spectral type. This translates to a flux ratio $\lesssim 0.35$\,\% in the $H$ band ($\Delta H \sim 6.1$\,mag). From the PM vectors anomaly, \citet{Kervella_2018__0} detected the signature of orbital motion, likely due to this companion.
	
	Our interferometric observations do not reveal this possible companion. We found no significant detection at more than $1.6\sigma$ (see Table~\ref{table__fitted_parameters}). From our derived $3\sigma$ sensitivity limits, we noticed that this dataset provides a dynamic range of $\Delta H \sim 5$\,mag (see Table~\ref{table__limits}).
	
	\subsection{S~Mus}
	
	This 9.66\,d period Cepheid is a known spectroscopic binary. The duplicity was first suspected by \citet{Walraven_1964_06_0} from multicolor photometry, and \citet{Lloyd-Evans_1968__0} from RV measurements. It was later confirmed by \citet{Stobie_1970__0}. Several other detections were later reported from IUE spectra and RVs, providing some characteristics of the companion and the spectroscopic orbital solutions \citep{Lloyd-Evans_1982_06_0,Bohm-Vitense_1985_09_0,Bohm-Vitense_1986_04_0,Bohm-Vitense_1990_01_0,Evans_1990_05_0,Petterson_2004_05_0}. This companion has one of the shortest known orbital period for binary Cepheid systems, which is $\sim 505$\,d.
	
	IUE and HST spectra allowed to estimate the mass ratio by measuring the orbital velocity amplitude of both components \citep{Bohm-Vitense_1986_04_0,Evans_1994_12_0,Bohm-Vitense_1997_03_0}. The latest refined value is $M_\mathrm{cep} / M_\mathrm{comp} = 1.14$. The spectral type of the companion was also estimated to be between B3V and B5V, with an average of B3.8V, which would correspond to a $\sim 5.2\,M_\odot$ star. This therefore provides an estimate of the Cepheid mass of $5.9\,M_\odot$ \citep{Bohm-Vitense_1997_03_0}. The spectral type was then refined to B3V by \citet{Evans_2004_05_0} using FUSE spectra, providing a new estimate of the Cepheid mass of $6.0\,M_\odot$. The expected flux ratio in $H$ for a B3V companion would be $\sim 1.2$\,\%. However, from the PM vectors, \citet{Kervella_2018__0} derived an approximate mass for the companion of $2.2\,M_\odot$, which is not consistent with a B3V star, but with a later spectral type.
	
	We obtained three observing epochs with PIONIER (see Table~\ref{table__journal}), and the companion is detected for all of them at more than $5\sigma$. Our fitted astrometric positions are reported in Table~\ref{table__fitted_parameters}. Our measured flux ratio of $\sim 0.9$\,\% is in agreement with the expected value, and would correspond to a slightly later spectral type. A more detailed discussion is presented in Sect.~\ref{section__discussion}.
	
	After removing the companion, we derived the detection limit for a possible third component at each epoch. We can exclude an additional companion with a flux ratio higher than 0.3\,\% within 50\,mas.

	\subsection{S~Nor}
	
	This Cepheid is possibly a member of a multiple system. One companion has been spatially resolved by \citet{Evans_2013_10_0} at a separation $\sim 0.9\arcsec$, with an approximate orbital period of $\sim 8700$\,yrs. Another possible wider and hotter component would be located at $\sim 36\arcsec$, but the physical association is still uncertain. Orbital motion in RVs is also suspected, maybe linked to another component, but still needs to be confirmed. \citet{Szabados_1989_01_0} estimated a range of orbital period for this spectroscopic companion between 3300 and 6350\,days, but that could not be confirmed by \citet{Bersier_1994_11_0} who did not find orbital motion larger than $\sim 0.3\,\mathrm{km~s^{-1}}$ from high-precision spectroscopic measurements. \citet{Groenewegen_2008_09_0} gathered literature data and estimated an orbital period of 3584\,days, but fixing the eccentricity to zero. This value is in the range given by \citet{Szabados_1989_01_0}. Using the spectroscopic orbit of \citet{Groenewegen_2008_09_0} with PM vectors, \citet{Kervella_2018__0} estimated an approximate mass of $1.5\,M_\odot$ for a close-in companion, which would correspond to an $\sim$F0V star. From IUE spectra, \citet{Evans_1992_04_0} detected a hot companion with a spectral type equivalent to a B9.5V star attributed to one of the wide components. A B9.5V star would correspond to a flux ratio $f \sim 0.3$\,\%, but the wide components are out of the interferometric field of view. If instead this bright companion corresponds to the spectroscopic one, we might be able to detect it from interferometry, otherwise the F0V companion would be more challenging to detect as we would have $f \sim 0.1$\,\%.
	
	We have three epochs of interferometric observations with PIONIER separated by about one year (see Table~\ref{table__journal}). Unfortunately, we do not have any detection at more than $2.3\sigma$ (see Table~\ref{table__fitted_parameters}). According to our detection limits listed in Table~\ref{table__limits}, a companion with a flux ratio $>0.6$\,\% would have been detected at more than $3\sigma$ from our data. The best detection level is achieved with the CPs by combining the two datasets of 2017, separated by about a month. This is justified by the fact that the orbital period of the spectroscopic component  should be between 3300 and 6350\,days \citep{Szabados_1989_01_0}, which would lead to an orbital variation $< 1$\,\%.

	\subsection{W~Sgr}

	This 7.6\,d Cepheid is a member of a triple system, composed of a wide and a spectroscopic component. The wide companion was discovered from speckle interferometry by \citet{Morgan_1978_06_0} at a separation of 116\,mas, and later detected at a separation of 160\,mas with the HST/STIS by  \citet{Evans_2009_03_0}. The same authors identified this wide companion to be the hottest component of the system, previously detected from IUE observations \citep{Bohm-Vitense_1985_09_0,Evans_1991_05_0}. Its spectral type is A0V. The closest companion was discovered by \citet{Babel_1989_06_0} by combining several RV observations obtained over years and new accurate measurements from the CORAVEL instrument. The first spectroscopic orbital elements have been determined, including an orbital period of 1780\,d. The orbit has then been refined with years \citep{Bersier_1994_11_0,Petterson_2004_05_0,Groenewegen_2008_09_0}, the latest value being 1651\,d. This companion was also detected with the Fine Guidance Sensor (FGS) on board the HST from the orbital reflex motion of the Cepheid \citep{Benedict_2007_04_0}. This enabled the determination of the orbital inclination and semi-major axis (12.9\,mas). They also provide an estimate for the Cepheid mass of $6.5\,M_\odot$ assuming a mass from the companion spectral type. However, \citet{Evans_2009_03_0} set an upper limit for the spectral type of this companion to be later than F5V, which in turn set an upper limit to the Cepheid mass of $5.4\,M_\odot$. An F5V companion would correspond to a contrast of $\sim 7.7$\,mag ($f \sim 0.1$\,\%), and will be challenging to detect from interferometry. The approximate mass derived by \citet[$0.5\,M_\odot$]{Kervella_2018__0} is probably too small as it would correspond to an ever later spectral type of M0V.

	We observed W~Sgr with PIONIER as backup target in 2017 (see Table~\ref{table__journal}). Unfortunately, the companion is not detected from these observations, with a most significant detection level at $2.7\sigma$ (see Table~\ref{table__fitted_parameters}). Our estimated contrast limits shows that with this dataset we were only sensitive to companions brighter than $f \sim 0.8$\,\%.

	\subsection{X~Sgr}

	The binary nature of this system was discovered by \citet{Szabados_1990_01_0} who gathered RV data from the literature, although \citet{Lloyd-Evans_1968__0} already noticed a large scatter in the RVs. Due to the low amplitude of the systemic velocity ($K \sim 2-3\,\mathrm{km~s^{-1}}$), the spectroscopic orbit determination is difficult, they however estimated an orbital period to be $\sim 507$\,d. This makes the system among the shortest known orbital period binaries containing a Cepheid component in our Galaxy. \citet{Evans_1992_01_0} did not detect this companion from IUE spectra, and set an upper limit for the spectral type to be A0V. Two additional RV observations were obtained by \citet{Bersier_2002_06_0}, but  insufficient to better constrain the orbit. Since then, no new RV measurements have been published. \citet{Groenewegen_2008_09_0} recompiled all the RVs from the literature to better constrain the orbital elements, forcing the eccentricity to zero, and derived a period $P_\mathrm{orb} = 573.6$\,d. From HST/FGS observations, \citet{Benedict_2007_04_0} measured the parallax of X~Sgr to be $\pi = 3.00 \pm 0.18$\,mas ($d = 333 \pm 20$\,pc), but no detectable perturbation in the Cepheid orbit due to the companion has been noticed. \citet{Kervella_2018__0} detected the signature of a companion from the PM vectors, and estimated an approximate mass for the companion of $0.5\,M_\odot$. \citet{Li-Causi_2013_01_0} argued the detection of the companion at a separation of 10.7\,mas with VLTI/AMBER, with a flux ratio of $f_\mathrm{K} \sim 0.6$\,\%, but there is no indication about the detection level. Such contrast is difficult to achieve with the current state-of-the art interferometric recombiner, while AMBER is not designed to reach high-dynamic range. This detection is just at the limit of what this instrument can do, so this detection is uncertain. In addition, their angular diameter estimates are not in agreement with the diameter variation \citep{Breitfelder_2016_03_0}, where the minimum is around phase 0.8, while they found the opposite at both the minimum and maximum phases. We therefore suspect a spectral calibration problem instead of a binary detection. Furthermore, the expected contrast should be $\lesssim 0.4$\,\% in $K$ (about the same in $H$), and should be higher at the maximum diameter. 

	We performed one observation with PIONIER (see Table~\ref{table__journal}), and such companion should be detectable. We have a detection at $5.8\sigma$ with the CPs only, while there is no detection using all observables. Although the flux ratio is consistent with what expected (0.4\,\%, see Table~\ref{table__fitted_parameters}), it is worth mentioning that two other locations are also possible with similar detection level (4.9 and $5.2\sigma$), so unfortunately we still need confirmation with additional data. Note that these additional locations are spuriously produced by the non-optimal $(u, v)$ coverage, as the telescopes configuration was almost linear (as for V636~Sco, we were also in a non-standard configuration because of strong wind preventing the relocation).

	From our estimated $3\sigma$ detection limit (Table~\ref{table__limits}), and after removing the possible companion, we can however exclude a companion with a flux ratio larger than 0.7\,\%.

	\subsection{V350~Sgr}
	
	This 8.82\,d Cepheid has a well known spectroscopic companion. It was confirmed by \citet{Gieren_1982_05_0} from RV measurements, but was already suspected by \citet{Lloyd-Evans_1980__0}. \citet{Szabados_1990_01_0} gave a first estimate of the orbital period (1129\,d) by gathering all RVs from the literature for this star. Supplementary RVs over the years enabled the determination of the full spectroscopic orbital element and to refine the period to 1477\,d \citep[see e.g.][]{Evans_2011_09_0}. From IUE observations and spectral template comparison, \citet{Evans_1992_04_0} found the spectral type being best matched by a B9V star. This is consistent with the approximate mass of the companion ($3.4 \pm0.4\,M_\odot$) derived by \citet{Kervella_2018__0} from the proper motion vectors. Using spectra taken around minimum and maximum orbital phase with the Goddard High Resolution Spectrograph (GHRS) on the HST and assuming a mass for the companion (from a mass-luminosity relation), \citet{Evans_2011_07_0} estimated a mass for this Cepheid to be $5.0 \pm 0.7\,M_\odot$. This is in agreement with the latest estimate from \citet{Petterson_2004_05_0} who derived $6.0 \pm 0.9\,M_\odot$ with the same method but using new RV measurements. Recently, \citet{Evans_2018_10_0} refined the mass to $5.2 \pm 0.3\,M_\odot$ using the same method but from new HST/STIS spectra. 
	
	According to its spectral type, we should expect a $H$-band flux ratio of $\sim 0.8$\,\% for this system. This is detectable with PIONIER. We have three observations with this instrument, which are listed in Table~\ref{table__journal}. The companion is likely to be detected in the 2013 data, but not in 2017, probably because of the lower quantity of data acquired. The observations of 2013 July 10 have a detection at $4.5\sigma$ only with all observables, but not with the closure phases only. We also noticed other possible locations with similar detection levels. The observations of 2013 July 14 give a detection with both observables and only the CPs, which is consistent with the detection on 10 July, giving more confidence in this detection. As the orbital change is very small in four days ($< 0.4$\,\%), we combined the data and fit all the CPs. We found the same location with a detection level of $\sim 4\sigma$. This is reported in Table~\ref{table__fitted_parameters}. The measured flux ratio of 0.55\,\% is in agreement with expectations, although slightly lower. For the 2017 observations, we do not have a detection at more than $1.5\sigma$ whatever the observable. This might be due to a lower degree of freedom or the location of the companion too close to the Cepheid to be detected. In any case, additional observations are necessary to fully conclude and have an astrometric orbit.

	\subsection{V636~Sco}
	
	The spectroscopic binary nature was first noticed by \citet{Feast_1967__0}, but no orbital parameters were derived at that time. The same conclusion was reached by \citet{Lloyd-Evans_1968__0} who suggested an orbital period of $\sim 3.5$\,yrs. This was later confirmed by \citet{Lloyd-Evans_1980__0}, who then followed with the determination of the orbital elements \citep{Lloyd-Evans_1982_06_0}, including an orbital period of 1318\,days. Since then, the orbital elements have been updated \citep[see e.g.][]{Bohm-Vitense_1998_10_0,Petterson_2004_05_0},  the latest value being 1362.4\,days for the orbital period. This companion was also detected from IUE spectra by \citet{Bohm-Vitense_1985_09_0}, from which they determined an approximate effective temperature of 9400\,K. From additional IUE observations, \citet{Evans_1992_04_0} was able to estimate the spectral type for the companion to be B9.5V. \citet{Bohm-Vitense_1998_10_0} obtained RV measurements around the maximum and minimum orbital velocity using the HST, from which they derived a mass for the Cepheid of $3.1\,M_\odot$ (assuming a $2.5\,M_\odot$ companion). However, as this is too small for such a 6.8\,d pulsating Cepheid, they concluded that the companion might be itself a binary, making the V636~Sco system a probable triple system. Until now, there is no additional evidence to support this. \citet{Kervella_2018__0} derive an approximate mass of $2.3\,M_\odot$ for the companion, which is in good agreement with the B9.5V companion (assuming a $5.1\,M_\odot$ Cepheid).
	
	A B9.5V companion would give a flux ratio in $H$ of $\sim 0.5$\,\% ($\Delta H \sim 6$\,mag). To detect this companion, we obtained data from 2013 to 2017 (see Table~\ref{table__journal}), but we suffered from suboptimal weather conditions.  The observations of 2013 do not provide any detection above $2.3\sigma$ level, probably due to the low quantity of data obtained (see Table~\ref{table__fitted_parameters}). In 2015, no signal is detected above $2.5\sigma$. The 2016 data suffered from strong seeing variations (0.8-1.5\arcsec), and do not reveal either this companion. 2017 observations give a detection at a $3.8\sigma$ level, although we were in a non-standard telescope configuration because of strong wind preventing the relocation. Note that when using all observables, another position gives a similar detection level. So this detection should be considered preliminary.

	\subsection{AH~Vel}

	The binarity of this 4.23\,d period Cepheid has been suspected for decades \citep{Lloyd-Evans_1968__0,Gieren_1977_05_0,Lloyd-Evans_1982_06_0,Szabados_1989_01_0} from some variations in the RVs, however the orbital period is still unknown. No obvious trend in the residual of the RVs is seen. \citet{Bersier_2002_06_0} obtained more precise RV measurements, but only four points which were not enough to confirm the orbit. Since then there has not been additional information about this possible binary system. Recently, \citet{Kervella_2018__0} detected a strong signal in the residuals of the mean proper motion, which is attributed to the close-in component.

	We obtained three observing epochs with PIONIER with about one month interval (see Table~\ref{table__journal}). We searched for a companion in each epoch individually, but there is no detection at more than $2.3\sigma$. These observations were performed in service mode with only a few consecutive observations for the first and third observations (more are needed to detect faint companions), so this non-detection is somewhat expected. The second observation (2017~Jan.~1), which benefited from a longer observing sequence (2\,h), does not reveal this companion either. As it is probably on a very long orbital period, we can combine all epoch of closure phase data (not the $V^2$ as the star has different diameter at each epoch). Unfortunately, there is still no significant detection at more than $1.2\sigma$. 

	We estimate the detection limit from the method explained previously, and are listed in Table~\ref{table__limits} for the combined data. We conclude that no companion with a contrast lower than 1:140 is orbiting AH~Vel within 50\,mas.

	\section{Discussion}
	\label{section__discussion}
	
	In this section we discuss our previous detections and set upper limits for the spectral type of undetected components. We also present new high-precision RV measurements, which we used to revise the orbital and pulsation parameters.

	\paragraph{U~Aql:}

	\begin{figure*}[]
	\centering
	\resizebox{\hsize}{!}{\includegraphics{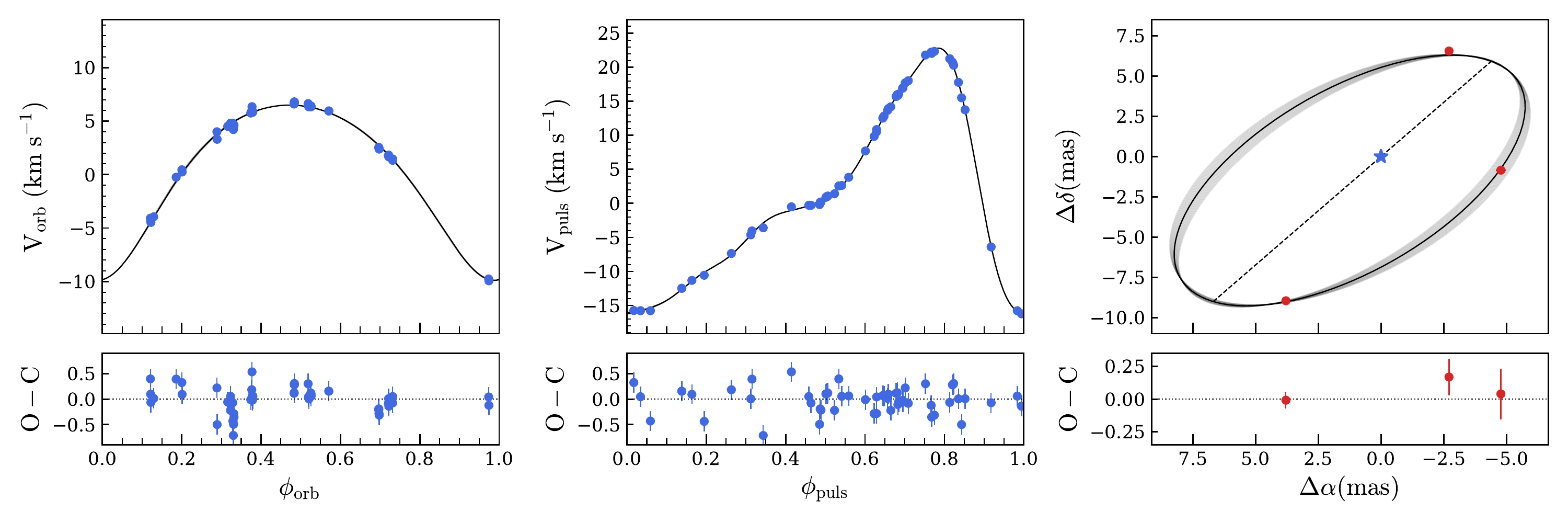}}
	\caption{Result of our combined spectroscopic and interferometric fit for U~Aql. Left: fitted (solid lines) and measured primary (blue dots) orbital velocity. Middle: fitted (solid line) and measured (blue dots) radial pulsation velocity. Right: relative astrometric orbit of U~Aql Ab.}
	\label{image__orbit_uaql}
	\end{figure*}

	\begin{table}[]
		\centering
		\caption{Final estimated parameters of the U~Aql system.}
		\begin{tabular}{ccc} 
			\hline
			\hline
			\multicolumn{3}{c}{\textbf{Pulsation}}																	\\
			$P_\mathrm{puls}$ (days)							  &  \multicolumn{2}{c}{$7.02414 \pm 0.00004$}\\
			$T_0$ (JD) \tablefootmark{a}						  &  \multicolumn{2}{c}{2~447~755.25}		\\
			$A_1$ ($\mathrm{km~s^{-1}}$)												  &	\multicolumn{2}{c}{$-7.47 \pm0.56$}			\\
			$B_1$ ($\mathrm{km~s^{-1}}$)												  &	\multicolumn{2}{c}{$-12.84 \pm0.35$}		\\
			$A_2$ ($\mathrm{km~s^{-1}}$)												   &	\multicolumn{2}{c}{$-7.71 \pm0.13$}		\\
			$B_2$ ($\mathrm{km~s^{-1}}$)												   &	\multicolumn{2}{c}{$-1.36 \pm0.69$}		\\
			$A_3$ ($\mathrm{km~s^{-1}}$)												  &	\multicolumn{2}{c}{$-1.76 \pm0.35$}			\\
			$B_3$ ($\mathrm{km~s^{-1}}$)												  &	\multicolumn{2}{c}{$2.52 \pm0.22$}		\\
			$A_4$ ($\mathrm{km~s^{-1}}$)												   &	\multicolumn{2}{c}{$-0.29 \pm0.24$}		\\
			$B_4$ ($\mathrm{km~s^{-1}}$)												   &	\multicolumn{2}{c}{$1.37 \pm0.06$}		\\
			$A_5$ ($\mathrm{km~s^{-1}}$)												   &	\multicolumn{2}{c}{$0.67 \pm0.10$}		\\
			$B_5$ ($\mathrm{km~s^{-1}}$)												   &	\multicolumn{2}{c}{$0.47 \pm0.16$}		\\
			$A_6$ ($\mathrm{km~s^{-1}}$)												   &	\multicolumn{2}{c}{$0.30 \pm0.03$}		\\
			$B_6$ ($\mathrm{km~s^{-1}}$)												   &	\multicolumn{2}{c}{$-0.02 \pm0.08$}		\\
			\hline
			\multicolumn{3}{c}{\textbf{Orbit}}  																			\\
			&  We87\tablefootmark{b}		& This work					 \\
			$P_\mathrm{orb}$ (days)						& $1856.4 \pm 4.3$			& $1831.4 \pm 6.5$ 		\\
			$T_\mathrm{p}$ (JD)								& $2~442~754\pm 38$		& $2~457~575.3 \pm 8.4$  \\
			$e$																& $0.165 \pm 0.027$     &  $0.193 \pm 0.005$  	\\
			$\omega$	($^\circ$)									&$190.5 \pm 7.7$ &	$167.1 \pm 1.9$		\\
			$K_1$ ($\mathrm{km~s^{-1}}$)				& $7.81 \pm 0.22$	  &	$8.41 \pm 0.04$		\\
			$K_2$ ($\mathrm{km~s^{-1}}$)\tablefootmark{c}				&	--	  &	$24.05 \pm 1.24$	\\
			$v_\gamma$	($\mathrm{km~s^{-1}}$)		&	$1.15 \pm 0.15$		&	$1.31 \pm 0.06$	\\
			$\Omega$	($^\circ$)									& -- &	$133.8 \pm 4.4$		\\
			$i$ ($^\circ$)												& --	&	$115.4 \pm 0.7$		\\
			$a$ (mas)													& --	 &	$10.06 \pm 0.16$		\\
			$a$ (au)													& --	 &	$5.94 \pm 0.22$		\\
			$M_1$ ($M_\odot$)									& --		&	$6.2 \pm 0.8$		\\
			$M_2$ ($M_\odot$)									& --		&	$2.2 \pm 0.2$		\\
			$d$ (pc)\tablefootmark{d}														& --	&	$592\pm19$ 	\\
			\hline
		\end{tabular}
		\tablefoot{$P_\mathrm{orb}$: orbital period. $T_\mathrm{p}$: time passage through periastron. $e$: eccentricity. $K_1, K_2$: RV semi-amplitude of the primary and secondary. $v_\gamma$: systemic velocity. $\omega$: argument of periastron. $\Omega$: position angle of the ascending node. $a$: semi-major axis. $i$: orbital inclination. $M_1, M_2$: mass of primary and secondary. $P_\mathrm{puls}$: pulsation period. $T_0$: reference epoch of maximum light. $A_\mathrm{i}, B_\mathrm{i}$: Fourier parameters.\\
			\tablefoottext{a}{held fixed during the fitting process.}
			\tablefoottext{b}{\citet{Welch_1987_07_0}.}
			\tablefoottext{c}{derived from the orbital elements and the assumed distance.}
			\tablefoottext{d}{assumed from the P-L relation of \citet{Storm_2011_10_0}.}
		}
		\label{table_orbit_uaql}
	\end{table}
	
	Our three epochs are consistent in terms of flux ratio, i.e. it decreases when the Cepheid diameter increases, making us confident about the detections. The average flux ratio $0.64 \pm 0.14$\,\% (mean and rms of the three values) corresponds to a B8-B9V companion\footnote{as previously, interpolated from the grid of \citet{Pecaut_2013_09_0}} , also in agreement with the IUE observations \citep{Evans_1992_04_0}. We can exclude from our $3\sigma$ contrast limits (see Table~\ref{table__limits}) any other companion earlier than a B9V star (after analytically removing the detected one).
	
	Although the listed astrometric positions still need to be confirmed with additional observations, we performed a preliminary orbital fit. As in \citet{Gallenne_2013_04_0}, we combined our astrometry with new high-precision single-line RV data to solve for all the orbital elements. We used new RVs obtained from 2012 to 2016 with the SOPHIE, CORALIE and HARPS spectrographs to better constrain the orbit with high-precision data. Data analysis and RV determination are explained in Appendix~\ref{appendix__new_rv}, and listed in Table~\ref{table_uaql_rv}.
		
	The orbital reflex motion of the Cepheid is simultaneously fitted with the radial pulsation using Fourier series of order $n$ ($n = 1, 2, ...$ and depends on the shape of the Cepheid velocity curve). We then followed the formalism detailed in \citet{Gallenne_2018_11_0}, who used a linear parametrization technique to solve for the orbital and pulsation parameters. The latter is defined with
	\begin{equation}
	\label{equation_vpuls}
	V_\mathrm{puls} = \sum_{i=1}^n [A_i \cos(2\pi i \phi_\mathrm{puls}) + B_i \sin(2\pi i \phi_\mathrm{puls})].
	\end{equation}		
		Briefly, we used a Markov chain Monte Carlo (MCMC) fitting routine\footnote{Using the Python package \texttt{emcee} developed by \citet{Foreman-Mackey_2013_03_0}} using the set of nonlinear parameters $P_\mathrm{orb}$, $T_\mathrm{p}$ and $e$ with two other linear parameters (related to $K_1, v_\gamma$ and $\omega$), together with the pulsation parameters ($P_\mathrm{puls}, T_0$, the Fourier parameters $A_\mathrm{n}, B_\mathrm{n}$) and the astrometric Thiele-Innes constants (parametrized in term of the orbital elements $a, \omega, \Omega$ and $i$). The names of these variables are defined in Table~\ref{table_orbit_uaql}. We adopted as best-fit parameters the median values of the distributions, and used the maximum value between the 16th and 84th percentiles as uncertainty estimates (although the distributions were roughly symmetrical about the median values). Zero point difference was corrected as explained in Appendix~\ref{appendix__new_rv}. First guess parameters were taken from \citet[][for the orbit, except $T_\mathrm{p}$ where we used our median time value]{Welch_1987_07_0} and \citet[][for the pulsation]{Samus_2017_01_0}. The final result is plotted in Fig.~\ref{image__orbit_uaql} and the derived parameters are listed in Table~\ref{table_orbit_uaql}. The systematic uncertainty from the wavelength calibration of the interferometric data was taken into account and was added quadratically to the error of the semi-major axis. We did not use previous RVs measurements from the literature for several reasons: 1) they are usually not very precise, 2) we wanted a dataset as uniform as possible (i.e. RVs estimated in a homogeneous way), 3) the effect on the RVs of a possible third component is reduced, and 4) we also avoid possible bias from the changing pulsation period of the Cepheid by limiting the time range. Although we find a slightly shorter orbital period, our revised orbital values are in rather good agreement with the previous determination of \citet{Welch_1987_07_0} from less precise RVs. The systemic velocity is in agreement within $1\sigma$, although the value from \citet{Welch_1987_07_0} has a large uncertainty.
	
	Unfortunately, the distance and masses of both components are degenerate as RVs of the companion are still missing. However, we can have a first estimate of the masses if we assume the distance $d = 592 \pm 19$\,pc from the Cepheid period-luminosity (P-L) relation of \citet{Storm_2011_10_0}. Masses are reported in Table~\ref{table_orbit_uaql}, with the uncertainties estimated from the MC simulations including the uncertainty on the distance with a normal distribution centred on 592\,pc with a standard deviation of 19\,pc. The Gaia parallax from the second data release \citep[GDR2,][]{Gaia-Collaboration_2018_08_0} was not adopted as its value is very inconsistent with what expected (the value $1042\pm114$\,pc is about a factor of two larger than expected). This might be due to the binarity and the changing color and brightness over their pulsation cycle that is not properly taken into account in the GDR2 astrometric pipeline processing.
	
	Our derived mass of the companion is slightly smaller than the one derived by \citet[2.3\,$M_\odot$]{Evans_1992_04_0} from a B9.8V spectral type, but still in agreement within $1\sigma$. Our value would be more consistent with an $\sim$A0.5V companion. Note however that the distance we used was estimated from an infrared surface-brightness method from the Cepheid photometry, and can be biased by several effects, as for instance the ignored photometric contribution of the companion or the value of the projection-factor. Our preliminary estimate of the Cepheid mass is in agreement with what we expect from evolution models \citep[$\sim 5.7\,M_\odot$,][]{Evans_2013__0}, although our uncertainty is still large. The mass of $5.1\pm0.7\,M_\odot$ inferred by \citet{Evans_1998_03_0} from measuring the orbital velocity amplitude is also consistent with our estimate.
	
	\begin{figure}[]
	\centering
	\resizebox{\hsize}{!}{\includegraphics{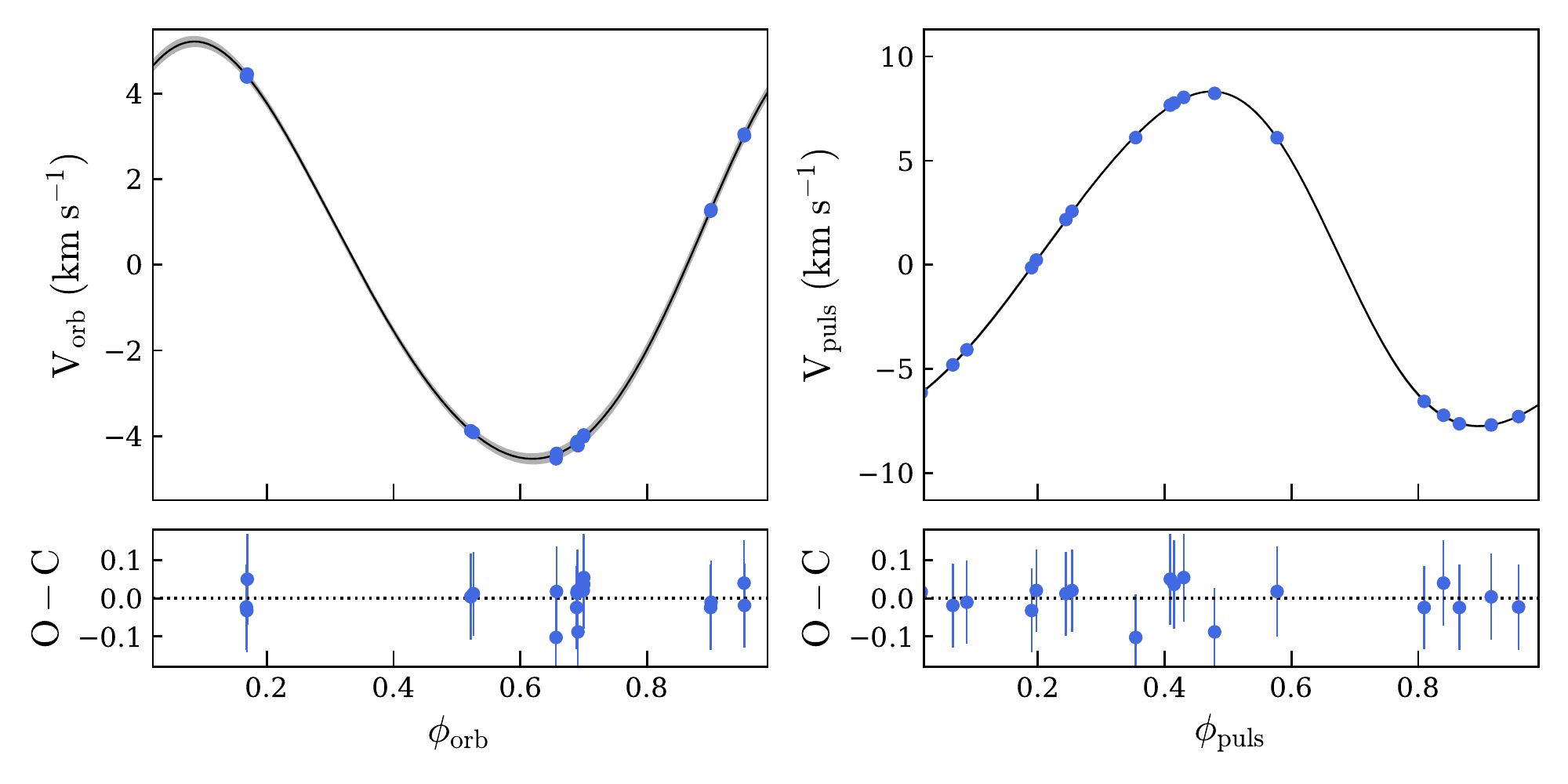}}
	\caption{Result of our combined fit of the orbital and radial pulsation velocity for FF~Aql. Left: fitted (solid lines) and measured primary (blue dots) orbital velocity. Right: fitted (solid line) and measured (blue dots) radial pulsation velocity.}
	\label{image__orbit_ffaql}
\end{figure}

	\paragraph{FF~Aql:}

\begin{table}[]
	\centering
	\caption{Pulsation and orbital parameters of the FF~Aql system.}
	\begin{tabular}{ccc} 
		\hline
		\hline
		\multicolumn{3}{c}{\textbf{Pulsation}}																	\\
		$P_\mathrm{puls}$ (days)                              &  \multicolumn{2}{c}{$4.471036 \pm 0.000015$}\\
		$T_0$ (JD)\tablefootmark{a}                          &  \multicolumn{2}{c}{2~436~792.539}     \\  
		$A_1$ ($\mathrm{km~s^{-1}}$)                                          & \multicolumn{2}{c}{$-8.01 \pm 0.10$} \\
		$B_1$ ($\mathrm{km~s^{-1}}$)                                          & \multicolumn{2}{c}{$0.26 \pm 0.88$} \\
		$A_2$ ($\mathrm{km~s^{-1}}$)                                          & \multicolumn{2}{c}{$0.39 \pm 0.21$} \\
		$B_2$ ($\mathrm{km~s^{-1}}$)                                          & \multicolumn{2}{c}{$0.91 \pm 0.11$} \\
		$A_3$ ($\mathrm{km~s^{-1}}$)                                          & \multicolumn{2}{c}{$0.27 \pm 0.04$} \\
		$B_3$ ($\mathrm{km~s^{-1}}$)                                          & \multicolumn{2}{c}{$-0.08 \pm 0.10$} \\
		$A_4$ ($\mathrm{km~s^{-1}}$)                                          & \multicolumn{2}{c}{$-0.02 \pm 0.03$} \\
		$B_4$ ($\mathrm{km~s^{-1}}$)                                          & \multicolumn{2}{c}{$-0.06 \pm 0.02$} \\
		\hline
		\multicolumn{3}{c}{\textbf{Orbit}}                             \\
		&  Go95\tablefootmark{c} 	& This work					 \\
		$P_\mathrm{orb}$ (days)                              &    $1433\pm 5$  &  $1430.3 \pm 2.6$  \\
		$T_\mathrm{p}$ (JD)        & $2~445~381\pm10$ &  $2~458~297.0\pm13.5$  \\ 
		$e$                                                   			&     $0.09\pm0.02$	&   $0.061 \pm 0.007$     \\
		$\omega$    ($^\circ$)                                &  327\tablefootmark{b} &   $316.0 \pm 4.0$      \\
		$K_1$ ($\mathrm{km~s^{-1}}$)                    & $5\pm0.5$      &  $4.824 \pm 0.008$           \\
		$v_\gamma$  ($\mathrm{km~s^{-1}}$)           &  $-16\pm0.5$          &  $-16.67 \pm 0.04$        \\
		$a_1~\sin{i}$ (au)											&	$0.66\pm0.07$ &		$0.633\pm0.002$						\\
		$f(M)$ ($M_\odot$)										&	$0.018\pm0.006$	 &		$0.0166\pm0.0001$	\\
		\hline
	\end{tabular}
	\tablefoot{Parameters are the same as in Table~\ref{table_orbit_uaql}. $a_1$: semi-major axis of the Cepheid's orbit relative to the center of mass. $f(M)$: the mass function.\\
		\tablefoottext{a}{Not fitted.}
		\tablefoottext{b}{Uncertainty not given by the authors.}
		\tablefoottext{c}{\citet{Gorynya_1995_05_0}.}
	}
	\label{table_orbit_ffaql}
	\end{table}

	The companion is detected with a low confidence level of $\sim 2\sigma$. FF~Aql was observed with only one bracket (i.e. one cal-sci-cal sequence), and such high-contrast companion usually needs several hours of observation to be strongly detected. Based on our measured $H$-band flux ratio and using the distance $d = 356$\,pc \citep{Benedict_2007_04_0}, a first estimate of the companion's spectral type would be between B9.5V-F1V. This is consistent with the expected spectral type of the companion A9V-F3V from \citet{Evans_1990_05_0}, although our estimate still depends on the exact Cepheid brightness at that pulsation phase. Note that we assumed main-sequence companions because from an evolutionary time-scale point of view, most of the companions orbiting Cepheids should be stars close to the main sequence.
	
	From our average $3\sigma$ contrast limits reported in Table~\ref{table__limits}, we can also exclude any companion orbiting FF~Aql within 50\,mas with a spectral type earlier than a B9V star. This is compatible with \citet{Evans_1992_01_0} who set an upper limit of A1V from an IUE spectrum.
	
	We also revised the spectroscopic orbit using new high-precision RV measurements obtained from 2013 to 2017 with the CORALIE, SOPHIE and HERMES spectrographs. Data analysis and RV determination are explained in Appendix~\ref{appendix__new_rv}, and listed in Table~\ref{table_ffaql_rv}. We also used the formalism detailed in \citet{Gallenne_2018_11_0}, but without the astrometric part. We also did not use previous RV measurements from the literature for the same reasons as explained before. As first guess parameters, we used the values from \citet[][for the orbit, except $T_\mathrm{p}$ where we used our median time value]{Gorynya_1995_05_0} and \citet[][for the pulsation]{Samus_2017_01_0}. $T_0$ cannot be properly determined from RVs by definition, so we did not fit this parameter. We corrected for the zero point difference to put the RVs consistent with the CORALIE system, as explained in Appendix~\ref{appendix__new_rv}. The final result is plotted in Fig.~\ref{image__orbit_ffaql} and the derived parameters are listed in Table~\ref{table_orbit_ffaql}. The final r.m.s. is small with only $90$\,m~s$^{-1}$. Our revised orbital parameters are in rather good agreement with the literature, within $1.5\sigma$ \citep{Gorynya_1995_05_0, Evans_1990_05_1,Abt_1959_11_0}.
	
	We noticed a slight difference in the systemic velocity between our estimate and the one from \citet{Gorynya_1995_05_0}. This might be linked to the wide companion, however, this might also be due to other non astrophysical effects. We should be cautious when studying long term variations of $v_\gamma$, unless a clear pattern is observed. Differences in $v_\gamma$ of the order of 0.5-1\,km~s$^{-1}$ might be caused, for instance, to the way RVs are determined (cross-correlation, bisector, ...), the mask used, the instrument zero points, etc... In this paper we did not perform such long term study as this is out of the scope of this paper, and would require a complete analysis of all available literature data.

	\paragraph{U~Car:}

	The companion orbiting this Cepheid is below our detection level. Our datasets enable however to exclude any companion with a flux ratio higher than 0.4\,\%, which would correspond to a spectral type earlier than B2V (see Table~\ref{table__limits}).

	\paragraph{Y~Car:}

	Our possibly detected companion has a flux ratio of $\sim 0.94$\,\%, and would correspond to a $\sim$A0V spectral type, in rather good agreement with the $\sim$B9V derived by \citet{Evans_1992_02_0}. Although additional data are still necessary to confirm, this possible detection seems in a very close orbit as we measured a projected separation of $\sim 2.5$\,mas ($\sim 3.5$\,au), as for S~Mus. 
	
	We first estimated the $3\sigma$ detection limit using all observables and removing the possible companion (because we have a detection), and  second using only the CPs but without removing it (because there is no detection). We noticed that the closure phases alone provide a contrast limit lower than 4\,mag (see Table~\ref{table__limits}), which explains the non-detection with only this observable. With all observables we add more constraints in this case, which provide contrast limit of 5\,mag (5.1\,mag within 25\,mas). This enables us to exclude at $3\sigma$ any companion with a spectral type earlier than A0V.
	
	\paragraph{YZ~Car:}
	
	The faint companion orbiting this Cepheid is not detected from our observations. Our $3\sigma$ detection limits (see Table~\ref{table__limits}) show that any companion with a spectral type later than B3V would not have been detected with this dataset, which is consistent with the expected range of spectral type for the companion (B8V-A0V). 
	
\begin{figure}[]
	\centering
	\resizebox{\hsize}{!}{\includegraphics{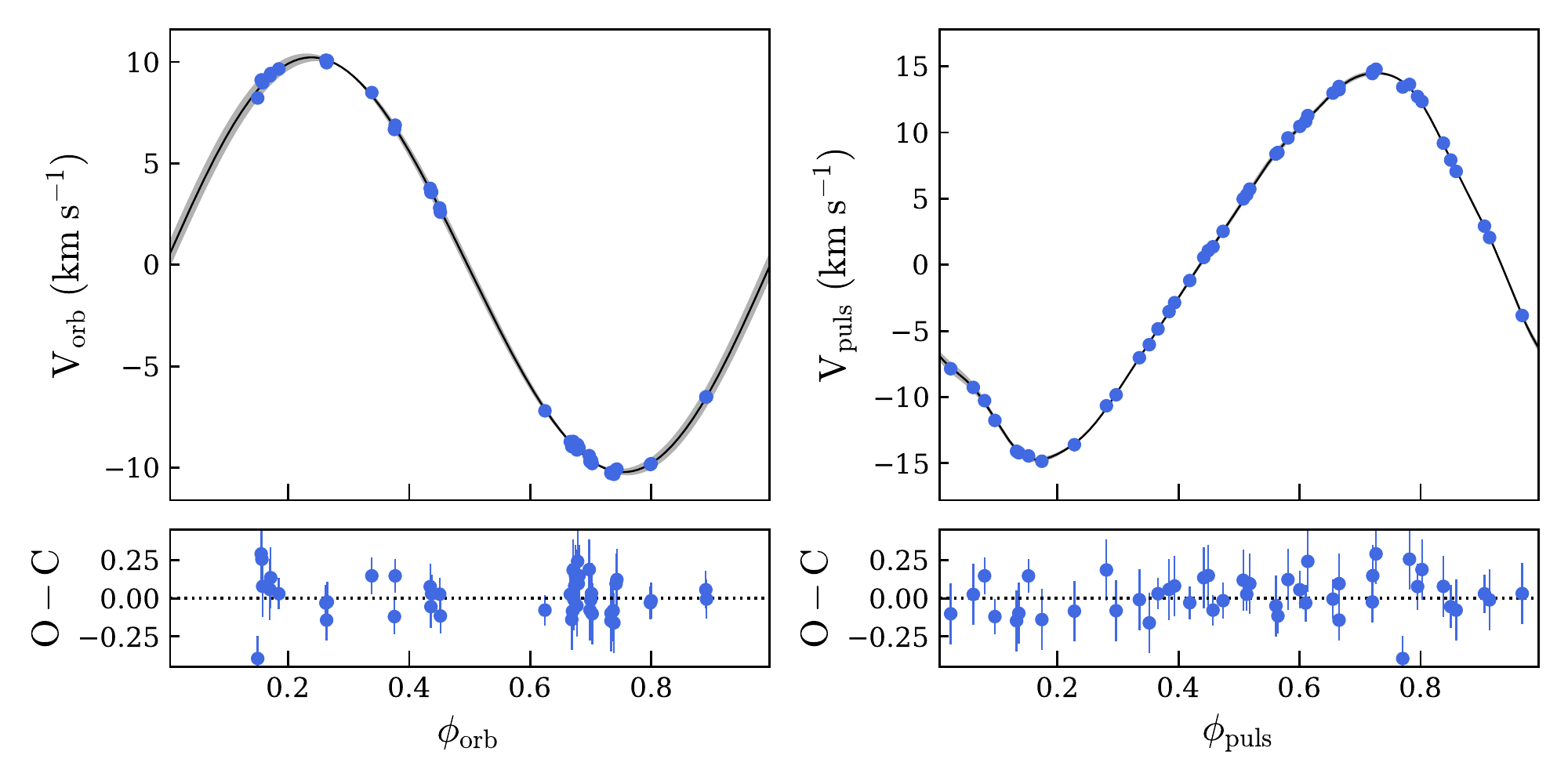}}
	\caption{Same as Fig.~\ref{image__orbit_ffaql} but for YZ~Car.}
	\label{image__orbit_yzcar}
\end{figure}

	New RV data were collected using the HARPS and CORALIE instruments (details in Appendix~\ref{appendix__new_rv}, and RVs listed in Table~\ref{table_yzcar_rv}), which span from 2013 to 2015. We used the same formalism as for FF~Aql to fit both the pulsation and orbital velocities. The pulsation phase coverage is not optimal to well constrain the pulsation and orbital fits, so we added additional velocities from \citet{Anderson_2016_10_0}, also obtained with CORALIE from 2014 to 2016. Before combining the data, we compared the systemic velocity of Anderson's data with ours, and we noticed a positive shift of $0.8\,\mathrm{km~s^{-1}}$. We therefore corrected for it and simultaneously fitted all dataset. Results are displayed in Fig.~\ref{image__orbit_yzcar} and listed in Table~\ref{table_orbit_yzcar}, with a final r.m.s of $90\,\mathrm{m~s^{-1}}$ ($190\,\mathrm{m~s^{-1}}$ without correcting for the shift). Note that we rescaled the velocity uncertainties of Anderson's data to $0.19\,\mathrm{km~s^{-1}}$ (r.m.s. of the residual without rescaling) to compensate for the use of a different binary mask in the RV determination. Our revised orbital solutions are in good agreement with \citet{Anderson_2016_10_0}. We therefore confirm the orbital period of $\sim 830\,$d, in disagreement with the 657\,d estimated by \citet{Petterson_2004_05_0}, as also noted by \citet{Anderson_2016_10_0}. As mentioned previously, \citet{Petterson_2004_05_0} used different dataset spanning several decades which can lead to biases if, for instance, period change is not taken into account.

	\paragraph{BP~Cir:}

\begin{table}[]
	\centering
	\caption{Final estimated parameters of the YZ~Car system.}
	\begin{tabular}{ccc} 
		\hline
		\hline
		\multicolumn{3}{c}{\textbf{Pulsation}}																	\\
		$P_\mathrm{puls}$ (days)							  &  \multicolumn{2}{c}{$18.1661 \pm0.0002$}\\
		$T_0$ (JD) \tablefootmark{a}						  &  \multicolumn{2}{c}{2~452~655.37}		\\ 
		$A_1$ ($\mathrm{km~s^{-1}}$)                                                  & \multicolumn{2}{c}{$-4.57 \pm 0.29$}            \\
		$B_1$ ($\mathrm{km~s^{-1}}$)                                                  & \multicolumn{2}{c}{$-13.26 \pm 0.09$}       \\
		$A_2$ ($\mathrm{km~s^{-1}}$)                                                   &    \multicolumn{2}{c}{$-0.71 \pm 0.03$}        \\
		$B_2$ ($\mathrm{km~s^{-1}}$)                                                   &    \multicolumn{2}{c}{$-0.81 \pm 0.02$}     \\
		$A_3$ ($\mathrm{km~s^{-1}}$)                                                  & \multicolumn{2}{c}{$-0.02 \pm 0.04$}            \\
		$B_3$ ($\mathrm{km~s^{-1}}$)                                                  & \multicolumn{2}{c}{$0.65 \pm 0.01$}     \\
		$A_4$ ($\mathrm{km~s^{-1}}$)                                                   &    \multicolumn{2}{c}{$-0.06 \pm 0.04$}     \\
		$B_4$ ($\mathrm{km~s^{-1}}$)                                                   &    \multicolumn{2}{c}{$0.36 \pm 0.01$}     \\
		$A_5$ ($\mathrm{km~s^{-1}}$)                                                   &    \multicolumn{2}{c}{$-0.15 \pm 0.01$}     \\
		$B_5$ ($\mathrm{km~s^{-1}}$)                                                   &    \multicolumn{2}{c}{$0.01 \pm 0.02$}        \\
		$A_6$ ($\mathrm{km~s^{-1}}$)                                                   &    \multicolumn{2}{c}{$-0.24 \pm 0.01$}     \\
		$B_6$ ($\mathrm{km~s^{-1}}$)                                                   &    \multicolumn{2}{c}{$-0.07 \pm 0.03$}        \\
		$A_7$ ($\mathrm{km~s^{-1}}$)                                                   &    \multicolumn{2}{c}{$-0.17 \pm 0.01$}     \\
		$B_7$ ($\mathrm{km~s^{-1}}$)                                                   &    \multicolumn{2}{c}{$-0.06 \pm 0.03$}        \\
		$A_8$ ($\mathrm{km~s^{-1}}$)                                                   &    \multicolumn{2}{c}{$-0.12 \pm 0.01$}     \\
		$B_8$ ($\mathrm{km~s^{-1}}$)                                                   &    \multicolumn{2}{c}{$-0.01 \pm 0.02$}        \\
		\hline
		\multicolumn{3}{c}{\textbf{Orbit}}  																			\\
		&  An16\tablefootmark{b}		& This work					 \\
		$P_\mathrm{orb}$ (days)						& $830.22 \pm 0.34$			& $831.6 \pm0.9$ 		\\
		$T_\mathrm{p}$ (JD)	& $2~453~422 \pm 29$		& $2~453~565.8 \pm 13.7$  \\
		$e$																& $0.041 \pm 0.010$     &  $0.035 \pm0.003$  	\\
		$\omega$	($^\circ$)									&$195 \pm 12$		 &	$260.5 \pm 6.8$		\\
		$K_1$ ($\mathrm{km~s^{-1}}$)                    & $10.26 \pm 0.82$      &  $10.249 \pm0.019$           \\
		$v_\gamma$  ($\mathrm{km~s^{-1}}$)           &  $0.84 \pm 0.06$          &  $0.02 \pm0.03$        \\
		$a_1~\sin{i}$ (au)											&	$0.783 \pm 0.063$ &		$0.783 \pm0.002$						\\
		$f(M)$ ($M_\odot$)										&	$0.093 \pm 0.041$	 &		$0.0926 \pm0.0005$	\\
		\hline																
	\end{tabular}
	\tablefoot{\tablefoottext{a}{Not fitted.} \tablefoottext{b}{\citet{Anderson_2016_10_0}.}
	}
	\label{table_orbit_yzcar}
\end{table}

	\begin{figure*}[!ht]
	\centering
	\resizebox{\hsize}{!}{\includegraphics{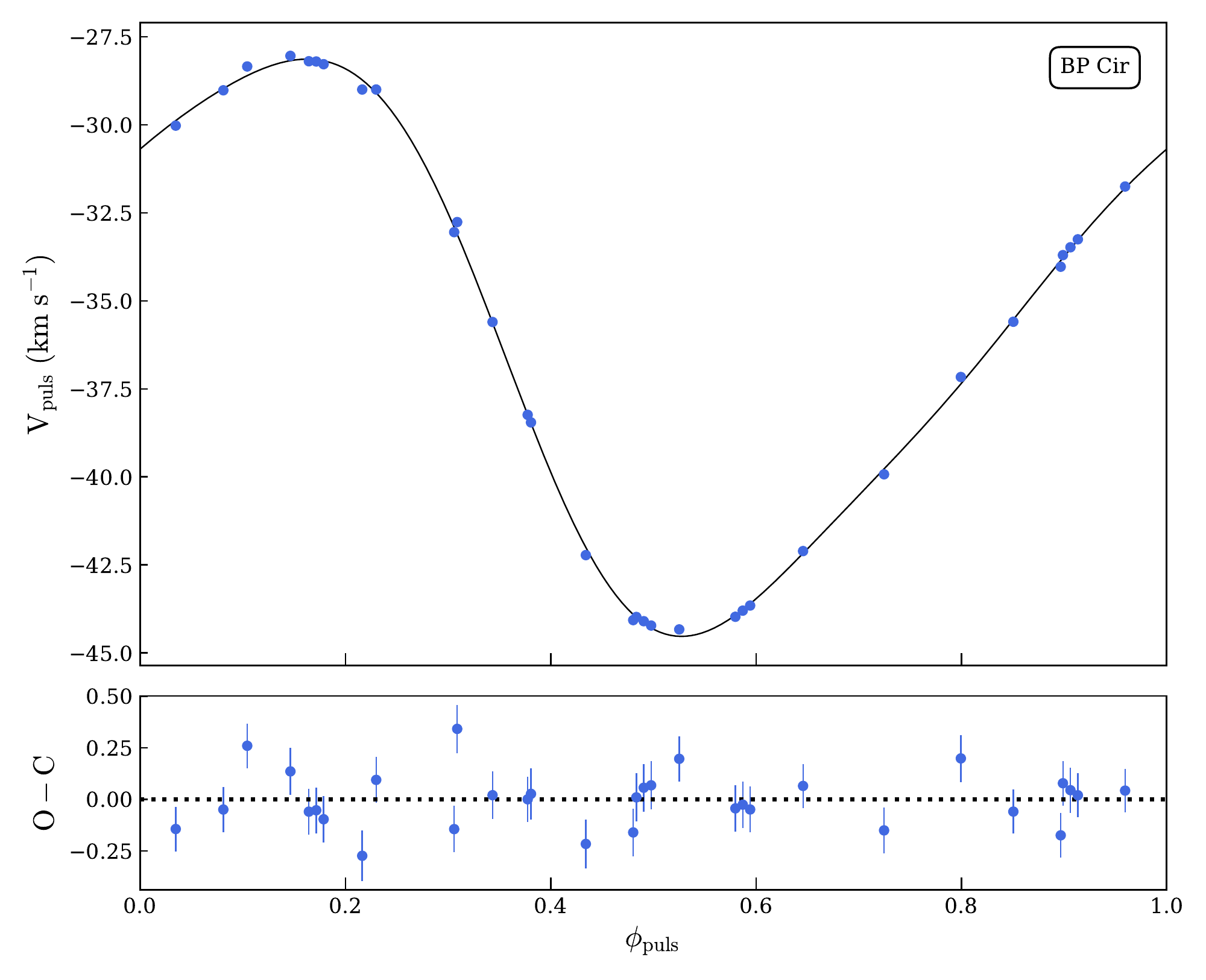}\includegraphics{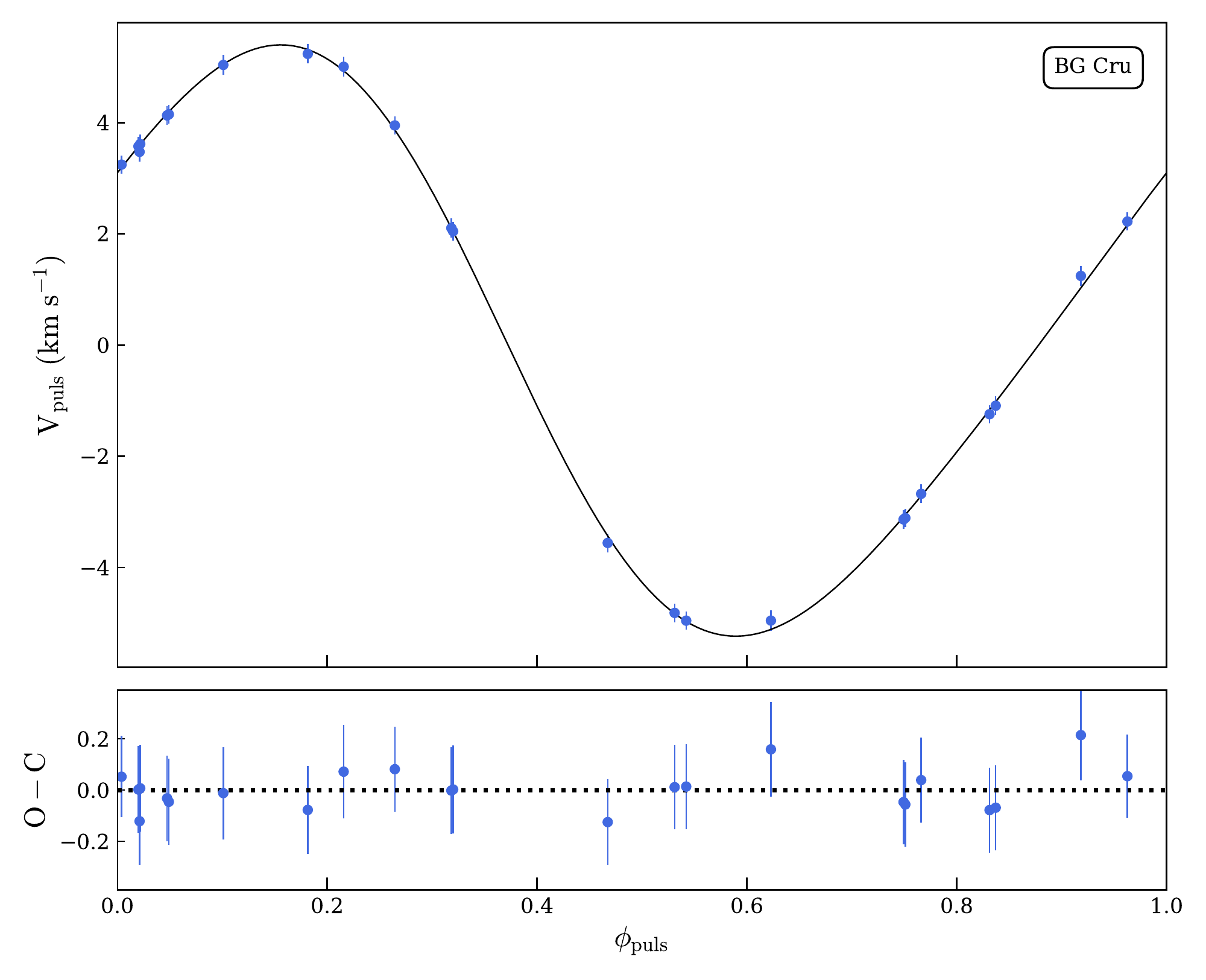}}
	\resizebox{\hsize}{!}{\includegraphics{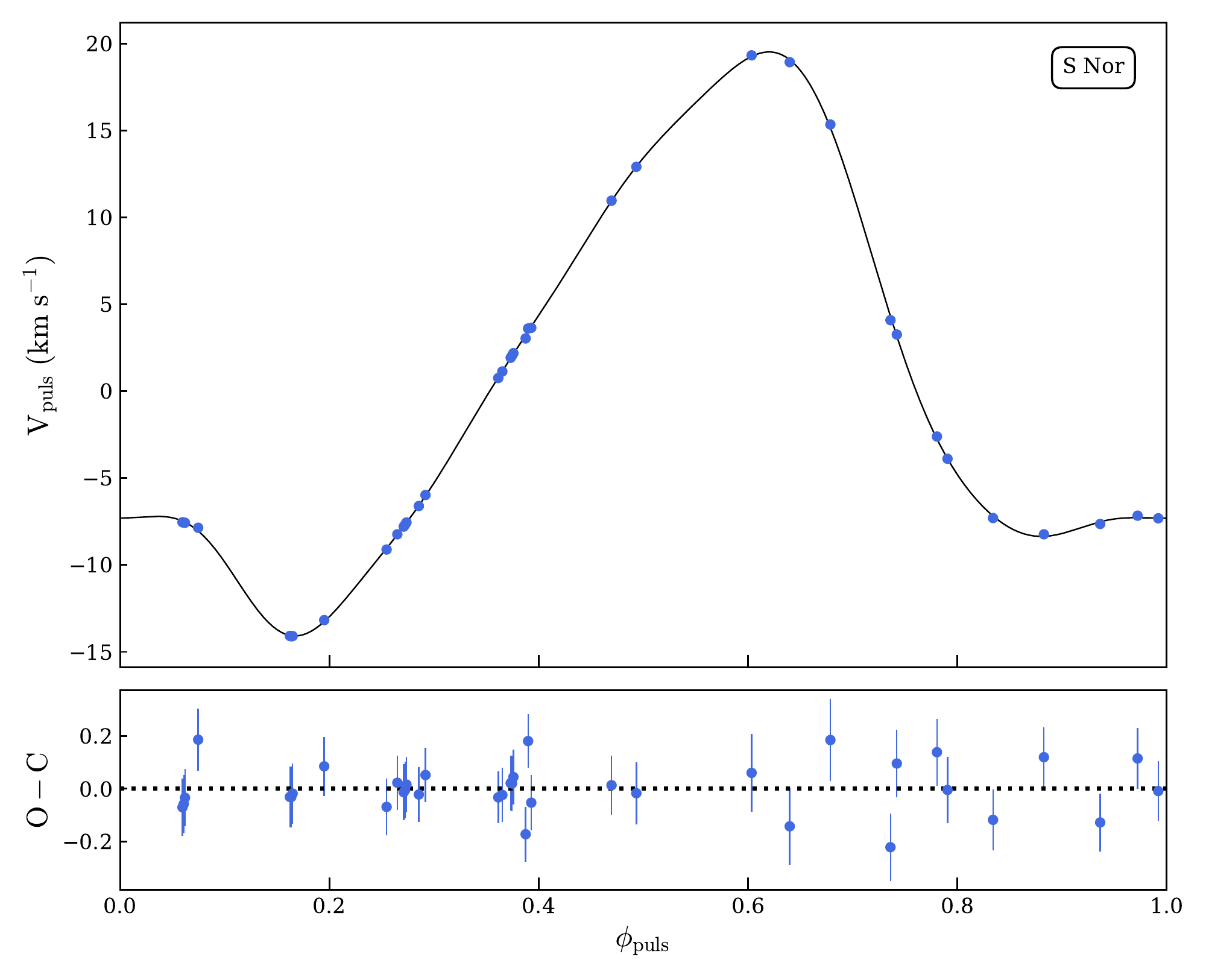}\includegraphics{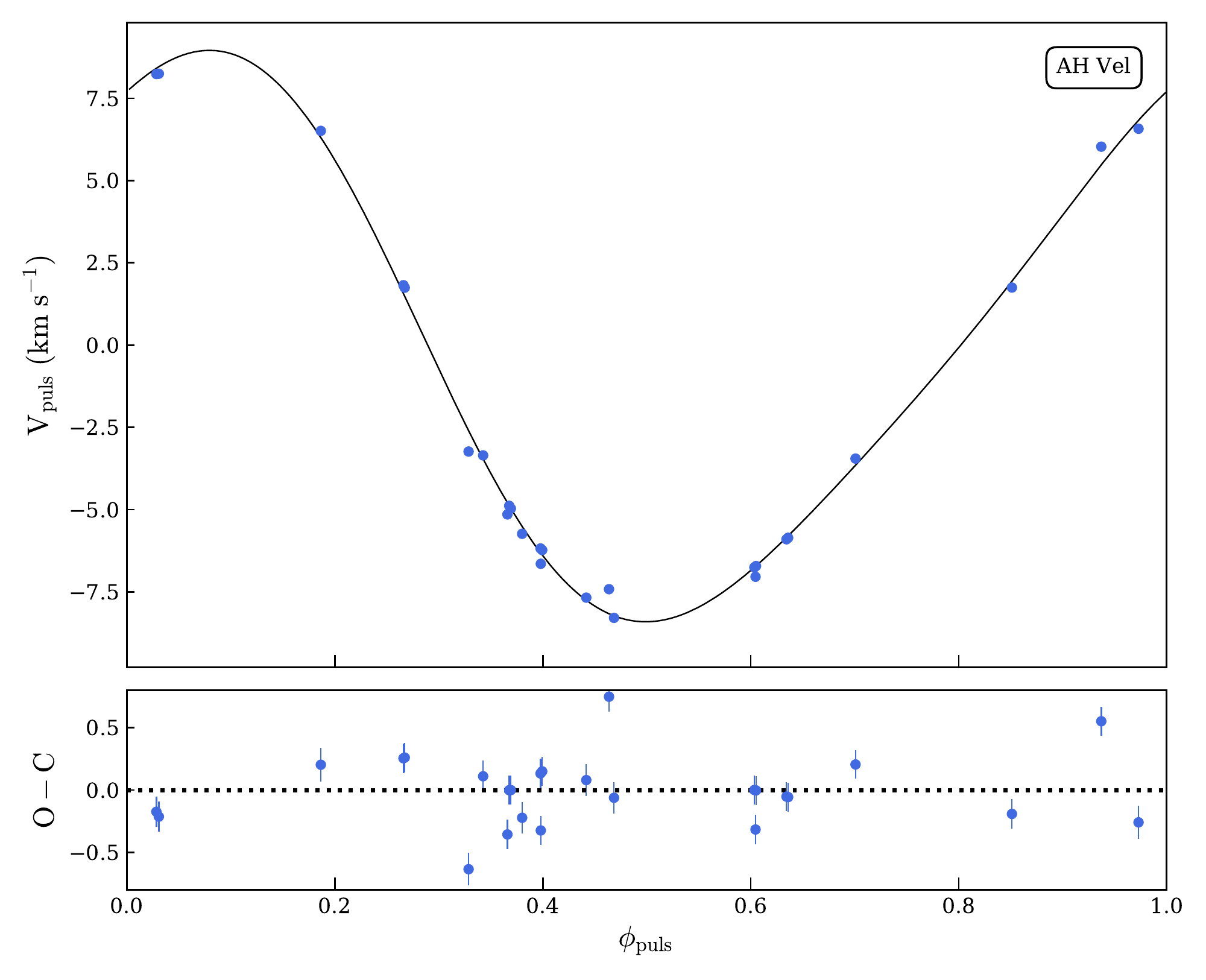} }
	\caption{Fitted radial pulsation velocity curve of BP~Cir, BG~Cru, S~Nor and AH~Vel.}
	\label{image__pulsation}
\end{figure*}

	\begin{table}[!h]
	\centering
	\caption{Final estimated pulsation parameters of BP~Cir, BG~Cru, S~Nor and AH~Vel.}
	\begin{tabular}{ccc} 
		\hline
		\hline
		\multicolumn{3}{c}{\textbf{BP~Cir}}																\\
		&	Pe04\tablefootmark{c}		& This work \\
		$P_\mathrm{puls}$ (days)                   &    $2.39819 \pm 0.00005$       &  $2.398131 \pm 0.000001$\\
		$T_0$ (JD)              							&    $2~444~297.03$     & $2~456~828.21 \pm 0.03$     \\
		$A_1$ ($\mathrm{km~s^{-1}}$)        &   --     & $-0.45 \pm 0.59$ \\
		$B_1$ ($\mathrm{km~s^{-1}}$)         &  --     & $7.80 \pm 0.05$ \\
		$A_2$ ($\mathrm{km~s^{-1}}$)       &   --   & $0.47 \pm 0.23$ \\
		$B_2$ ($\mathrm{km~s^{-1}}$)         &  --   & $-1.54 \pm 0.09$ \\
		$A_3$ ($\mathrm{km~s^{-1}}$)       &   --   & $-0.26 \pm 0.074$ \\
		$B_3$ ($\mathrm{km~s^{-1}}$)         &  --   & $0.33 \pm 0.07$ \\
		$v_\gamma$ ($\mathrm{km~s^{-1}}$)	&	$19.0$ 	& $18.00 \pm 0.11$ \\
		\hline
		\multicolumn{3}{c}{\textbf{BG~Cru}}																\\
		&	Sz89		& This work \\
		$P_\mathrm{puls}$ (days)                   &    $3.34272 \pm 0.00001$       &  $3.342536 \pm 0.000003$\\
		$T_0$ (JD)              							&    $2~440~393.660 \pm 0.08$     & $2~456~753.93 \pm 0.09$     \\
		$A_1$ ($\mathrm{km~s^{-1}}$)        &   --     & $3.48 \pm 0.68$ \\
		$B_1$ ($\mathrm{km~s^{-1}}$)         &  --     & $3.86 \pm 0.52$ \\
		$A_2$ ($\mathrm{km~s^{-1}}$)       &   --   & $-0.58 \pm 0.03$ \\
		$B_2$ ($\mathrm{km~s^{-1}}$)         &  --   & $0.02 \pm 0.19$ \\
		$v_\gamma$ ($\mathrm{km~s^{-1}}$)	&	$-19.6 \pm 0.3$\tablefootmark{a} 	& $-20.00 \pm 0.18$ \\
		\hline
		\multicolumn{3}{c}{\textbf{S~Nor}}																\\
		&	Me87\tablefootmark{d}		& This work \\
		$P_\mathrm{puls}$ (days)                   &    $9.7544 \pm 0.0005$       &  $9.75473 \pm 0.00001$\\
		$T_0$ (JD)              							&    $2~445~397.507$     & $2~444~018.13 \pm 0.05$     \\
		$A_1$ ($\mathrm{km~s^{-1}}$)        &   --     & $-12.25 \pm 0.20$ \\
		$B_1$ ($\mathrm{km~s^{-1}}$)         &  --     & $-6.85 \pm 0.37$ \\
		$A_2$ ($\mathrm{km~s^{-1}}$)       &   --   & $3.39 \pm 0.20$ \\
		$B_2$ ($\mathrm{km~s^{-1}}$)         &  --   & $3.53 \pm 0.22$ \\
		$A_3$ ($\mathrm{km~s^{-1}}$)       &   --   & $2.57 \pm 0.03$ \\
		$B_3$ ($\mathrm{km~s^{-1}}$)         &  --   & $-0.24 \pm 0.23$ \\
		$A_4$ ($\mathrm{km~s^{-1}}$)        &   --     & $-0.33 \pm 0.01$ \\
		$B_4$ ($\mathrm{km~s^{-1}}$)         &  --     & $-0.10 \pm 0.04$ \\
		$A_5$ ($\mathrm{km~s^{-1}}$)       &   --   & $-0.36 \pm 0.09$ \\
		$B_5$ ($\mathrm{km~s^{-1}}$)         &  --   & $0.58 \pm 0.05$ \\
		$A_6$ ($\mathrm{km~s^{-1}}$)       &   --   & $-0.05 \pm 0.01$ \\
		$B_6$ ($\mathrm{km~s^{-1}}$)         &  --   & $-0.01 \pm 0.01$ \\
		$A_7$ ($\mathrm{km~s^{-1}}$)       &   --   & $-0.24 \pm 0.03$ \\
		$B_7$ ($\mathrm{km~s^{-1}}$)         &  --   & $-0.17 \pm 0.05$ \\
		$A_8$ ($\mathrm{km~s^{-1}}$)       &   --   & $-0.04 \pm 0.02$ \\
		$B_8$ ($\mathrm{km~s^{-1}}$)         &  --   & $-0.09 \pm 0.01$ \\
		$v_\gamma$ ($\mathrm{km~s^{-1}}$)	&	$5.85 \pm 0.06$ 	& $5.14 \pm 0.12$ \\
		\hline
		\multicolumn{3}{c}{\textbf{AH~Vel}}																	\\
		&	Sz89\tablefootmark{b}		& This work \\
		$P_\mathrm{puls}$ (days)                   &    $4.227231 \pm 0.000007$       &  $4.227527 \pm 0.000001$\\
		$T_0$ (JD)              							&    $2~442~035.703 \pm 0.07$    & $2~456~605.05 \pm 0.08$     \\
		$A_1$ ($\mathrm{km~s^{-1}}$)        &   --     & $7.99 \pm 0.36$ \\
		$B_1$ ($\mathrm{km~s^{-1}}$)         &  --     & $2.50 \pm 0.93$ \\
		$A_2$ ($\mathrm{km~s^{-1}}$)       &   --   & $-0.42 \pm 0.26$ \\
		$B_2$ ($\mathrm{km~s^{-1}}$)         &  --   & $1.15 \pm 0.13$ \\
		$v_\gamma$ ($\mathrm{km~s^{-1}}$) &	$24.5 \pm 0.3$\tablefootmark{a} 	& $24.68 \pm 0.13$ \\
		\hline
	\end{tabular}
	\tablefoot{\tablefoottext{a}{From \citet{Gieren_1977_05_0} for AH~Vel, from \citet{Stobie_1979_12_0} for BG~Cru.} \tablefoottext{b}{\citet{Szabados_1989_01_0}.}
		\tablefoottext{c}{\citet{Petterson_2004_05_0}.} \tablefoottext{d}{\citet{Mermilliod_1987_09_0}.}
	}
	\label{table_pulsation}
\end{table}

	We confirm the presence of a companion orbiting the Cepheid. Our measured flux ratio is in very good agreement with the detection from IUE spectra \citep{Evans_1994_11_0} for a B6V star. From our measured projected separation, $\rho$, and the Kepler's third law, we can estimate a lower limit for the orbital period as we know that the semi-major axis $a > \rho$. Adopting a mass $M_2 = 4.7\,M_\odot$ for the companion, $M_1 = 4.9\,M_\odot$ for the Cepheid \citep{Evans_2013_10_0}, the distance $d = 850$\,pc and 15\,\% uncertainty on those values, we found $P_\mathrm{orb} \gtrsim 40$\,years. 
	
	There is no existing infrared light curve for BP Cir to estimate its magnitude at our given pulsation phase. We therefore took the value $m_\mathrm{H} = 5.58 \pm 0.04$\,mag given by \citet[2MASS catalog][]{Cutri_2003_03_0}. To take into account the phase mismatch with the mean magnitude, we also quadratically added a conservative uncertainty of 0.06\,mag, which corresponds to half the amplitude of the light curve in the $I$ band \citep[][probably smaller in $H$]{Berdnikov_2008_04_0}. We estimated the magnitudes $m_\mathrm{H} (\mathrm{comp}) = 9.35 \pm 0.10$\,mag and $m_\mathrm{H} (\mathrm{cep}) = 5.61 \pm 0.07$\,mag. 
	
	According to our estimated detection limits listed in Table~\ref{table__limits}, we did not detect additional companions within 50\,mas with a flux ratio larger than 1.3\,\%, corresponding to an upper limit of approximately A2V for the spectral type (assuming $d = 850$\,pc).
	
	Spectra for this Cepheid were also obtained with the HARPS and SOPHIE spectrographs from 2013 to 2015 (see Table~\ref{table_bpcir_rv}). Differently to our previous analysis, for which we had first guesses of the orbital parameters, we first analysed our RVs by fitting only the pulsation of the Cepheid with Fourier series as described in Eq.~(\ref{equation_vpuls}), in which we added the systemic velocity $v_\gamma$. The curve is displayed in Fig.~\ref{image__pulsation}, and the fitted parameters are listed in Table~\ref{table_pulsation}. Here, we performed a simple Monte Carlo simulation by randomly creating 1~000 synthetic RVs around our best fit solutions. We used normal distributions with standard deviations corresponding to the measurement uncertainties. We then took the median value of the distributions and used the maximum value between the 16th and 84th percentiles as uncertainty estimates. We used only $n = 3$ Fourier coefficients as the fit is not improved by using an order 4. The residuals are very small with a r.m.s of $\mathrm{0.14\,km~s^{-1}}$. This confirms a very long orbital period, which is also consistent with the location of our detected component. To search for a sign of modulation, we generated a Lomb-Scargle periodogram\footnote{We used the \emph{astroML} python module and its bootstrapping package to find the significance levels  \citep{VanderPlas_2012_10_0}} including older RV measurements \citep[][and ignoring a possible period change]{Balona_1981_12_0,Petterson_2004_05_0,Petterson_2005_10_0}. We restricted our period search to twice the longest time span of the data (i.e. $P_\mathrm{max} = 2~(t_\mathrm{max} - t_\mathrm{min})$, to cover at least half an orbital period), while the minimum period is set to 400\,days \citep[about the shortest possible orbital period for Cepheids in binary systems, according to][]{Neilson_2015_02_0}. We found several peaks with a high probability (false alarm probability FAP $ < 0.1$\,\%), the strongest being at a period of $\sim 14680$\,days, then 6790 and 4500\,days. The two last values are too small to be consistent with our measured astrometric position, but the first one is in agreement with the lower limit estimated above. We stress that such analysis assumes a zero eccentricity and ignores pulsation period change. We unfortunately cannot yet determine the orbital period only from spectroscopy, but additional astrometric measurements will better constrain the orbit in a few years.

	\paragraph{BG~Cru:}
	
	The expected companion should have a spectral type later than a A1V star according to the detection limit set by \citet{Evans_1992_01_0}. This means that the flux ratio should be $\lesssim 0.5$\,\%. Our measured flux ratio for this possible detection is $0.53\pm0.12$\,\%, in agreement with this detection limit, and corresponds to a companion with a spectral type in the range B9V-A4V.
	
	From our detection limits, we reached the same conclusion as \citet{Evans_1992_01_0} within 50\,mas from the Cepheid, i.e. there is no companion with a spectral type earlier than A2V.
	
	We performed the same analysis as for BP~Cir with new high-precision RV data (see Appendix~\ref{appendix__new_rv}). The pulsation curve is displayed in Fig.~\ref{image__pulsation} and the fitted parameters in Table~\ref{table_pulsation}. The residuals are small ($0.8\,\mathrm{km~s^{-1}}$), suggesting no spectroscopic companion, a very long orbital period or a high orbital inclination. The periodogram for these data do not show sign of any significant peak (FAP = 9.1\,\%). We included older (less precise) RVs collected from the literature \citep{Lloyd-Evans_1980__0,Stobie_1979_12_1,Usenko_2014_07_0} and calculated the periodogram. There is a significant peak (with a FAP $< 0.1$\,\%) at a period of $\sim 5365$\,days. This period would be consistent with \citet{Szabados_1989_01_0} who found a $\sim 5000\,$day pattern from the light time effect. However we cannot exclude other peaks with similar significant levels (with FAP $< 1\,$\%). Additional high-precision RVs will be necessary to confirm this period.

	\paragraph{T~Mon:} 

	The spectroscopic companion is also below our sensitivity limit. Our estimated detection level is $\Delta H \sim 4.7$\,mag, well below the 8.1\,mag required to detect it. However, although a brighter component is unlikely, we can still rule out an orbiting companion brighter than $H \sim 7.4$\,mag, which would correspond to a companion earlier than a B1V star (see Table~\ref{table__limits}).

	\paragraph{R~Mus:} 
	
	We cannot confirm from our interferometric observations the presence of the spectroscopic companion. Its expected flux ratio is $f_\mathrm{H} \lesssim 0.4$\,\%, while our detection limits show that with this dataset we were sensitive to the companion having $f_\mathrm{H} > 1$\,\%. We can however reject the presence of any companion brighter than a B8V star within 50\,mas.
	
	\paragraph{S~Mus:} 
	
	We confirm the presence of a close-in companion. The contrast is slightly higher than the expected value, a $\sim$~B6V being more appropriate, but detections in more photometric bands are necessary to better constrain the spectral type. Our estimated detection limits enable us to exclude any possible third component with a spectral type earlier than B9V.
	
\begin{figure*}[]
		\centering
		\resizebox{\hsize}{!}{\includegraphics{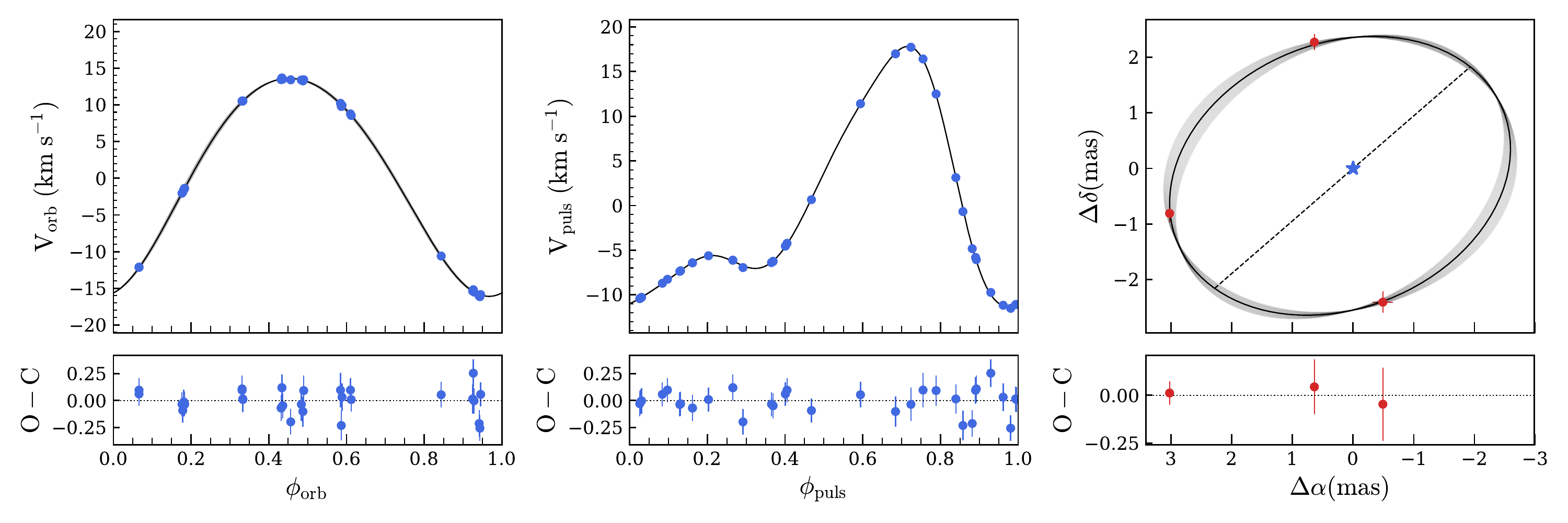}}
		\caption{Result of our combined fit for S~Mus. Left: fitted (solid lines) and measured primary (blue dots) orbital velocity. Middle: fitted (solid line) and measured (blue dots) radial pulsation velocity. Right: relative astrometric orbit of S~Mus Ab.}
		\label{image__orbit_smus}
	\end{figure*}

	As for U~Aql, we performed a preliminary MCMC orbital fit by combining our astrometry with single-line RVs to solve for all the orbital elements. The resulting parameters are listed in Table~\ref{table_orbit_smus} and plotted in Fig~\ref{image__orbit_smus}. We remove the degeneracy between the mass and the distance using a P-L relation \citep{Storm_2011_10_0}, giving $d = 858\pm17$\,pc. First guess values were taken from \citet{Petterson_2004_05_0} and \citet{Samus_2017_01_0}.
	
	Our estimated mass ratio is in agreement with the 0.88 derived by \citet{Bohm-Vitense_1997_03_0}. The Cepheid mass is in slight agreement with the $6.0\pm0.4\,M_\odot$ estimate by \citet{Evans_2004_05_0} from the FUSE spectra, but this is expected as they used $q = 0.88$. Our derived companion mass is also smaller than the one for a B3V star ($\sim 5.3\,M_\odot$), and would be more compatible with a B6V star, which is also consistent with our measured flux ratio. However, this is still preliminary and the astrometric orbit will have a better coverage soon.
	
	Finally, let us also note that from our average $3\sigma$ contrast limits (see Table~\ref{table__limits}), we exclude any additional companion within 50\,mas from the Cepheid with a spectral type earlier than B9V.
	
	\begin{table}[]
		\centering
		\caption{Final estimated parameters of the S~Mus system.}
		\begin{tabular}{ccc} 
			\hline
			\hline
			\multicolumn{3}{c}{\textbf{Pulsation}}																	\\
			$P_\mathrm{puls}$ (days)							  &  \multicolumn{2}{c}{$9.65996\pm0.00006$} \\
			$T_0$ (JD) \tablefootmark{a}						  &  \multicolumn{2}{c}{2~440~299.42}		\\
			$A_1$ ($\mathrm{km~s^{-1}}$)					&	\multicolumn{2}{c}{$-6.94\pm0.82$}			\\
			$B_1$ ($\mathrm{km~s^{-1}}$)					&	\multicolumn{2}{c}{$-9.81\pm0.54$}		\\
			$A_2$ ($\mathrm{km~s^{-1}}$)				&	\multicolumn{2}{c}{$-4.53\pm0.67$}		\\
			$B_2$ ($\mathrm{km~s^{-1}}$)					&	\multicolumn{2}{c}{$3.87\pm0.79$}		\\
			$A_3$ ($\mathrm{km~s^{-1}}$)				&	\multicolumn{2}{c}{$-0.60\pm0.34$}			\\
			$B_3$ ($\mathrm{km~s^{-1}}$)				&	\multicolumn{2}{c}{$1.46\pm0.15$}		\\
			$A_4$ ($\mathrm{km~s^{-1}}$)			&	\multicolumn{2}{c}{$0.86\pm0.10$}		\\
			$B_4$ ($\mathrm{km~s^{-1}}$)			&	\multicolumn{2}{c}{$0.23\pm0.24$}		\\
			$A_5$ ($\mathrm{km~s^{-1}}$)			&	\multicolumn{2}{c}{$0.21\pm0.04$}		\\
			$B_5$ ($\mathrm{km~s^{-1}}$)			&	\multicolumn{2}{c}{$-0.05\pm0.10$}		\\
			\hline
			\multicolumn{3}{c}{\textbf{Orbit}}  																			\\
			&	Pe04 \tablefootmark{b}		& This work	\\
			$P_\mathrm{orb}$ (days)						&	$504.9\pm0.07$		& $506.3\pm0.5$ 	\\
			$T_\mathrm{p}$ (JD)								&	$2~448~590 \pm 5$	& $2~457~165.9\pm4.4$ \\
			$e$															&	$0.080 \pm 0.002$     &  $0.088 \pm 0.006$ \\
			$\omega$	($^\circ$)								&  $206 \pm 5$	 &	$194.8 \pm 3.3$	  \\
			$K_1$ ($\mathrm{km~s^{-1}}$)		&	$14.7 \pm 0.2$		  &	$14.85 \pm 0.03$ \\
			$K_2$ ($\mathrm{km~s^{-1}}$)\tablefootmark{c}		&		--	  &	$16.63 \pm 3.27$	 \\
			$v_\gamma$	($\mathrm{km~s^{-1}}$)&	  $-0.5 \pm 0.5$		&	$-1.65 \pm 0.11$		\\
			$\Omega$	($^\circ$)								&	-- &	$99.6 \pm 14.4$	 \\
			$i$ ($^\circ$)											&	--	&	$144.7 \pm 2.8$	  \\
			$a$ (mas)												&	--	 &	$2.95 \pm 0.09$	  \\
			$a$ (au)												&	--	 &	$2.53 \pm 0.09$	  \\
			$M_1$ ($M_\odot$)						&		--			&	$4.44\pm0.91$	  \\
			$M_2$ ($M_\odot$)						&		--			&	$3.98\pm0.21$	  \\
			$d$ (pc)\tablefootmark{d}										&		--			&	$858\pm17$			\\
			\hline
		\end{tabular}
		\tablefoot{\tablefoottext{a}{held fixed with the value from \citet{Samus_2017_01_0}.}
		\tablefoottext{b}{\citet{Petterson_2004_05_0}.}
		\tablefoottext{c}{derived from the orbital elements and the assumed distance.}
		\tablefoottext{d}{assumed from the P-L relation of \citet{Storm_2011_10_0}.}
		}
		\label{table_orbit_smus}
	\end{table}
	
	\paragraph{S~Nor:} 
	
	The possible spectroscopic component is expected to have a contrast $\sim 0.3$\,\% ($\Delta H \sim 6.4$\,mag), but this is just below the sensitivity limit of our datasets (see Table~\ref{table__limits}), which is consistent with our non-detection. We are therefore not able to confirm the presence of a close-in companion. From our detection limit, we can exclude a companion brighter than $H = 5.6$\,mag within 25\,mas, and brighter than 5.3\,mag within 25-50\,mas, which correspond to stars with spectral type earlier than B7V and B5V, respectively.
	
	New RVs were also collected for this Cepheid from 2013 to 2018 using both the CORALIE and HARPS spectrographs (see Appendix~\ref{appendix__new_rv} for details). RVs are listed in Table~\ref{table_snor_rv}. These observations span half the period of $\sim 10$\,yrs given by \citet{Groenewegen_2008_09_0}, so the orbital motion should be detected from our new observations. As previously, we first fitted the pulsation of the star. The parameters and the pulsation curve are in Table~\ref{table_pulsation} and Fig.~\ref{image__pulsation}, respectively. The r.m.s of the fit is $\mathrm{90\,m~s^{-1}}$. After correcting for the pulsation, the residual periodogram does not show any significant peak, the highest being at a too short period of 760\,d with a FAP = 3\,\%. We increased the time span by including RVs from \citet[][with a zero point correction as given by \citealt{Udry_1999__0}]{Bersier_1994_11_0}, but there is also no significant peak in the residual periodogram, the highest being at 836\,d with a FAP = 14\,\%. We suggest that S~Nor is probably not a spectrosopic binary. The other possibility would be that this hypothetical companion has a very low mass, and might have a high orbital inclination.

	\paragraph{W~Sgr:}

	\begin{figure}[]
	\centering
	\resizebox{\hsize}{!}{\includegraphics{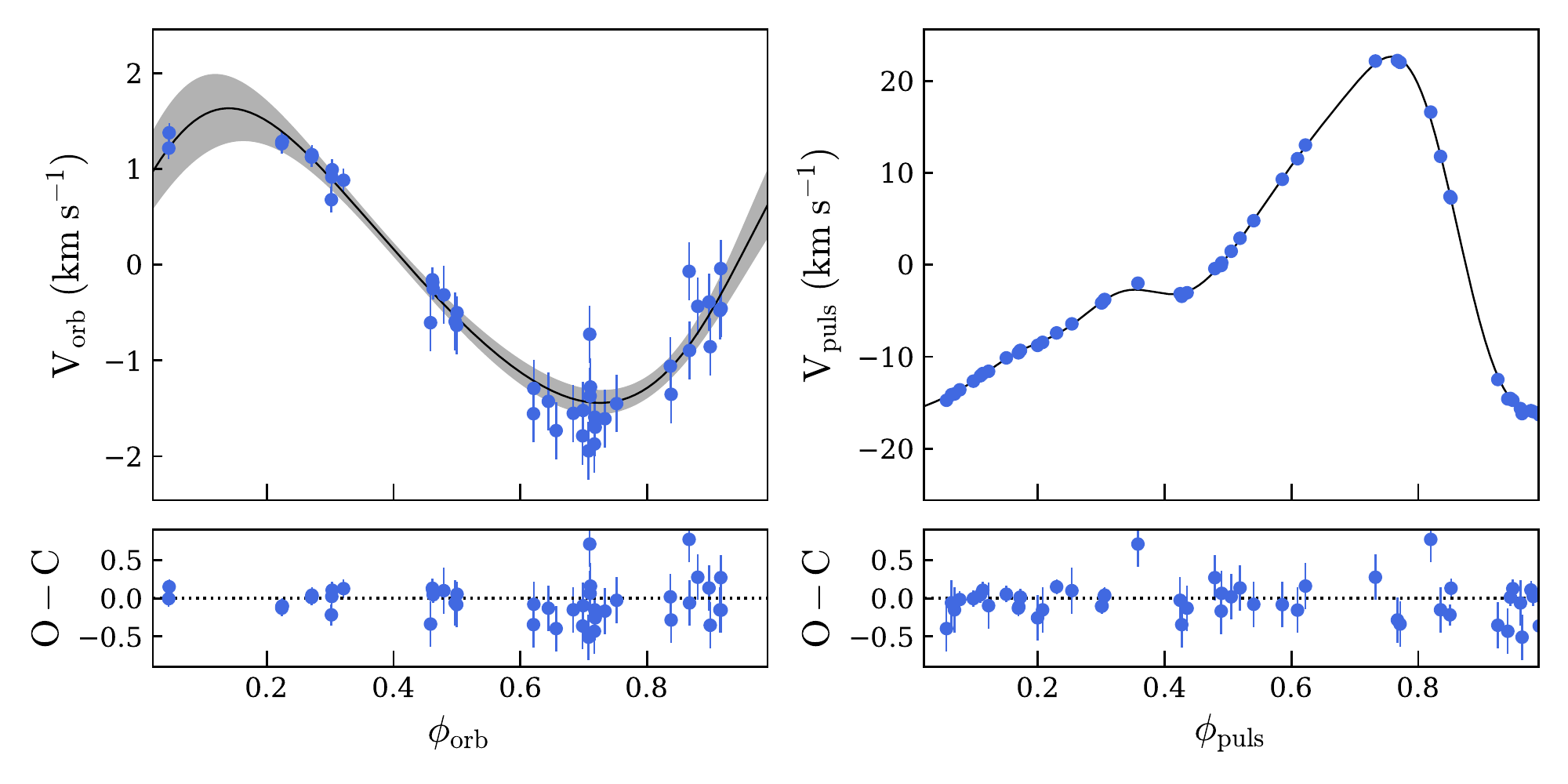}}
	\caption{Same as Fig.~\ref{image__orbit_ffaql} but for W~Sgr.}
	\label{image__orbit_wsgr}
\end{figure}

\begin{table}[]
	\centering
	\caption{Final estimated parameters of the W~Sgr system.}
	\begin{tabular}{ccc} 
		\hline
		\hline
		\multicolumn{3}{c}{\textbf{Pulsation}}																	\\
		$P_\mathrm{puls}$ (days)							  &  \multicolumn{2}{c}{$7.59508 +/- 0.00002$}\\
		$T_0$ (JD) \tablefootmark{a}						  &  \multicolumn{2}{c}{2~443~374.77}		\\ 
		$A_1$ ($\mathrm{km~s^{-1}}$)												  &	\multicolumn{2}{c}{$-9.86 \pm 0.33$}			\\
		$B_1$ ($\mathrm{km~s^{-1}}$)												  &	\multicolumn{2}{c}{$-10.70 \pm 0.31$}		\\
		$A_2$ ($\mathrm{km~s^{-1}}$)												   &	\multicolumn{2}{c}{$-7.47 \pm 0.23$}		\\
		$B_2$ ($\mathrm{km~s^{-1}}$)												   &	\multicolumn{2}{c}{$3.38 \pm 0.46$}		\\
		$A_3$ ($\mathrm{km~s^{-1}}$)												  &	\multicolumn{2}{c}{$0.31 \pm 0.24$}			\\
		$B_3$ ($\mathrm{km~s^{-1}}$)												  &	\multicolumn{2}{c}{$2.59 \pm 0.04$}		\\
		$A_4$ ($\mathrm{km~s^{-1}}$)												   &	\multicolumn{2}{c}{$1.12 \pm 0.17$}		\\
		$B_4$ ($\mathrm{km~s^{-1}}$)												   &	\multicolumn{2}{c}{$1.30 \pm 0.15$}		\\
		$A_5$ ($\mathrm{km~s^{-1}}$)												   &	\multicolumn{2}{c}{$0.40 \pm 0.09$}		\\
		$B_5$ ($\mathrm{km~s^{-1}}$)												   &	\multicolumn{2}{c}{$-0.57 \pm 0.06$}		\\
		$A_6$ ($\mathrm{km~s^{-1}}$)												   &	\multicolumn{2}{c}{$0.17 \pm 0.08$}		\\
		$B_6$ ($\mathrm{km~s^{-1}}$)												   &	\multicolumn{2}{c}{$-0.43 \pm 0.04$}		\\
		$A_7$ ($\mathrm{km~s^{-1}}$)												   &	\multicolumn{2}{c}{$0.03 \pm 0.01$}		\\
		$B_7$ ($\mathrm{km~s^{-1}}$)												   &	\multicolumn{2}{c}{$-0.21 \pm 0.04$}		\\
		$A_8$ ($\mathrm{km~s^{-1}}$)												   &	\multicolumn{2}{c}{$-0.05 \pm 0.02$}		\\
		$B_8$ ($\mathrm{km~s^{-1}}$)												   &	\multicolumn{2}{c}{$0.06 \pm 0.01$}		\\
		$A_9$ ($\mathrm{km~s^{-1}}$)												   &	\multicolumn{2}{c}{$0.12 \pm 0.02$}		\\
		$B_9$ ($\mathrm{km~s^{-1}}$)												   &	\multicolumn{2}{c}{$0.06 \pm 0.03$}		\\
		$A_{10}$ ($\mathrm{km~s^{-1}}$)												   &	\multicolumn{2}{c}{$0.06 \pm 0.01$}		\\
		$B_{10}$ ($\mathrm{km~s^{-1}}$)												   &	\multicolumn{2}{c}{$-0.04 \pm 0.01$}		\\
		\hline
		\multicolumn{3}{c}{\textbf{Orbit}}  																			\\
		&  Pe04\tablefootmark{b}		& This work					 \\
		$P_\mathrm{orb}$ (days)						& $1582 \pm 9$			& $1615.5 \pm 11.0$ 		\\
		$T_\mathrm{p}$ (JD)							&   2~448~286			 &	$2~457~992.4 \pm 19.5$	 \\ 
		$e$																& $0.41 \pm 0.05$     &  $0.197 \pm 0.018$  	\\
		$\omega$	($^\circ$)									&$328 \pm 5$		 &	$288.4 \pm 5.7$		\\
		$K_1$ ($\mathrm{km~s^{-1}}$)                    & $1.08 \pm 0.03$      &  $1.562 \pm 0.011$           \\
		$v_\gamma$  ($\mathrm{km~s^{-1}}$)           &  $-26.0 \pm 0.1$          &  $-27.87 \pm 0.02$        \\
		$a_1~\sin{i}$ (au)											&	$0.143$ &		$0.228 \pm 0.003$						\\
		$f(M)$ ($M_\odot$)										&	$0.157\,(\times10^3)$	 &		$0.60 \pm 0.02\,(\times10^3$)	\\
		\hline																
	\end{tabular}
	\tablefoot{\tablefoottext{a}{Not fitted.} \tablefoottext{b}{\citet{Petterson_2004_05_0}.}
	}
	\label{table_orbit_wsgr}
\end{table}

	The spectroscopic component is not detected from our interferometric observations. We estimated the $3\sigma$ detection level to be $\Delta H = 5.1$\,mag (see Table~\ref{table__limits}), which is not enough to detect the F5V companion ($\Delta H \sim 7.7$\,mag). However, this limit enables us to rule out any component with a spectral type earlier than B8V.
	
	New RVs were collected for W~Sgr from the HARPS, CORALIE and HERMES instruments (details in Appendix~\ref{appendix__new_rv} and RVs are given in Table~\ref{table_wsgr_rv}), spanning from 2013 to 2017. We did not use RVs measurements from the literature as explained previously. The combined fit is displayed in Fig.~\ref{image__orbit_wsgr}, and the revised orbit in Table~\ref{table_orbit_wsgr}. The same fitting formalism as for FF~Aql was applied. The residual standard deviation is $80\,\mathrm{m~s^{-1}}$. Our newly derived orbit is in slight agreement with previous determinations \citep{Babel_1989_06_0,Bersier_1994_11_0,Albrow_1996_06_0}. However, such low-amplitude orbit is difficult to constrain as it needs continuous high-precision measurements to allow a good determination of the orbit. Combining several datasets is usually not optimal as several additional effects can alter the results, as for instance the presence of a third component, the pulsation period change, or the method used to estimate the RVs. To well constrain such low amplitude binary, this is critical to control such effects.
	
	\paragraph{X~Sgr:}
	
	The spectroscopic component is possibly detected at a separation of $\sim 14$\,mas with the closure phase signal only, but additional observations are necessary to firmly conclude. Our measured flux ratio is compatible with a B9-A2V star, in agreement with the limit set by \citet{Evans_1992_01_0}. From our estimated detection limit (Table~\ref{table__limits}), we are also able to exclude any additional component with a contrast lower than $\sim 1:140$, which would correspond to companions with a spectral type earlier than B8V.

	\paragraph{V350~Sgr:}

	\begin{figure}[]
	\centering
	\resizebox{\hsize}{!}{\includegraphics{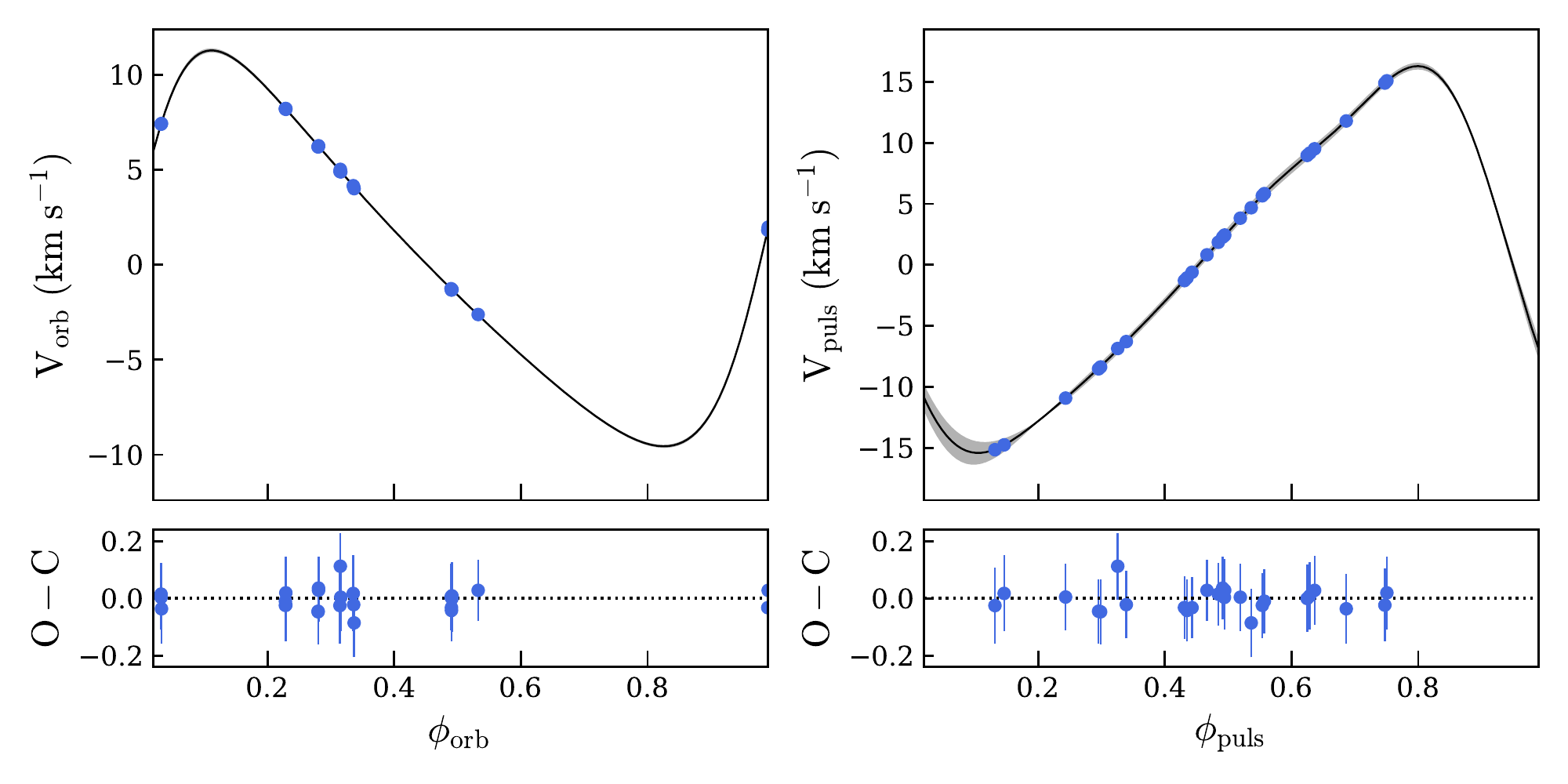}}
	\caption{Same as Fig.~\ref{image__orbit_ffaql} but for V350~Sgr.}
	\label{image__orbit_v350sgr}
\end{figure}

\begin{table}[]
	\centering
	\caption{Final estimated parameters of the V350~Sgr system.}
	\begin{tabular}{ccc} 
		\hline
		\hline
		\multicolumn{3}{c}{\textbf{Pulsation}}																	\\
		$P_\mathrm{puls}$ (days)							  &  \multicolumn{2}{c}{$5.154151 \pm 0.000008$}\\
		$T_0$ (JD) \tablefootmark{a}						  &  \multicolumn{2}{c}{2~440~146.6156}		\\ 
		$A_1$ ($\mathrm{km~s^{-1}}$)												  &	\multicolumn{2}{c}{$-5.37 \pm0.42$}			\\
		$B_1$ ($\mathrm{km~s^{-1}}$)												  &	\multicolumn{2}{c}{$-14.00 \pm0.17$}		\\
		$A_2$ ($\mathrm{km~s^{-1}}$)												   &	\multicolumn{2}{c}{$-3.76 \pm0.25$}		\\
		$B_2$ ($\mathrm{km~s^{-1}}$)												   &	\multicolumn{2}{c}{$-4.22 \pm0.23$}		\\
		$A_3$ ($\mathrm{km~s^{-1}}$)												  &	\multicolumn{2}{c}{$-2.72 \pm0.10$}			\\
		$B_3$ ($\mathrm{km~s^{-1}}$)												  &	\multicolumn{2}{c}{$-0.97 \pm0.25$}		\\
		$A_4$ ($\mathrm{km~s^{-1}}$)												   &	\multicolumn{2}{c}{$-1.26 \pm0.02$}		\\
		$B_4$ ($\mathrm{km~s^{-1}}$)												   &	\multicolumn{2}{c}{$0.20 \pm0.14$}		\\
		$A_5$ ($\mathrm{km~s^{-1}}$)												   &	\multicolumn{2}{c}{$-0.39 \pm 0.06$}		\\
		$B_5$ ($\mathrm{km~s^{-1}}$)												   &	\multicolumn{2}{c}{$0.42 \pm 0.07$}		\\
		$A_6$ ($\mathrm{km~s^{-1}}$)												   &	\multicolumn{2}{c}{$-0.05 \pm 0.05$}		\\
		$B_6$ ($\mathrm{km~s^{-1}}$)												   &	\multicolumn{2}{c}{$0.29 \pm 0.01$}		\\
		$A_7$ ($\mathrm{km~s^{-1}}$)												   &	\multicolumn{2}{c}{$0.11 \pm 0.03$}		\\
		$B_7$ ($\mathrm{km~s^{-1}}$)												   &	\multicolumn{2}{c}{$0.14 \pm 0.02$}		\\
		\hline
		\multicolumn{3}{c}{\textbf{Orbit}}  																			\\
		& Ev11\tablefootmark{b}		& This work					 \\
		$P_\mathrm{orb}$ (days)						& $1472.91 \pm 1.33$			& $1465.3 \pm 0.4$ 		\\
		$T_\mathrm{p}$ (JD)								& $2~450~526.63\pm 6.60$ & $2~450~554.6\pm2.2$ \\
		$e$																& $0.369 \pm 0.011$     &  $0.336 \pm 0.003$  	\\
		$\omega$	($^\circ$)									&$279.03 \pm 1.72$ &	$283.7 \pm 0.7$		\\
		$K_1$ ($\mathrm{km~s^{-1}}$)                    & $10.59\pm0.10$      &  $10.38 \pm 0.03$           \\
		$v_\gamma$  ($\mathrm{km~s^{-1}}$)           &  $11.36\pm0.07$          &  $10.11 \pm 0.02$        \\
		$a_1~\sin{i}$ (au)											&	$1.330\pm0.014$ &		$1.317 \pm 0.004$						\\
		$f(M)$ ($M_\odot$)										&	$0.146\pm0.005$	 &		$0.142 \pm 0.001$	\\
		\hline																
	\end{tabular}
	\tablefoot{\tablefoottext{a}{not fitted.} \tablefoottext{b}{\citet{Evans_2011_09_0}.}
	}
	\label{table_orbit_v350sgr}
	\end{table}

	Our candidate companion has an $H$-band flux ratio of $0.55\pm0.11$\,\% which corresponds to approximately a B9V-A1V star. This is in agreement with the B9V spectral type estimated by \citet[][see also \citealt{Evans_2018_10_0}]{Evans_1992_04_0}. Additional astrometric data are still necessary to confirm.
	
	The $3\sigma$ detection limits of the 2013 and 2017 observations are listed in Table~\ref{table__limits}. The 2017 observations give a B8V limit, which is consistent with our non-detection. The 2013 data give, after removing the detected component, a detection level $\Delta H= 5.3$\,mag, and enable us to exclude additional companions earlier than B8V.
	
	A new set of RVs has been obtained with the CORALIE and HARPS spectrographs (details in Appendix~\ref{appendix__new_rv}), collected from 2013 to 2015. Due to our limited dataset which has a poor orbital and pulsation phase coverage, we also collected RVs from \citet{Evans_2011_09_0}. By fitting the two datasets separately, we noticed a systemic velocity difference of 1.45\,$\mathrm{km~s^{-1}}$, which we subtracted from Evans' data. We took as first guess values for the orbit and pulsation the ones derived by \citet{Evans_2011_09_0}, and applied the same fitting formalism as for FF~Aql. In Fig.~\ref{image__orbit_v350sgr} are displayed the pulsation and orbital velocity curves, and our revised parameters in Table~\ref{table_orbit_v350sgr}. Our values are in slight agreement with previous works \citep{Evans_2011_09_0,Petterson_2004_05_0}, with a final r.m.s. of the residual of $330\,\mathrm{m~s^{-1}}$. However, here we only collected the data provided by Evans (taken from 2005 by Eaton) and we did not combine with additional data from the literature.
	
	\paragraph{V636~Sco:}

	Despite several epoch observations, we did not succeed in detecting the companion. Such faint companions (i.e. $f \lesssim 0.5$\,\%) are difficult to detect and need optimal observing conditions. Table~\ref{table__limits} lists the $3\sigma$ detection limits for our observations. Within 50\,mas, we can rule out a companion with a contrast lower than $\Delta H = 5.8$\,mag (i.e. with a flux ratio $> 0.5$\,\%), which would correspond to spectral types earlier than B9V.
	
\begin{figure}[]
	\centering
	\resizebox{\hsize}{!}{\includegraphics{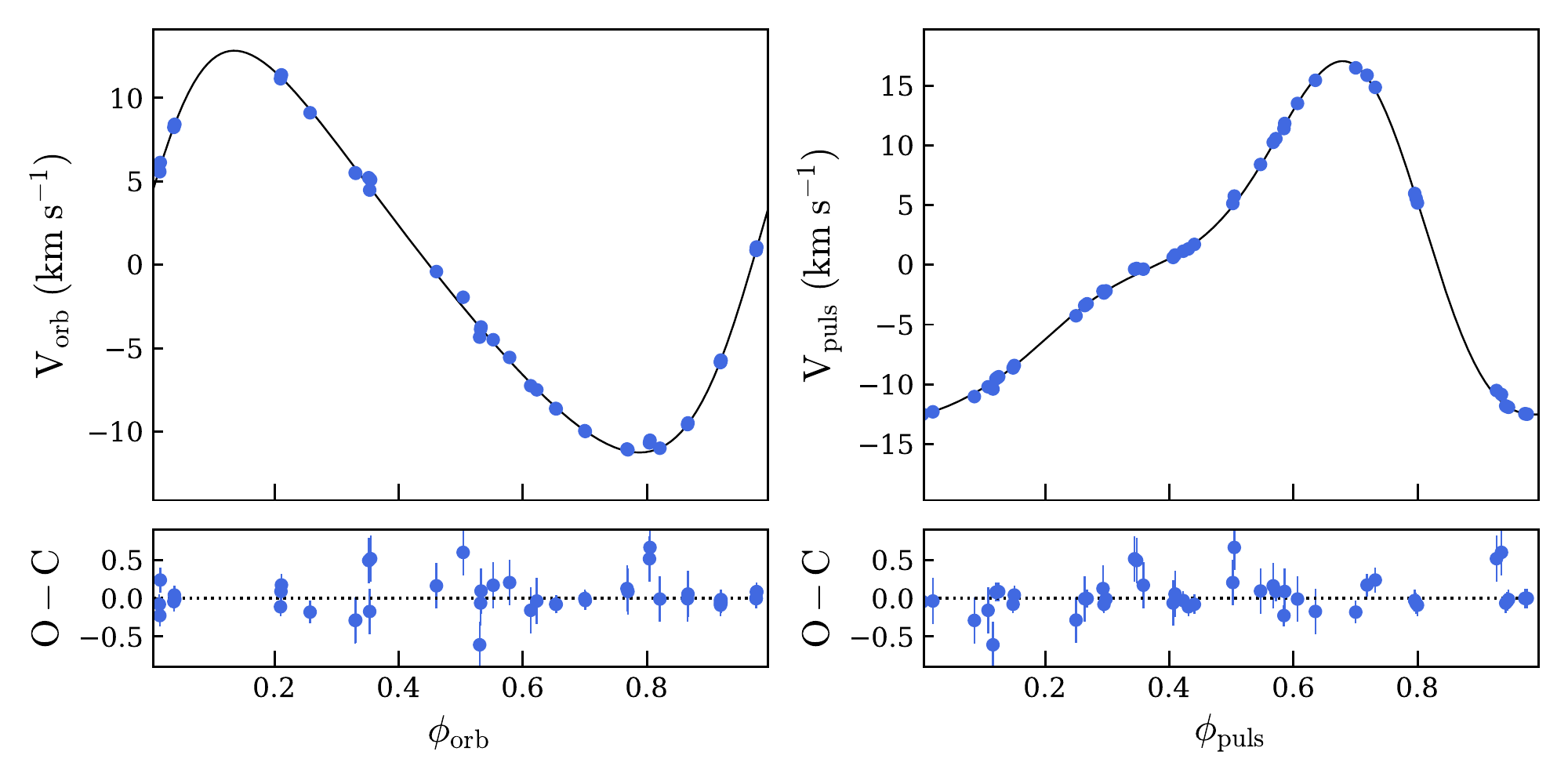}}
	\caption{Same as Fig.~\ref{image__orbit_ffaql} but for V636~Sco.}
	\label{image__orbit_v636sco}
\end{figure}

	New RVs have been obtained with the CORALIE and HARPS spectrographs (details in Appendix~\ref{appendix__new_rv}, and RVs are given in Table~\ref{table_v636sco_rv}), spanning from 2013 to 2015. As our dataset is limited and did not have a sufficient phase coverage, we combined them with RVs from \citet{Petterson_2004_05_0}. We noticed that their zero point is shifted by $1\,\mathrm{km~s^{-1}}$, which we corrected for our combined fit. We took as first guess values to fit the pulsation and orbit the ones derived by \citet[][except $T_\mathrm{p}$ where we used our median time value]{Bohm-Vitense_1998_10_0}. The pulsation and orbital velocity curves are displayed in Fig.~\ref{image__orbit_v636sco}, and our fitted parameters in Table~\ref{table_orbit_v636sco}. We used the same fitting formalism as for FF~Aql. Our revised orbit is in good agreement with previous works \citep{Bohm-Vitense_1998_10_0,Lloyd-Evans_1982_06_0}, with a final r.m.s. of the residual of $230\,\mathrm{m~s^{-1}}$. However, the orbital period of 1362\,days given by \citet{Petterson_2004_05_0} is not in agreement, and the fit does not converge if we use this value as first guess. As a first test, we re-derived the pulsation and orbital parameters using only the RVs from \citet[][post-1997 data only]{Petterson_2004_05_0}. We found a period of 1323.4\,days, more consistent with our previous estimate, as for the other parameters, except the systemic velocity which is positively shifted by $1\,\mathrm{km~s^{-1}}$. A possible explanation for the orbital period mismatch is the pulsation period change which was not taken into account in \citet{Petterson_2004_05_0} as they combined their data with imprecise old measurements \citep{Stibbs_1955__0,Lloyd-Evans_1968__0}. 

\begin{table}[]
	\centering
	\caption{Final estimated parameters of the V636~Sco system.}
	\begin{tabular}{ccc} 
		\hline
		\hline
		\multicolumn{3}{c}{\textbf{Pulsation}}																	\\
		$P_\mathrm{puls}$ (days)							  &  \multicolumn{2}{c}{$6.79703 \pm 0.00002$}\\
		$T_0$ (JD) \tablefootmark{a}						  &  \multicolumn{2}{c}{2~434~906.47}		\\ 
		$A_1$ ($\mathrm{km~s^{-1}}$)												  &	\multicolumn{2}{c}{$-9.44 \pm 0.45$}			\\
		$B_1$ ($\mathrm{km~s^{-1}}$)												  &	\multicolumn{2}{c}{$-7.77 \pm 0.57$}		\\
		$A_2$ ($\mathrm{km~s^{-1}}$)												   &	\multicolumn{2}{c}{$-4.36 \pm 0.35$}		\\
		$B_2$ ($\mathrm{km~s^{-1}}$)												   &	\multicolumn{2}{c}{$2.69 \pm 0.50$}		\\
		$A_3$ ($\mathrm{km~s^{-1}}$)												  &	\multicolumn{2}{c}{$0.99 \pm 0.21$}			\\
		$B_3$ ($\mathrm{km~s^{-1}}$)												  &	\multicolumn{2}{c}{$1.11 \pm 0.21$}		\\
		$A_4$ ($\mathrm{km~s^{-1}}$)												   &	\multicolumn{2}{c}{$0.30 \pm 0.06$}		\\
		$B_4$ ($\mathrm{km~s^{-1}}$)												   &	\multicolumn{2}{c}{$-0.04 \pm 0.09$}		\\
		\hline
		\multicolumn{3}{c}{\textbf{Orbit}}  																			\\
		&  Bo98\tablefootmark{b}		& This work					 \\
		$P_\mathrm{orb}$ (days)						& $1323.6 \pm 1.2$			& $1320.6 \pm 1.3$ 		\\
		$T_\mathrm{p}$ (JD)		& $2~430~411.4\pm 20.9$ & $2~456~865.5\pm 5.7$  \\
		$e$																& $0.213 \pm 0.020$     &  $0.250 \pm 0.004$  	\\
		$\omega$	($^\circ$)									&$294 \pm 52$		 &	$288.0 \pm 2.5$		\\
		$K_1$ ($\mathrm{km~s^{-1}}$)                    & $12.19\pm0.22$      &  $11.98 \pm 0.06$           \\
		$v_\gamma$  ($\mathrm{km~s^{-1}}$)           &  $9.09\pm0.17$          &  $9.0 \pm 0.06$        \\
		$a_1~\sin{i}$ (au)											&	$1.451\pm0.027$ &		$1.409 \pm 0.007$						\\
		$f(M)$ ($M_\odot$)										&	$0.230\pm0.013$	 &		$0.214 \pm 0.003$	\\
		\hline																
	\end{tabular}
	\tablefoot{\tablefoottext{a}{Not fitted.} \tablefoottext{b}{\citet{Bohm-Vitense_1998_10_0}.}
	}
	\label{table_orbit_v636sco}
\end{table}

	\paragraph{AH~Vel:}

	We did not detect any companion from our interferometric observations. However, our derived detection limits allow us to set an upper limit on the spectral type. According to our estimate of $\Delta H_{3\sigma} = 5.3$\,mag, we can rule out any component earlier than an A0V star within 50\,mas. 
	
	New spectroscopic observations were also collected from 2012 to 2015 with the spectrograph HARPS and CORALIE (see Appendix~\ref{appendix__new_rv} and Tables~\ref{table_uaql_rv}-\ref{table_ahvel_rv}). As for BP~Cir, we first analysed our RVs by fitting only the pulsation of the Cepheid with Fourier series. The curve is displayed in Fig.~\ref{image__pulsation}, and the fitted parameters are listed in Table~\ref{table_pulsation}. Although we see no trend in the residual (r.m.s of $0.28\,\mathrm{km~s^{-1}}$), we calculated the periodogram. We did not find any significant periodic signal in the power spectrum, the highest peak having a FAP of $\sim 8$\,\%.. Ignoring possible period change, we performed the same analysis by including old RV measurements \citep{Lloyd-Evans_1968__0,Gieren_1977_05_0,Lloyd-Evans_1980__0,Bersier_2002_06_0} to have a longer time span. We identified a significant peak (with a FAP $< 0.1$\,\%) at a period of $\sim 7060$\,days. We also noticed two other peaks with significant levels at periods $\sim 3100$ and $\sim 14000$\,days (with FAP $< 1\,$\%). We stress that such analysis assume a zero eccentricity and ignore pulsation period change. This preliminary $\sim 7060$\,days orbital period needs confirmation with additional high-precision contemporaneous RV measurements.

	\section{Conclusion}
	\label{section__conclusion}
	
	We reported new multi-telescope interferometric observations for 16 Galactic classical Cepheids. We used the \texttt{CANDID} algorithm \citep{Gallenne_2015_07_0} to search for high-contrast companions within a relative distance to the Cepheid of 50\,mas. We also report detection limit for undetected components (secondary or tertiary).
	
	The components orbiting U~Aql, BP~Cir and S~Mus are clearly detected. They are located at projected separations of 2-40\,mas, and with flux ratios in the range 0.4-3.5\,\%. We have preliminary detections for FF~Aql, Y~Car, BG~Cru, X~Sgr, V350~Sgr and V636~Sco, but a confirmation is needed due to our low detections levels or several possible locations. For U~Car, YZ~Car, T~Mon, R~Mus, S~Nor, W~Sgr and AH~Vel we have no detection, however we set upper limits on the companion' spectral types. Upper limits on the spectral type of possible tertiary component were also estimated for the other targets.
	
	We present preliminary complete astrometric and spectroscopic orbits for the Cepheids U Aql and S~Mus, combining astrometric and single-line velocity measurements. We derived preliminary dynamical masses for these Cepheids assuming a distance from a P-L relation \citep{Storm_2011_10_0}. We found $\mathrm{M = 4.97 \pm 0.62\,M_\odot}$ and $\mathrm{M = 4.63 \pm 0.99\,M_\odot}$, respectively for U~Aql and S~Mus.
	
	Based on new high-precision spectroscopic observations with the SOPHIE, CORALIE and HARPS spectrographs, we revised the pulsation and spectroscopic orbital parameters for FF~Aql, YZ~Car, W~Sgr, V350~Sgr, and V636~Sco, while only the pulsation parameters of BP~Cir, BG~Cru, S~Nor and AH~Vel were updated.
	
	Our interferometric observations also provided angular diameter measurements for all targets, and can be used for instance in a Baade-Wesselink method analysis \citep[see e.g.][]{Kervella_2004_03_0,Gallenne_2012_03_0,Breitfelder_2016_03_0}.
	
	Our interferometric program is promising to independently and accurately determine the mass and distance of Cepheids, as demonstrated by \citet{Gallenne_2018_11_0} who determined the distance of the Cepheid V1334~Cyg at a 1\,\% accuracy level, providing the most accurate independent distance for a Cepheid. They also derived the mass of both components at $< 3$\,\% precision, which is also unique for a Galactic Cepheid. We now have additional Cepheids for which we detected the companion and measured their astrometric position \citep[U~Aql, \object{RT~Aur}, \object{AX~Cir}, BP~Cir, S~Mus, and \object{AW~Per},][]{Gallenne_2015_07_0,Gallenne_2014_01_0,Gallenne_2013_04_0}. The same analysis as that of V1334~Cyg can be applied if we have RVs for the companions. An astrometric follow-up is on-going to secure a better orbital coverage of these systems, and perform a combined fit with RV measurements.
	
	
	\begin{acknowledgements}
		The authors would like to thank the CHARA Array and Mount Wilson Observatory staff for their support. The CHARA Array is supported by the National Science Foundation under Grant No. AST-1211929, AST-1411654, AST-1636624, and AST-1715788.  Institutional support has been provided from the GSU College of Arts and Sciences and the GSU Office of the Vice President for Research and Economic Development. The authors also thank all the people involved in the VLTI project. We acknowledge the support of the French Agence Nationale de la Recherche (ANR-15-CE31-0012-01, project Unlock-Cepheids). WG and GP gratefully acknowledge financial support from the BASAL Centro de Astrofisica y Tecnologias Afines (CATA, AFB-170002). WG also acknowledges financial support from the Millenium Institute of Astrophysics (MAS) of the Iniciativa Cientifica Milenio del Ministerio de Economia, Fomento y Turismo de Chile (project IC120009). We acknowledge financial support from the Programme National de Physique Stellaire (PNPS) of CNRS/INSU, France. Support from the Polish National Science Centre grants MAESTRO UMO-2017/26/A/ST9/00446 and from the IdP II 2015 0002 64 grant of the Polish Ministry of Science and Higher Education is also acknowledged. The research leading to these results has received funding from the European Research Council (ERC) under the European Union's Horizon 2020 research and innovation programme (grant agreement No. 695099 and 639889). NRE acknowledges support from the Chandra X-ray Center NASA (contract NAS8-03060) and the HST grants GO-13454.001-A and  GO-14194.002. JDM acknowledges HST/STScI grant support: HST-GO-13454.07-A, HST -GO-13841.006-A, HST-GO-14194.008-A. This work is based upon observations obtained with the Georgia State University Center for High Angular Resolution Astronomy Array at Mount Wilson Observatory. BP acknowledges financial support from the Polish National Science Center grant SONATA 2014/15/D/ST9/02248. This research is based on observations made with the SOPHIE spectrograph on the 1.93-m telescope at Observatoire de Haute-Provence (program IDs 13A.PNPS.GALL, 14A.PNPS.GALL, 15A.PNPS.GALL, 13B.PNPS.KER, 16B.PNPS.KER) and the CORALIE spectrograph on the Euler telescope at La Silla Observatory (program IDs CNTAC CN2013A, CN2014A and CN2015A). The Swiss 1.2\,m Euler telescope is supported by the Swiss National Science Foundation. This research is based on observations made with the Mercator Telescope, operated on the island of La Palma by the Flemish Community, at the Spanish Observatorio del Roque de los Muchachos of the Instituto de Astrof\'isica de Canarias. HERMES is supported by the Fund for Scientific Research of Flanders (FWO), Belgium; the Research Council of K.U.Leuven, Belgium; the Fonds National de la Recherche Scientifique (F.R.S.- FNRS), Belgium; the Royal Observatory of Belgium; the Observatoire de Gen\`eve,Switzerland, and the Th\"uringer Landessternwarte, Tautenburg, Germany.
	\end{acknowledgements}
	
	
	\bibliographystyle{aa}   
	\bibliography{/Users/agallenn/Sciences/Articles/bibliographie}

	
	\begin{appendix} 
		
		\section{Parameters of the calibrators used for MIRC and PIONIER interferometric observations}
		\label{appendix__calibrators}

	The parameters of the calibrators we used are listed in Table~\ref{table_calibrators}. They were collected with the SearchCal software.

			\begin{table*}[!ht]
			\centering
			\caption{Calibrators used for our observations.}
			\begin{tabular}{ccccc|ccccc} 
				\hline
				\hline
				HD		&	Sp.~type	&	$V$  & $H$ 	&  $\theta_\mathrm{UD}$  &  HD		&	Sp.~type	&	$V$  & $H$ 	&  $\theta_\mathrm{UD}$	\\
							&					&	(mag)	&	(mag)	&	(mas)				&			&		&	(mag)	&	(mag)	&	(mas)   \\
				\hline
43299  &  K3III-IV  &  6.84  & 4.44  & $0.720\pm0.051$  &  132209  &  A9/F0IV/V  &  6.56  & 5.83  & $0.248\pm0.017$  \\
45317  &  K0III  &  6.87  & 4.60  & $0.628\pm0.045$  &  144230  &  K2III  &  8.67  & 5.41  & $0.465\pm0.011$  \\
66080  &  G6III  &  7.44  & 5.48  & $0.398\pm0.028$  &  145361  &  F2IV/V  &  5.77  & 4.89  & $0.449\pm0.032$  \\
70195  &  G8/K0III  &  7.08  & 5.08  & $0.492\pm0.035$  &  145883  &  K2III  &  8.45  & 5.39  & $0.457\pm0.010$  \\
73075  &  K1III  &  7.34  & 5.16  & $0.467\pm0.033$  &  146247  &  K2III  &  7.21  & 4.61  & $0.653\pm0.046$  \\
85253  &  K1(III)  &  8.69  & --  & $0.320\pm0.023$  &  147075  &  K1III  &  7.90  & 5.27  & $0.465\pm0.033$  \\
89517  &  K1/2III  &  8.59  & 5.87  & $0.366\pm0.026$  &  147422  &  K0III  &  8.04  & 5.34  & $0.445\pm0.011$  \\
89839  &  F7V  &  7.64  & 6.47  & $0.225\pm0.016$  &  148679  &  K0III  &  6.94  & 4.69  & $0.609\pm0.043$  \\
90074  &  G6III  &  6.35  & 4.52  & $0.652\pm0.046$  &  149835  &  K0III  &  7.30  & 5.07  & $0.499\pm0.035$  \\
90246  &  K0III  &  8.25  & 5.82  & $0.349\pm0.025$  &  151005  &  K0III  &  7.21  & 4.69  & $0.595\pm0.042$  \\
90980  &  K0III  &  6.74  & 4.42  & $0.657\pm0.047$  &  151337  &  K0IV/V  &  7.38  & 5.39  & $0.425\pm0.030$  \\
92156  &  G0IV/V  &  8.03  & 6.73  & $0.197\pm0.014$  &  152272  &  K1III  &  7.35  & 4.91  & $0.555\pm0.039$  \\
93307  &  G0V  &  7.78  & 6.45  & $0.229\pm0.016$  &  154250  &  K0III  &  8.00  & 5.72  & $0.365\pm0.026$  \\
94256  &  K0III  &  7.95  & 5.73  & $0.357\pm0.025$  &  154486  &  K2/3III  &  6.94  & 3.50  & $1.056\pm0.015$  \\
96068  &  G8III  &  6.52  & 4.37  & $0.709\pm0.090$  &  155019  &  K1III  &  8.99  & 5.24  & $0.475\pm0.034$  \\
97744  &  K0/1III  &  8.30  & 5.93  & $0.336\pm0.024$  &  156992  &  K3III  &  6.36  & 3.12  & $1.240\pm0.017$  \\
98692  &  K2III  &  7.45  & 4.99  & $0.568\pm0.040$  &  159217  &  A0V  &  4.59  & 4.66  & $0.373\pm0.026$  \\
98732  &  K0III  &  7.02  & 4.95  & $0.580\pm0.041$  &  159285  &  K1III  &  7.97  & 5.44  & $0.416\pm0.030$  \\
98897  &  G8III  &  6.63  & 4.31  & $0.603\pm0.045$  &  159941  &  M0III  &  7.82  & 3.73  & $1.081\pm0.015$  \\
99048  &  K2III  &  7.13  & 4.28  & $0.673\pm0.048$  &  160113  &  G5V  &  7.28  & 5.69  & $0.333\pm0.024$  \\
100078  &  K2III  &  7.33  & 4.19  & $0.836\pm0.011$  &  162415  &  K5III  &  6.94  & 3.66  & $1.003\pm0.072$  \\
101805  &  F8V  &  6.47  & 5.30  & $0.375\pm0.026$  &  163652  &  G8III  &  5.74  & 3.64  & $0.900\pm0.064$  \\
102534  &  K1III  &  6.76  & 4.66  & $0.643\pm0.046$  &  166230  &  A8III  &  5.10  & 4.66  & $0.409\pm0.029$  \\
102969  &  G8III  &  7.66  & 5.28  & $0.460\pm0.033$  &  166295  &  K2III/IV  &  6.68  & 2.94  & $1.266\pm0.017$  \\
105939  &  K0III  &  7.05  & 4.70  & $0.601\pm0.043$  &  169236  &  K0III  &  6.14  & 3.69  & $0.890\pm0.063$  \\
107013  &  K1III  &  7.97  & 5.77  & $0.343\pm0.024$  &  171960  &  K3III  &  7.29  & 3.50  & $1.121\pm0.016$  \\
107720  &  K1III  &  7.12  & 4.78  & $0.586\pm0.042$  &  174774  &  K4III  &  7.56  & 3.83  & $1.103\pm0.015$  \\
109761  &  G6III  &  7.41  & 5.26  & $0.442\pm0.031$  &  177067  &  K0III  &  6.91  & 4.64  & $0.634\pm0.045$  \\
110532  &  K0/1III  &  6.42  & 4.16  & $0.783\pm0.011$  &  178218  &  K0III  &  6.88  & 4.55  & $0.658\pm0.047$  \\
110924  &  K0II/III  &  6.64  & 4.15  & $0.745\pm0.053$  &  182807  &  F7V  &  6.20  & 4.93  & $0.457\pm0.032$  \\
112124  &  G8III  &  7.21  & 4.87  & $0.536\pm0.038$  &  184985  &  F7V  &  5.45  & 4.42  & $0.589\pm0.041$  \\
115669  &  K2/3III  &  6.91  & 4.13  & $0.766\pm0.055$  &  185124  &  F3IV/V  &  5.68  & 4.58  & $0.515\pm0.036$  \\
121901  &  F0/2III/IV  &  6.47  & 5.72  & $0.284\pm0.020$  &  188844  &  K0III/IV  &  6.57  & 4.50  & $0.660\pm0.047$  \\
125136  &  K0III  &  7.44  & 5.27  & $0.440\pm0.031$  &  196870  &  K0III  &  6.61  & 4.32  & $0.685\pm0.049$  \\
130551  &  F5V  &  8.75  & 5.96  & $0.276\pm0.019$  &  198001  &  B9.5V  &  3.77  & 3.71  & $0.510\pm0.036$  \\
				\hline
			\end{tabular}
			\label{table_calibrators}
		\end{table*}
%
		\section{New radial velocities from the CORALIE, SOPHIE, HARPS and HERMES spectrographs}
		\label{appendix__new_rv}
		We collected several spectra from 2012 to 2017 with the fiber-fed SOPHIE spectrograph \citep[$R \sim 75~000$][]{Bouchy_2006_02_0}, mounted on the 1.93\,m telescope of the Observatoire de Haute Provence (France), the CORALIE spectrograph \citep[$R \sim 60~000$][]{Queloz_2001_09_0} at the Swiss 1.2\,m Euler telescope located at La Silla Observatory, the HARPS spectrograph \citep[$R \sim 115~000$][]{Pepe_2002_12_0} mounted on the 3.6\,m ESO telescope at La Silla Observatory, and the HERMES spectrograph \citep[$R \sim 85~000$][]{Raskin_2011_02_0} mounted on the Flemish 1.2\,m telescope of the Roque de los Muchachos Observatory. These four instruments cover the visible wavelength. Exposure times of a few minutes allowed a signal-to-noise ratio $> 10$ per pixel at 550nm. All data were reduced using the dedicated instrument pipeline.
		
		Radial velocities were estimated using the cross-correlation method. We created our own weighted binary mask by selecting unblended lines from high-resolution synthetic spectra ($R \sim 120~000$) covering the wavelength range 4500-6800\,\AA. The cross-correlation function is then fitted by a Gaussian whose minimum value gives an estimate of the RV. Uncertainties include photon noise and internal drift. Zero point difference was set to the HARPS or CORALIE system (when there are no HARPS data) following the table of \citet{Soubiran_2013_04_0} for CORALIE, HARPS and SOPHIE, and \citet{Gallenne_2018_11_0} for HERMES. RV measurements are listed from Table~\ref{table_ffaql_rv} to \ref{table_yzcar_rv} (without correction for the zero point).

		\begin{table}[]
			\centering
			\caption{Radial velocity measurements of U~Aql.}
			\begin{tabular}{cccc} 
				\hline
				\hline
				MJD		&	RV	&	$\sigma_\mathrm{RV}$	&  Inst.	\\
				(days)			&	($\mathrm{km~s^{-1}}$)	&	($\mathrm{km~s^{-1}}$)								&		 \\
				\hline
				56213.985087  &  $-3.15$  &  0.11  &  HARPS \\
				56239.989431  &  $-14.08$  &  0.13  &  HARPS \\
				56241.017804  &  $-9.62$  &  0.12  &  HARPS \\
				56448.452412  &  23.07  &  0.12  &  HARPS \\
				56553.149712  &  15.01  &  0.11  &  HARPS \\
				56908.128645  &  $-5.14$  &  0.12  &  HARPS \\
				57635.059769  &  2.57  &  0.12  &  HARPS \\
				57636.026251  &  13.93  &  0.15  &  HARPS \\
				57901.307348  &  $-$0.37  &  0.11  &  HARPS \\
				57902.184474  &  11.02  &  0.12  &  HARPS \\
				57903.270561  &  18.38  &  0.16  &  HARPS \\
				57916.219488  &  11.24  &  0.12  &  HARPS \\
				56471.085989  &  0.01  &  0.12  &  SOPHIE \\
				56472.078955  &  $-$9.35  &  0.11  &  SOPHIE \\
				56473.031282  &  $-$4.13  &  0.11  &  SOPHIE \\
				56474.075913  &  2.86  &  0.10  &  SOPHIE \\
				56475.073713  &  6.28  &  0.10  &  SOPHIE \\
				56476.039045  &  16.31  &  0.11  &  SOPHIE \\
				56477.057247  &  28.52  &  0.14  &  SOPHIE \\
				56557.801666  &  $-$0.03  &  0.10  &  SOPHIE \\
				56558.863991  &  6.84  &  0.10  &  SOPHIE \\
				56559.878414  &  11.18  &  0.11  &  SOPHIE \\
				56560.831546  &  24.27  &  0.12  &  SOPHIE \\
				56561.821282  &  25.15  &  0.14  &  SOPHIE \\
				56562.862657  &  $-$8.40  &  0.12  &  SOPHIE \\
				56814.104828  &  29.67  &  0.14  &  SOPHIE \\
				56816.084294  &  $-$7.90  &  0.12  &  SOPHIE \\
				56818.036831  &  3.23  &  0.10  &  SOPHIE \\
				57179.053565  &  21.38  &  0.13  &  SOPHIE \\
				57180.059515  &  17.09  &  0.14  &  SOPHIE \\
				57181.057767  &  $-$12.90  &  0.12  &  SOPHIE \\
				56399.324683  &  22.89  &  0.12  &  CORALIE \\
				56400.321920  &  20.68  &  0.13  &  CORALIE \\
				56461.289864  &  7.59  &  0.10  &  CORALIE \\
				56461.417069  &  8.81  &  0.10  &  CORALIE \\
				56462.285588  &  20.36  &  0.12  &  CORALIE \\
				56462.411588  &  22.04  &  0.12  &  CORALIE \\
				56462.424401  &  22.25  &  0.12  &  CORALIE \\
				56750.368062  &  23.61  &  0.12  &  CORALIE \\
				56750.379174  &  23.73  &  0.12  &  CORALIE \\
				56750.390205  &  23.89  &  0.12  &  CORALIE \\
				56751.364938  &  28.62  &  0.14  &  CORALIE \\
				56751.376051  &  28.39  &  0.14  &  CORALIE \\
				56751.387024  &  28.17  &  0.14  &  CORALIE \\
				56826.390217  &  8.62  &  0.10  &  CORALIE \\
				56826.401190  &  8.68  &  0.10  &  CORALIE \\
				56826.412220  &  8.73  &  0.10  &  CORALIE \\
				56826.423204  &  8.80  &  0.10  &  CORALIE \\
				56827.397793  &  20.23  &  0.11  &  CORALIE \\
				56827.408801  &  20.40  &  0.11  &  CORALIE \\
				56827.419785  &  20.54  &  0.11  &  CORALIE \\
				57135.350389  &  4.33  &  0.10  &  CORALIE \\
				57135.363875  &  4.35  &  0.10  &  CORALIE \\
				57136.347787  &  14.64  &  0.11  &  CORALIE \\
				57137.372212  &  26.42  &  0.14  &  CORALIE \\
				57473.393066  &  2.32  &  0.11  &  CORALIE \\
				57476.419085  &  $-$22.77  &  0.11  &  CORALIE \\
				57477.398320  &  $-$17.75  &  0.10  &  CORALIE \\
				\hline
			\end{tabular}
			\label{table_uaql_rv}
		\end{table}

		\begin{table*}[]
	\centering
	\caption{Radial velocity measurements of FF~Aql.}
	\begin{tabular}{cccc|cccc} 
		\hline
		\hline
		MJD		&	RV	&	$\sigma_\mathrm{RV}$	&  Inst. & MJD		&	RV	&	$\sigma_\mathrm{RV}$	&  Inst.	\\
		(days)			&	($\mathrm{km~s^{-1}}$)	&	($\mathrm{km~s^{-1}}$)	& &	(days)			&	($\mathrm{km~s^{-1}}$)	&	($\mathrm{km~s^{-1}}$)								&	 \\
		\hline
        56399.372658  &  $-$15.05  &  0.11  &  CORALIE &         56446.118545  &  $-$27.33  &  0.11  &  SOPHIE \\
56400.369897  &  $-$15.04  &  0.11  &  CORALIE &          56447.036907  &  $-$26.93  &  0.11  &  SOPHIE \\
56461.266962  &  $-$20.53  &  0.11  &  CORALIE &          56447.041583  &  $-$26.90  &  0.11  &  SOPHIE \\
56462.238574  &  $-$12.98  &  0.11  &  CORALIE &          56448.104962  &  $-$18.19  &  0.11  &  SOPHIE \\
56462.306712  &  $-$12.71  &  0.11  &  CORALIE &          56449.111335  &  $-$12.52  &  0.12  &  SOPHIE \\
56750.401128  &  $-$23.07  &  0.11  &  CORALIE &          57644.844908  &  $-$28.18  &  0.11  &  SOPHIE \\
56751.397916  &  $-$19.50  &  0.11  &  CORALIE &          57650.786739  &  $-$18.38  &  0.11  &  SOPHIE \\
56826.298032  &  $-$20.93  &  0.11  &  CORALIE &          56855.100210  &  $-$9.29  &  0.11  &  HERMES \\
56827.307666  &  $-$18.49  &  0.11  &  CORALIE &          56856.044422  &  $-$4.94  &  0.11  &  HERMES \\
57135.333071  &  $-$19.57  &  0.11  &  CORALIE &          56857.030988  &  $-$15.49  &  0.11  &  HERMES \\
57136.368934  &  $-$12.44  &  0.11  &  CORALIE &          56858.028974  &  $-$20.61  &  0.11  &  HERMES \\
57137.346922  &  $-$4.65  &  0.11  &  CORALIE &          56859.028970  &  $-$14.50  &  0.11  &  HERMES \\
57470.401531  &  $-$25.73  &  0.11  &  CORALIE &          56860.039263  &  $-$5.82  &  0.11  &  HERMES \\
57472.399768  &  $-$12.02  &  0.11  &  CORALIE &          56861.040693  &  $-$8.50  &  0.11  &  HERMES \\
57473.403685  &  $-$11.72  &  0.11  &  CORALIE &          56862.042044  &  $-$20.34  &  0.11  &  HERMES \\
57476.412521  &  $-$16.04  &  0.11  &  CORALIE &          56864.028492  &  $-$9.42  &  0.11  &  HERMES \\
57477.405108  &  $-$9.90  &  0.11  &  CORALIE &          56865.033989  &  $-$4.88  &  0.11  &  HERMES \\
57478.417082  &  $-$19.10  &  0.11  &  CORALIE &          56866.028913  &  $-$16.11  &  0.11  &  HERMES \\
57498.390839  &  $-$19.93  &  0.10  &  CORALIE &          57178.049664  &  $-$4.67  &  0.11  &  HERMES \\
57500.395837  &  $-$14.09  &  0.11  &  CORALIE &          57180.029113  &  $-$20.13  &  0.11  &  HERMES \\
57504.428585  &  $-$10.85  &  0.11  &  CORALIE &          57181.025161  &  $-$13.44  &  0.10  &  HERMES \\
57507.391957  &  $-$19.65  &  0.10  &  CORALIE &          57182.016061  &  $-$5.44  &  0.11  &  HERMES \\
57508.324257  &  $-$11.52  &  0.11  &  CORALIE &          57183.018201  &  $-$9.02  &  0.11  &  HERMES \\
57508.409506  &  $-$11.28  &  0.11  &  CORALIE &          57184.015859  &  $-$20.38  &  0.11  &  HERMES \\
57526.363465  &  $-$11.24  &  0.11  &  CORALIE &          57185.010898  &  $-$17.70  &  0.10  &  HERMES \\
57529.360015  &  $-$23.38  &  0.11  &  CORALIE &          57186.000617  &  $-$8.77  &  0.10  &  HERMES \\
57534.333270  &  $-$18.96  &  0.11  &  CORALIE &          57188.001022  &  $-$16.84  &  0.11  &  HERMES \\
57536.358535  &  $-$17.40  &  0.11  &  CORALIE &          57567.050781  &  $-$11.82  &  0.11  &  HERMES \\
57835.363211  &  $-$13.43  &  0.11  &  CORALIE &          57568.069138  &  $-$24.06  &  0.11  &  HERMES \\
57840.389289  &  $-$19.25  &  0.11  &  CORALIE &          57569.055557  &  $-$27.04  &  0.11  &  HERMES \\
57845.398729  &  $-$26.57  &  0.11  &  CORALIE &          57571.125510  &  $-$11.85  &  0.11  &  HERMES \\
57848.406122  &  $-$12.93  &  0.11  &  CORALIE &          57572.090768  &  $-$17.34  &  0.11  &  HERMES \\
57869.420962  &  $-$22.77  &  0.10  &  CORALIE &          57573.108456  &  $-$27.59  &  0.11  &  HERMES \\
57870.394944  &  $-$14.23  &  0.11  &  CORALIE &          57574.147203  &  $-$23.42  &  0.10  &  HERMES \\
57871.412147  &  $-$15.38  &  0.11  &  CORALIE &          57998.851481  &  $-$23.18  &  0.11  &  HERMES \\
57872.418526  &  $-$27.64  &  0.11  &  CORALIE &          57999.886736  &  $-$13.93  &  0.11  &  HERMES \\
57876.376142  &  $-$22.32  &  0.11  &  CORALIE &          58000.894608  &  $-$12.18  &  0.11  &  HERMES \\
58228.406034  &  $-$6.22  &  0.11  &  CORALIE &          58001.886619  &  $-$24.89  &  0.11  &  HERMES \\
58229.415646  &  $-$13.14  &  0.11  &  CORALIE &          58024.923068  &  $-$26.40  &  0.11  &  HERMES \\
58230.399781  &  $-$21.85  &  0.11  &  CORALIE &          58025.896626  &  $-$20.77  &  0.10  &  HERMES \\
58231.340071  &  $-$18.04  &  0.11  &  CORALIE &          58032.877642  &  $-$20.52  &  0.11  &  HERMES \\
58235.321229  &  $-$20.78  &  0.11  &  CORALIE &          58069.831394  &  $-$24.86  &  0.11  &  HERMES \\
58237.392442  &  $-$5.85  &  0.11  &  CORALIE &          58070.832411  &  $-$17.81  &  0.10  &  HERMES \\
		\hline
	\end{tabular}
	\label{table_ffaql_rv}
\end{table*}

		\begin{table}[]
	\centering
	\caption{Radial velocity measurements of YZ~Car.}
	\begin{tabular}{cccc} 
		\hline
		\hline
		MJD		&	RV	&	$\sigma_\mathrm{RV}$	&  Inst.	\\
		(days)			&	($\mathrm{km~s^{-1}}$)	&	($\mathrm{km~s^{-1}}$)								&		 \\
		\hline
		56240.378977  &  4.72  &  0.10  &  HARPS \\
		56448.004155  &  16.36  &  0.15  &  HARPS \\
		56448.999019  &  11.49  &  0.14  &  HARPS \\
		56450.012629  &  6.42  &  0.13  &  HARPS \\
		56605.365482  &  $-$5.81  &  0.10  &  HARPS \\
		56398.983650  &  $-$5.06  &  0.12  &  CORALIE \\
		56399.992636  &  $-$7.80  &  0.11  &  CORALIE \\
		56461.042669  &  7.99  &  0.11  &  CORALIE \\
		56461.979525  &  11.12  &  0.11  &  CORALIE \\
		56750.001563  &  $-$11.08  &  0.11  &  CORALIE \\
		56751.001637  &  $-$7.34  &  0.12  &  CORALIE \\
		56825.971994  &  3.81  &  0.12  &  CORALIE \\
		56826.967563  &  6.39  &  0.13  &  CORALIE \\
		57134.983140  &  20.87  &  0.12  &  CORALIE \\
		57135.985212  &  23.26  &  0.13  &  CORALIE \\
		57136.984309  &  24.46  &  0.13  &  CORALIE \\
		57198.012720  &  $-$2.02  &  0.12  &  CORALIE \\
		\hline
	\end{tabular}
	\label{table_yzcar_rv}
\end{table}	

		\begin{table}[]
	\centering
	\caption{Radial velocity measurements of BP~Cir.}
	\begin{tabular}{cccc} 
		\hline
		\hline
		MJD		&	RV	&	$\sigma_\mathrm{RV}$	&  Inst.	\\
		(days)			&	($\mathrm{km~s^{-1}}$)	&	($\mathrm{km~s^{-1}}$)								&		 \\
		\hline
		56399.093613  &  $-$10.40  &  0.11  &  CORALIE \\
		56399.394898  &  $-$11.05  &  0.11  &  CORALIE \\
		56400.102809  &  $-$26.39  &  0.11  &  CORALIE \\
		56400.392092  &  $-$24.16  &  0.11  &  CORALIE \\
		56461.097700  &  $-$13.81  &  0.10  &  CORALIE \\
		56462.100168  &  $-$20.29  &  0.11  &  CORALIE \\
		56750.129817  &  $-$26.03  &  0.12  &  CORALIE \\
		56750.147064  &  $-$26.15  &  0.12  &  CORALIE \\
		56750.164299  &  $-$26.27  &  0.12  &  CORALIE \\
		56751.126307  &  $-$15.76  &  0.11  &  CORALIE \\
		56751.143808  &  $-$15.53  &  0.11  &  CORALIE \\
		56751.161170  &  $-$15.31  &  0.11  &  CORALIE \\
		56826.104945  &  $-$10.25  &  0.11  &  CORALIE \\
		56826.122179  &  $-$10.26  &  0.11  &  CORALIE \\
		56826.139470  &  $-$10.33  &  0.11  &  CORALIE \\
		56827.101280  &  $-$26.03  &  0.11  &  CORALIE \\
		56827.118525  &  $-$25.85  &  0.11  &  CORALIE \\
		56827.135781  &  $-$25.71  &  0.11  &  CORALIE \\
		57135.152884  &  $-$12.08  &  0.11  &  CORALIE \\
		57135.264080  &  $-$11.07  &  0.11  &  CORALIE \\
		57136.110904  &  $-$24.27  &  0.12  &  CORALIE \\
		57136.221556  &  $-$26.12  &  0.12  &  CORALIE \\
		57137.109560  &  $-$17.65  &  0.11  &  CORALIE \\
		57137.220223  &  $-$16.09  &  0.11  &  CORALIE \\
		57198.154736  &  $-$15.10  &  0.11  &  CORALIE \\
		57198.244362  &  $-$17.66  &  0.12  &  CORALIE \\
		56876.984618  &  $-$20.48  &  0.12  &  HARPS \\
		56877.988780  &  $-$19.19  &  0.11  &  HARPS \\
		56878.988466  &  $-$11.02  &  0.12  &  HARPS \\
		56907.987903  &  $-$14.79  &  0.12  &  HARPS \\
		56908.985107  &  $-$21.95  &  0.11  &  HARPS \\
		56909.996458  &  $-$10.07  &  0.11  &  HARPS \\
		\hline
	\end{tabular}
	\label{table_bpcir_rv}
\end{table}	

		\begin{table}[]
			\centering
			\caption{Radial velocity measurements of BG~Cru.}
			\begin{tabular}{cccc} 
				\hline
				\hline
				MJD		&	RV	&	$\sigma_\mathrm{RV}$	&  Inst.	\\
				(days)			&	($\mathrm{km~s^{-1}}$)	&	($\mathrm{km~s^{-1}}$)								&		 \\
				\hline
				56399.020058  &  $-$17.80  &  0.16  &  CORALIE \\
				56399.158116  &  $-$16.77  &  0.16  &  CORALIE \\
				56400.028892  &  $-$16.07  &  0.17  &  CORALIE \\
				56461.085840  &  $-$24.84  &  0.17  &  CORALIE \\
				56461.122691  &  $-$24.98  &  0.17  &  CORALIE \\
				56462.089207  &  $-$21.26  &  0.17  &  CORALIE \\
				56462.108743  &  $-$21.11  &  0.17  &  CORALIE \\
				56750.178391  &  $-$16.45  &  0.17  &  CORALIE \\
				56750.183935  &  $-$16.40  &  0.17  &  CORALIE \\
				56751.175225  &  $-$17.91  &  0.17  &  CORALIE \\
				56751.180850  &  $-$17.97  &  0.17  &  CORALIE \\
				56826.151628  &  $-$23.16  &  0.17  &  CORALIE \\
				56826.157172  &  $-$23.13  &  0.17  &  CORALIE \\
				56827.147936  &  $-$15.89  &  0.17  &  CORALIE \\
				56827.153491  &  $-$15.87  &  0.17  &  CORALIE \\
				57135.109433  &  $-$14.78  &  0.17  &  CORALIE \\
				57136.065043  &  $-$23.58  &  0.17  &  CORALIE \\
				57137.063637  &  $-$22.70  &  0.17  &  CORALIE \\
				57198.081791  &  $-$16.55  &  0.17  &  CORALIE \\
				56242.383128  &  $-$14.95  &  0.18  &  HARPS \\
				56448.022480  &  $-$24.95  &  0.18  &  HARPS \\
				56449.009933  &  $-$18.75  &  0.18  &  HARPS \\
				56450.003818  &  $-$14.98  &  0.18  &  HARPS \\
				\hline
			\end{tabular}
			\label{table_bgcru_rv}
		\end{table}	
		
		\begin{table}[]
			\centering
			\caption{Radial velocity measurements of S~Mus.}
			\begin{tabular}{cccc} 
				\hline
				\hline
				MJD		&	RV	&	$\sigma_\mathrm{RV}$	&  Inst.	\\
				(days)			&	($\mathrm{km~s^{-1}}$)	&	($\mathrm{km~s^{-1}}$)								&		 \\
				\hline
				56399.031006  &  28.65  &  0.14  &  CORALIE \\
				56400.040077  &  24.12  &  0.14  &  CORALIE \\
				56460.979983  &  $-$1.28  &  0.11  &  CORALIE \\
				56462.000067  &  1.23  &  0.11  &  CORALIE \\
				56750.090093  &  $-$13.66  &  0.12  &  CORALIE \\
				56750.108184  &  $-$13.59  &  0.12  &  CORALIE \\
				56751.086165  &  $-$10.41  &  0.11  &  CORALIE \\
				56751.104279  &  $-$10.34  &  0.11  &  CORALIE \\
				56826.060596  &  2.91  &  0.12  &  CORALIE \\
				56826.078744  &  2.66  &  0.12  &  CORALIE \\
				56827.056896  &  $-$2.27  &  0.12  &  CORALIE \\
				56827.075009  &  $-$2.24  &  0.12  &  CORALIE \\
				57135.095919  &  $-$22.36  &  0.12  &  CORALIE \\
				57136.051396  &  $-$29.08  &  0.12  &  CORALIE \\
				57137.049998  &  $-$26.31  &  0.11  &  CORALIE \\
				57198.069001  &  $-$18.45  &  0.11  &  CORALIE \\
				57198.110956  &  $-$18.13  &  0.11  &  CORALIE \\
				56241.387451  &  $-$10.05  &  0.11  &  HARPS \\
				56242.380878  &  $-$2.81  &  0.11  &  HARPS \\
				56448.015262  &  24.85  &  0.16  &  HARPS \\
				56449.007513  &  7.67  &  0.14  &  HARPS \\
				56450.007856  &  $-$2.92  &  0.13  &  HARPS \\
				56876.979793  &  5.44  &  0.12  &  HARPS \\
				56876.979793  &  5.44  &  0.12  &  HARPS \\
				56877.984082  &  5.75  &  0.12  &  HARPS \\
				56877.984082  &  5.75  &  0.12  &  HARPS \\
				56878.984443  &  5.64  &  0.12  &  HARPS \\
				56878.984443  &  5.64  &  0.12  &  HARPS \\
				57902.192804  &  5.01  &  0.12  &  HARPS \\
				57916.029102  &  29.48  &  0.16  &  HARPS \\
				58098.313410  &  $-$0.92  &  0.12  &  HARPS \\
				58139.324547  &  $-$13.92  &  0.14  &  HARPS \\
				58140.194042  &  $-$26.83  &  0.13  &  HARPS \\
				58141.169979  &  $-$27.44  &  0.12  &  HARPS \\
				\hline
			\end{tabular}
			\label{table_smus_rv}
		\end{table}	
		
		\begin{table}[]
			\centering
			\caption{Radial velocity measurements of S~Nor.}
			\begin{tabular}{cccc} 
				\hline
				\hline
				MJD		&	RV	&	$\sigma_\mathrm{RV}$	&  Inst.	\\
				(days)			&	($\mathrm{km~s^{-1}}$)	&	($\mathrm{km~s^{-1}}$)								&		 \\
				\hline
				56399.195452  &  $-$1.50  &  0.10  &  CORALIE \\
				56400.190823  &  8.14  &  0.10  &  CORALIE \\
				56461.180856  &  24.03  &  0.15  &  CORALIE \\
				56462.178882  &  8.36  &  0.13  &  CORALIE \\
				56750.225028  &  $-$2.69  &  0.11  &  CORALIE \\
				56750.237714  &  $-$2.58  &  0.11  &  CORALIE \\
				56750.250377  &  $-$2.45  &  0.11  &  CORALIE \\
				56751.221880  &  7.03  &  0.10  &  CORALIE \\
				56751.234543  &  7.15  &  0.10  &  CORALIE \\
				56751.247229  &  7.29  &  0.10  &  CORALIE \\
				56826.200692  &  $-$2.44  &  0.11  &  CORALIE \\
				56826.212405  &  $-$2.46  &  0.11  &  CORALIE \\
				56826.224175  &  $-$2.47  &  0.11  &  CORALIE \\
				56827.201646  &  $-$8.98  &  0.12  &  CORALIE \\
				56827.214366  &  $-$8.99  &  0.12  &  CORALIE \\
				56827.227028  &  $-$8.99  &  0.12  &  CORALIE \\
				57135.196163  &  9.19  &  0.13  &  CORALIE \\
				57136.152244  &  $-$2.20  &  0.12  &  CORALIE \\
				57137.150896  &  $-$2.54  &  0.11  &  CORALIE \\
				57198.201542  &  $-$8.07  &  0.11  &  CORALIE \\
				56210.992794  &  $-$2.18  &  0.11  &  HARPS \\
				56448.031120  &  $-$0.84  &  0.10  &  HARPS \\
				56449.016947  &  8.78  &  0.10  &  HARPS \\
				56449.996498  &  18.04  &  0.12  &  HARPS \\
				57634.991198  &  $-$2.03  &  0.12  &  HARPS \\
				57635.989614  &  $-$2.72  &  0.12  &  HARPS \\
				57901.123581  &  $-$3.98  &  0.11  &  HARPS \\
				57901.223692  &  $-$3.10  &  0.10  &  HARPS \\
				57902.164485  &  5.89  &  0.10  &  HARPS \\
				57902.202431  &  6.27  &  0.10  &  HARPS \\
				57903.220158  &  16.09  &  0.11  &  HARPS \\
				57916.106421  &  1.25  &  0.13  &  HARPS \\
				58139.373600  &  20.48  &  0.15  &  HARPS \\
				58140.366871  &  2.52  &  0.13  &  HARPS \\
				58141.363640  &  $-$3.10  &  0.11  &  HARPS \\
				58146.310875  &  8.74  &  0.10  &  HARPS \\
				58148.394062  &  24.46  &  0.15  &  HARPS \\
				\hline
			\end{tabular}
			\label{table_snor_rv}
		\end{table}	

		\begin{table}[]
	\centering
	\caption{Radial velocity measurements of W~Sgr.}
	\begin{tabular}{cccc} 
		\hline
		\hline
		MJD		&	RV	&	$\sigma_\mathrm{RV}$	&  Inst.	\\
		(days)			&	($\mathrm{km~s^{-1}}$)	&	($\mathrm{km~s^{-1}}$)								&		 \\
		\hline
		56877.006483  &  $-$19.53  &  0.14  &  HARPS \\
		56878.004557  &  $-$42.93  &  0.12  &  HARPS \\
		56879.003313  &  $-$38.79  &  0.11  &  HARPS \\
		56908.139825  &  $-$41.83  &  0.12  &  HARPS \\
		56399.247308  &  $-$41.59  &  0.12  &  CORALIE \\
		56400.243681  &  $-$40.64  &  0.11  &  CORALIE \\
		56461.166435  &  $-$39.31  &  0.11  &  CORALIE \\
		56462.164196  &  $-$34.06  &  0.10  &  CORALIE \\
		56750.314961  &  $-$36.06  &  0.11  &  CORALIE \\
		56750.320587  &  $-$36.02  &  0.10  &  CORALIE \\
		56751.312083  &  $-$30.64  &  0.10  &  CORALIE \\
		56751.317686  &  $-$30.60  &  0.10  &  CORALIE \\
		56826.284416  &  $-$36.11  &  0.10  &  CORALIE \\
		56826.289832  &  $-$36.08  &  0.10  &  CORALIE \\
		56827.293751  &  $-$30.59  &  0.10  &  CORALIE \\
		56827.299353  &  $-$30.55  &  0.10  &  CORALIE \\
		57135.248675  &  $-$20.91  &  0.13  &  CORALIE \\
		57136.206267  &  $-$44.04  &  0.12  &  CORALIE \\
		57137.204904  &  $-$40.25  &  0.11  &  CORALIE \\
		57198.274594  &  $-$38.56  &  0.11  &  CORALIE \\
		57568.076693  &  $-$18.37  &  0.12  &  HERMES \\
		57569.063800  &  $-$45.39  &  0.11  &  HERMES \\
		57572.099017  &  $-$31.47  &  0.10  &  HERMES \\
		57573.116228  &  $-$28.28  &  0.10  &  HERMES \\
		57574.084807  &  $-$15.81  &  0.11  &  HERMES \\
		57627.855429  &  $-$8.93  &  0.12  &  HERMES \\
		57628.863024  &  $-$19.42  &  0.12  &  HERMES \\
		57629.850733  &  $-$45.49  &  0.11  &  HERMES \\
		57630.860779  &  $-$41.76  &  0.11  &  HERMES \\
		57635.845838  &  $-$7.05  &  0.13  &  HERMES \\
		57999.854550  &  $-$8.23  &  0.12  &  HERMES \\
		58001.862346  &  $-$42.47  &  0.11  &  HERMES \\
		58002.855074  &  $-$40.57  &  0.11  &  HERMES \\
		58026.818757  &  $-$34.08  &  0.10  &  HERMES \\
		58239.173008  &  $-$34.66  &  0.10  &  HERMES \\
		58246.104342  &  $-$38.20  &  0.11  &  HERMES \\
		58247.170152  &  $-$32.77  &  0.10  &  HERMES \\
		58248.155015  &  $-$28.62  &  0.10  &  HERMES \\
		\hline
	\end{tabular}
	\label{table_wsgr_rv}
\end{table}	

		\begin{table}[]
			\centering
			\caption{Radial velocity measurements of V350~Sgr.}
			\begin{tabular}{cccc} 
				\hline
				\hline
				MJD		&	RV	&	$\sigma_\mathrm{RV}$	&  Inst.	\\
				(days)			&	($\mathrm{km~s^{-1}}$)	&	($\mathrm{km~s^{-1}}$)								&		 \\
				\hline
				56399.281661  &  10.44  &  0.11  &  CORALIE \\
				56400.278886  &  20.66  &  0.12  &  CORALIE \\
				56461.344731  &  18.44  &  0.11  &  CORALIE \\
				56461.394616  &  18.99  &  0.11  &  CORALIE \\
				56462.385830  &  28.46  &  0.12  &  CORALIE \\
				56750.333340  &  23.09  &  0.11  &  CORALIE \\
				56750.349846  &  23.26  &  0.11  &  CORALIE \\
				56751.330291  &  32.30  &  0.13  &  CORALIE \\
				56751.346797  &  32.48  &  0.13  &  CORALIE \\
				56826.312569  &  6.92  &  0.11  &  CORALIE \\
				56826.329097  &  7.07  &  0.11  &  CORALIE \\
				56827.322199  &  17.69  &  0.11  &  CORALIE \\
				56827.338704  &  17.86  &  0.11  &  CORALIE \\
				57135.291045  &  $-$3.00  &  0.12  &  CORALIE \\
				57136.258205  &  6.61  &  0.11  &  CORALIE \\
				57136.279295  &  6.83  &  0.11  &  CORALIE \\
				57137.256942  &  16.85  &  0.12  &  CORALIE \\
				57137.278020  &  17.04  &  0.12  &  CORALIE \\
				57198.290947  &  7.36  &  0.11  &  CORALIE \\
				56877.010494  &  $-$0.99  &  0.13  &  HARPS \\
				56878.009348  &  7.28  &  0.12  &  HARPS \\
				56879.007162  &  17.93  &  0.12  &  HARPS \\
				56908.010120  &  $-$1.39  &  0.13  &  HARPS \\
				56909.004039  &  7.05  &  0.12  &  HARPS \\
				56910.020588  &  18.00  &  0.12  &  HARPS \\
				\hline
			\end{tabular}
			\label{table_v350sgr_rv}
		\end{table}	
		
		\begin{table}[]
			\centering
			\caption{Radial velocity measurements of V636~Sco.}
			\begin{tabular}{cccc} 
				\hline
				\hline
				MJD		&	RV	&	$\sigma_\mathrm{RV}$	&  Inst.	\\
				(days)			&	($\mathrm{km~s^{-1}}$)	&	($\mathrm{km~s^{-1}}$)								&		 \\
				\hline
				56876.996143  &  16.46  &  0.13  &  HARPS \\
				56877.977498  &  26.24  &  0.14  &  HARPS \\
				56878.977834  &  29.78  &  0.16  &  HARPS \\
				56908.004614  &  4.81  &  0.14  &  HARPS \\
				56909.009721  &  8.97  &  0.13  &  HARPS \\
				56910.013893  &  15.28  &  0.13  &  HARPS \\
				56399.218154  &  $-$8.14  &  0.12  &  CORALIE \\
				56400.213531  &  $-$1.90  &  0.12  &  CORALIE \\
				56461.202289  &  $-$4.18  &  0.12  &  CORALIE \\
				56462.255361  &  0.20  &  0.12  &  CORALIE \\
				56750.265874  &  9.22  &  0.14  &  CORALIE \\
				56750.280933  &  8.82  &  0.14  &  CORALIE \\
				56750.295981  &  8.43  &  0.14  &  CORALIE \\
				56751.262801  &  $-$8.49  &  0.13  &  CORALIE \\
				56751.277952  &  $-$8.55  &  0.13  &  CORALIE \\
				56751.293081  &  $-$8.60  &  0.13  &  CORALIE \\
				56826.239611  &  $-$2.58  &  0.13  &  CORALIE \\
				56826.252690  &  $-$2.60  &  0.13  &  CORALIE \\
				56826.265907  &  $-$2.61  &  0.13  &  CORALIE \\
				56827.244015  &  0.45  &  0.12  &  CORALIE \\
				56827.259130  &  0.54  &  0.12  &  CORALIE \\
				56827.274188  &  0.62  &  0.12  &  CORALIE \\
				57135.218268  &  21.59  &  0.12  &  CORALIE \\
				57136.176623  &  30.80  &  0.13  &  CORALIE \\
				57137.175246  &  36.07  &  0.15  &  CORALIE \\
				57198.224645  &  34.78  &  0.15  &  CORALIE \\
				\hline
			\end{tabular}
			\label{table_v636sco_rv}
		\end{table}	

		\begin{table}[]
	\centering
	\caption{Radial velocity measurements of AH~Vel.}
	\begin{tabular}{cccc} 
		\hline
		\hline
		MJD		&	RV	&	$\sigma_\mathrm{RV}$	&  Inst.	\\
		(days)			&	($\mathrm{km~s^{-1}}$)	&	($\mathrm{km~s^{-1}}$)								&		 \\
		\hline
		56398.967796  &  19.49  &  0.12  &  CORALIE \\
		56399.103520  &  17.99  &  0.12  &  CORALIE \\
		56399.976450  &  17.59  &  0.12  &  CORALIE \\
		56460.953595  &  32.87  &  0.12  &  CORALIE \\
		56460.963630  &  32.88  &  0.12  &  CORALIE \\
		56461.958118  &  26.45  &  0.12  &  CORALIE \\
		56461.963245  &  26.38  &  0.12  &  CORALIE \\
		56749.987206  &  18.45  &  0.12  &  CORALIE \\
		56749.993907  &  18.40  &  0.12  &  CORALIE \\
		56750.987012  &  18.73  &  0.12  &  CORALIE \\
		56750.993713  &  18.78  &  0.12  &  CORALIE \\
		56825.955373  &  19.75  &  0.12  &  CORALIE \\
		56825.962074  &  19.67  &  0.12  &  CORALIE \\
		56826.952059  &  17.88  &  0.12  &  CORALIE \\
		56826.958703  &  17.92  &  0.12  &  CORALIE \\
		57134.970601  &  17.21  &  0.12  &  CORALIE \\
		57135.972656  &  21.18  &  0.11  &  CORALIE \\
		57136.971725  &  30.66  &  0.12  &  CORALIE \\
		56213.389964  &  16.37  &  0.12  &  HARPS \\
		56240.372796  &  26.41  &  0.12  &  HARPS \\
		56242.391438  &  21.43  &  0.13  &  HARPS \\
		56448.036297  &  31.24  &  0.13  &  HARPS \\
		56450.018720  &  16.99  &  0.13  &  HARPS \\
		56605.358090  &  31.18  &  0.13  &  HARPS \\
		56606.176535  &  18.93  &  0.12  &  HARPS \\
		56910.399809  &  21.31  &  0.13  &  HARPS \\
		\hline
	\end{tabular}
	\label{table_ahvel_rv}
\end{table}

	\end{appendix}
	
\end{document}